\documentclass[pra,twocolumn,superscriptaddress,notitlepage,aps, floatfix]{revtex4-2}
\clearpage
\usepackage{etoolbox}

\pdfoutput=1
\usepackage{makeidx}
\usepackage[utf8]{inputenc}
\usepackage{graphicx}
\usepackage{amsmath}
\usepackage{amssymb}
\usepackage{mathrsfs}
\usepackage{dcolumn}

\usepackage{bbm}
\usepackage[dvipsnames]{xcolor}
\usepackage{esint}
\usepackage{lineno}
\usepackage{booktabs}

\usepackage{microtype}

\usepackage[english]{babel}
\usepackage{blindtext}
\blindmathtrue

\usepackage{xcolor}
\usepackage{ulem}

\usepackage{booktabs,makecell}

\usepackage{orcidlink}

\newcommand{\bra}[1]{\langle #1|}
\newcommand{\ket}[1]{|#1 \rangle}

\newcommand{\ketbra}[2]{|#1\rangle\langle#2|}

\usepackage{bm}
\newcommand{\norm}[1]{\left\lVert#1\right\rVert}

\newcommand{\figref}[1]{Fig.~\ref{#1}}
\renewcommand{\eqref}[1]{Eq.~(\ref{#1})}

\newcommand{\coloneq}{:=}

\begin{document}

\title{Transient entanglement generation in driven chiral networks beyond the secular approximation}

\author{Yan Xi Foo \orcidlink{0009-0001-0811-4673}}
\affiliation{Quantum Science and Engineering Centre,
Nanyang Technological University, Singapore 639798}

\author{Kian Hwee Lim \orcidlink{0000-0003-2154-4288}}
\affiliation{Centre for Quantum Technologies, National University of Singapore, Singapore 117543}

\author{Jia-Bin You \orcidlink{0000-0001-8815-1855}}
\affiliation{Institute of High Performance Computing, A*STAR (Agency for Science, Technology and Research), Singapore 138632}

\author{Leong Chuan Kwek}
\affiliation{Centre for Quantum Technologies, National University of Singapore, Singapore 117543}
\affiliation{National Institute of Education, Nanyang Technological University, Singapore 637616}

\author{Davit Aghamalyan}
\affiliation{SMT Pillar, Singapore University of Technology and Design, Singapore 487372}

\begin{abstract}

We study transient entanglement generation between two quantum nodes coupled through a chiral one-dimensional channel. In an emitter-only Born-Markov description, we show that continuous driving and an initial ground state can raise the maximum transient concurrence above the undriven $2/e$ benchmark associated with the effectively single-excitation model \cite{gonzalez2015chiral}. We then consider a more microscopic XX spin-chain channel with triangular plaquette couplings and compare a nonsecular time-convolutionless master equation (TCL-ME) with matrix-product-state (MPS) simulations. In the optimal driven regime, the nonsecular TCL-2 treatment reproduces the concurrence envelope and first transient peak qualitatively, while the remaining discrepancy is mainly attributable to beyond-Born system-bath correlations. The enhancement is traced to the failure of the secular approximation under strong driving, where nearby dressed transitions are not well separated on the dissipative timescale and nonsecular terms mix dressed-state coherences. Finally, we examine within TCL-2 the sensitivity of the protocol to positional disorder, imperfect chirality, and loss into nonguided modes. These results clarify when the familiar $2/e$ limitation ceases to apply and separate the roles of secular breakdown, Born-factorization error, and reduced-state memory in driven chiral entanglement generation; we believe that our study contributes to one of the first studies where the breakdown of the secular approximation is useful rather than detrimental. 

\end{abstract}

\maketitle

\section{Introduction}

Rapid long-distance entanglement generation between remote quantum nodes is a basic primitive for fundamental tasks in quantum computing, such as quantum teleportation and quantum key distribution, and it can also be leveraged for high-precision quantum measurements ~\cite{nielsen2002quantum,giovannetti2011advances}. In one-dimensional quantum networks, such nodes exchange quantum information through shared quantum channels which play the role of a
``quantum bus''~\cite{kimble2008quantum,cirac1997quantum,duan2001long}. Examples of this include information exchange via emission of optical, microwave, or acoustic waves propagating through a fiber~\cite{orieux2016recent}; superconducting
striplines~\cite{blais2004cavity} can also play the role of a waveguide. Alternatively, nodes in a quantum network can be coupled to an engineered spin chain~\cite{masson2020atomic,ramos2014quantum} which plays the role of a ``waveguide'' for magnons, thus mediating long-distance interactions between these
nodes~\cite{bose2007quantum, coish2008exponential, PhysRevA.57.120,
nikolopoulos2013quantum}. 

An outstanding challenge in this domain has been to design robust and high-fidelity entanglement generation protocols. One approach towards this goal relies on the use of chiral quantum
networks ~\cite{pichler2015quantum,lodahl2017chiral}, where nodes are connected via directional quantum channels. For instance, in the framework of waveguide quantum electrodynamics (QED), chirality emerges when the system's coupling to left- and right-propagating waveguide modes is asymmetric. Chiral light-matter interaction has played an important role in devising new quantum networks, where exotic many-body interactions arise ~\cite{pichler2015quantum,ramos2016non} due to control over the directionality of photon emission and absorption. This is particularly viable in sub-wavelength-diameter optical fibers~\cite{lodahl2017chiral,sollner2015deterministic,mahmoodian2016quantum,hauff2022chiral,petersen2014chiral,scheucher2016quantum,liedl2024observation,le2017nanofiber}, where strong
transverse confinement induces selection rules between the local polarization of light and the propagation direction. Many experiments have reported chiral systems with high directionality \cite{sollner2015deterministic,mahmoodian2016quantum,hauff2022chiral,petersen2014chiral,scheucher2016quantum,liedl2024observation}. Photonic crystals have turned out to be particularly promising, with directionality as high as $\sim 90\%$ being demonstrated in such devices \cite{sollner2015deterministic,mahmoodian2016quantum,hauff2022chiral}; decay into non-waveguide modes and similar parasitic losses can also be minimised by exploiting photonic bandgap effects \cite{arcari2014near}, leading to a near-unity $\beta$-factor (the ratio of the emission rate into the waveguide modes to the total emission rate).

As shown in Refs.~\cite{pichler2015quantum,ramos2016non}, chiral quantum networks need not be restricted to photonic waveguides: an engineered spin chain can play the role of a one-dimensional waveguide for magnons, while at the same time providing a microscopic and structured reservoir. Spin chains are therefore especially useful for the present work. Besides offering a tunable route to chirality through plaquette couplings pierced by a synthetic flux, they allow one to go beyond an emitter-only Born--Markov description and test explicitly the roles of finite bandwidth, propagation delay, and system--bath correlations. Chiral channels of this kind are already known to enhance transient entanglement generation relative to bidirectional settings~\cite{gonzalez2015chiral,mok2020microresonators,you2025generation,lim2024exponentially}, and related architectures have also been explored for quantum-state-transfer and entanglement-transfer protocols~\cite{mok2020microresonators,mok2020long}.

Motivated by this, we use the spin chain not merely as an alternative implementation of chirality, but as a microscopic model with which to test which features of transient entanglement generation survive beyond the usual emitter-only approximations. The familiar $2/e$ concurrence benchmark~\cite{gonzalez2015chiral} is derived for an undriven, effectively single-excitation, emitter-only Born-Markov description. Our starting point is therefore deliberately narrower than a search for a new universal bound: we ask instead which part of the $2/e$ limitation is tied to that specific approximation hierarchy, and which part persists once driving, structured baths, and explicit system-bath correlations are admitted.

The paper is organized around a distinction between three effects that are easy to conflate: breakdown of the secular approximation, breakdown of the Born approximation, and reduced-state non-Markovianity. In the strong-driving regime, nearby dressed transitions need not be well separated on the dissipative timescale, so non-secular coherence-population coupling terms can become important. At the same time, in a microscopic spin-chain realization, system-bath correlations can build up and invalidate Born factorization even when standard reduced dynamics witnesses still indicate behaviour close to Markovian.

We therefore proceed in two stages. Firstly, within the emitter-only Born--Markov master equation, we lift the single-excitation restriction and show that continuous driving can push the first transient concurrence above the undriven $2/e$ benchmark. Second, we turn to a microscopic spin-chain model and compare non-secular TCL-2 dynamics against MPS simulations in order to determine which features of that enhancement survive beyond the emitter-only reduction. Of note, the spin-chain model does not simply add back Born corrections but also changes the bath structure and coupling geometry, so its comparison with $2/e$ is qualitative rather than literal. Table~\ref{tab:benchmarkscope} summarizes the model hierarchy studied in this work, the controls optimized in each case, and the corresponding best transient concurrence values.

\begin{table*}[t]
\centering
\small
\renewcommand{\arraystretch}{1.15}

\begin{tabular*}{\textwidth}{@{\extracolsep{\fill}}lllll@{}}
\toprule
\makecell[l]{Model} &
\makecell[l]{Initial\\state} &
\makecell[l]{Controls\\optimized} &
\makecell[l]{Main assumptions /\\new ingredients} &
\makecell[l]{Best reported\\concurrence} \\
\midrule
\makecell[l]{Undriven chiral\\emitter-only model}
& $\ket{eg}$
& none
& \makecell[l]{Born--Markov, effectively\\single-excitation, no continuous driving}
& \makecell[l]{$2/e \approx 0.74$,\\see \cite{gonzalez2015chiral}} \\

\makecell[l]{Driven\\emitter-only model}
& $\ket{gg}$
& $\Omega_1,\Omega_2$
& \makecell[l]{Born--Markov, continuous driving,\\population leakage into $\ket{ee}$ allowed}
& $\approx 0.77$ \\

\makecell[l]{Microscopic spin-chain\\TCL-2 model}
& $\ket{gg}$
& $\Omega_1,\Omega_2,g_1,g_2$
& \makecell[l]{Structured finite-bandwidth bath,\\two-point plaquette coupling,\\non-secular dynamics}
& $\approx 0.78$ \\

\makecell[l]{Microscopic spin-chain\\MPS model}
& $\ket{gg}$
& \makecell[l]{Same set as\\TCL-2 comparison}
& \makecell[l]{Explicit bath evolution,\\beyond-Born correlations included}
& \makecell[l]{$\approx 0.80$,\\see Fig.~\ref{fig:fig6}} \\
\bottomrule
\end{tabular*}

\caption{Scope of the concurrence benchmarks discussed in this work, with differing bath properties and coupling being incorporated in each model.}
\label{tab:benchmarkscope}
\end{table*}

The paper is organized as follows. In Sec. \ref{section: ME}, we will first explain the chiral network model and give a detailed master equation description of the quantum nodes within the Born-Markov approximation. In Sec.
\ref{section: othermethods}, we first analyze the driven transient regime within that emitter-only description and then turn to a microscopic spin-chain model treated using TCL-2 and MPS methods in Sec. \ref{sec: spin-chain}. In Sec.~\ref{sec: robustness}, we investigate the sensitivity of the protocol to experimentally relevant imperfections, including positional disorder, detuning fluctuations, imperfect chirality, and loss into nonguided modes. Finally, in Sec.~\ref{section: conclusions}, we summarize the main results and discuss natural extensions.

\section{Master equation for the two emitters coupled to a 1D quantum network}
\label{section: ME}

\begin{figure}
  \centering
  \includegraphics[width=0.9\linewidth]{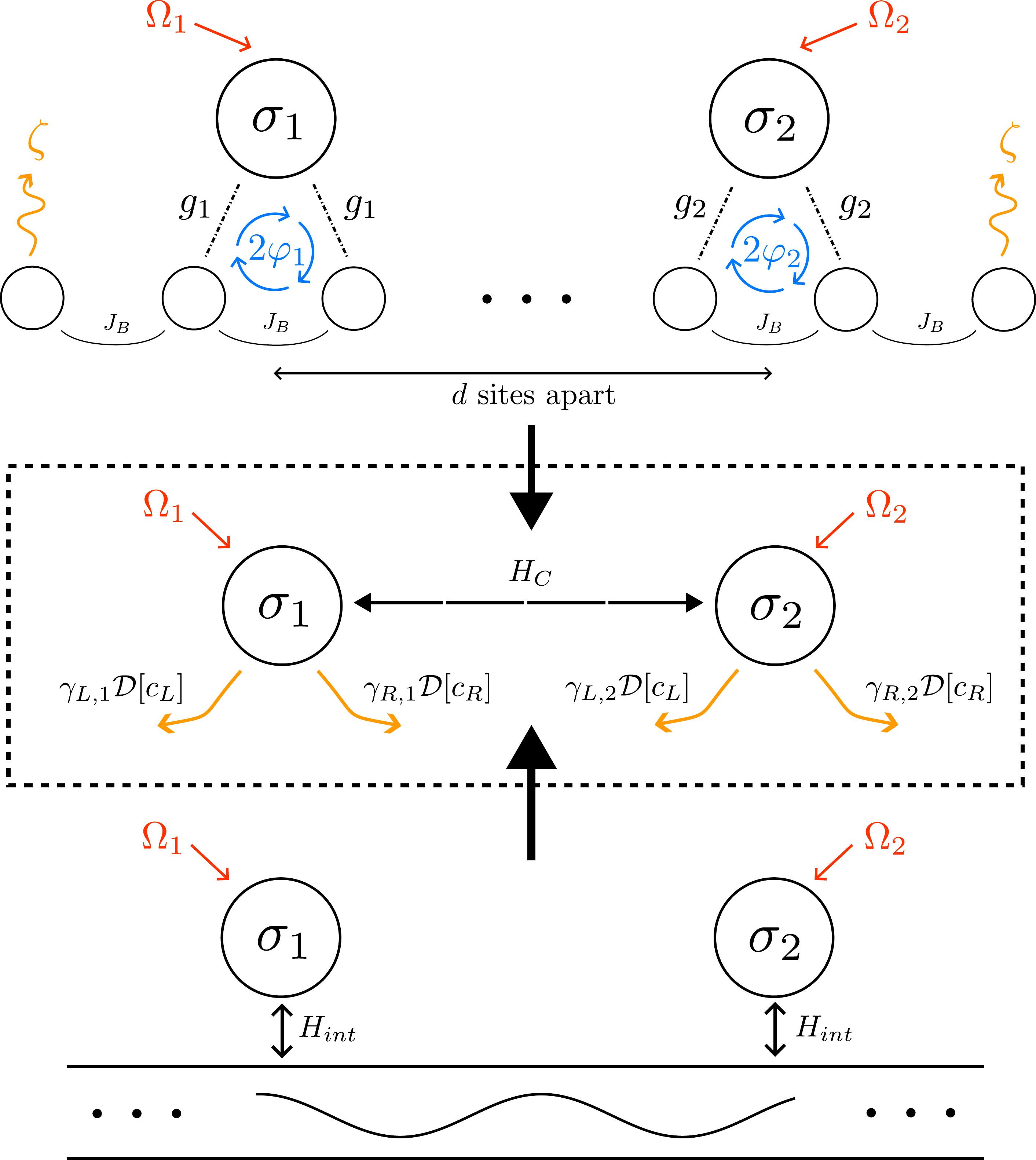}
  \caption
  {Two possible realizations of a quantum one-dimensional chiral network. In the upper panel, the channel is an engineered spin chain, where the boundaries of the chain are subject to a dissipative decay rate $\zeta$. Chirality is generated by coupling each node $i\in \{1,2\}$ to two neighbouring bath spins by coupling strength $g_i$ so as to form a triangular plaquette threaded by a synthetic magnetic flux $2\varphi$~\cite{ramos2016non}. Tracing out the spin chain within the Born--Markov approximation yields directional decay rates $\gamma_{L,i}$ and $\gamma_{R,i}$, both controlled by $\varphi_{i}$, together with a spin-chain-mediated coherent interaction $H_C$. In the lower panel, the channel is a chiral waveguide, and tracing out the waveguide yields the corresponding waveguide-mediated interaction $H_C$. Throughout, we set the Lamb-shifted detuning between the quantum nodes and the spin-chain or waveguide modes to zero and allow for coherent driving at rate $\Omega_i$.}
  \label{fig: microscopic description of model}
\end{figure}

For the rest of the paper, we take $\hbar = 1$. As shown in \figref{fig:
microscopic description of model}, two possible realizations of our quantum
1D chiral network are two two-level emitters coupled to a 1D chiral
waveguide or to a 1D chiral spin chain. Since the 1D chiral waveguide and
the 1D chiral spin chain have more degrees of freedom than all of the
two-level emitters,
the master equations for just the emitters (which are the quantum nodes in
our network) have been derived in both cases under the Born-Markov and rotating-wave assumptions by treating the waveguide or the
spin chain as the ``bath'' and tracing it
out~\cite{ramos2016non,pichler2015quantum}. The resultant master equations for both cases have
the same structure and take the general form
\begin{subequations}
\label{MasEq}
\begin{align}
  \dot{\rho}_{S}=&-i[H_{S}+H_{C} + H_{\text{drive}},\rho_{S}] \nonumber \\
  &+\gamma_{R}\mathcal{D}[c_{R}]\rho_{S}+\gamma_{L}\mathcal{D}[c_{L}]\rho_{S} \\
  c_{L}=&\sigma_{1}+e^{i \phi}\sigma_{2} \\
  c_{R}=&\sigma_{1}+e^{-i \phi}\sigma_{2},
\end{align}
\end{subequations}
where $\mathcal{D}[c]\rho_S \equiv c\rho_S c^\dagger - (\rho_S c^\dagger c +
c^\dagger c \rho_S)/2$ and where $\sigma_i$ is the lowering
operator for the $i$-th two-level emitter. With reference to \figref{fig:
microscopic description of model}, $\phi$ is the phase picked up by the
propagating photon or magnon in the waveguide or the spin chain as it travels
from one emitter to another emitter. With $k$ as the wavenumber of the
propagating photon or magnon, in the spin chain case, we have $\phi = kd$
where $d$ is the number of sites between the two emitters and in the
waveguide case, we have $\phi = k |x_2 - x_1|$ where $|x_2 - x_1|$ is the
real space distance between the two emitters. Hence, we see that $\gamma_L
\mathcal{D}[c_L] \rho_S$ and $\gamma_R \mathcal{D}[c_R] \rho_S$ model the
decay of the emitter excitations into a left-propagating mode and a
right-propagating mode respectively of the waveguide or the spin chain. Note that in the spin chain case, $\gamma_{L(R)}$ are not explicitly given but are functions of gauge flux $\varphi$; in the fully on-resonance case, this obtains a simple expression: $\gamma_{i,R/L} = \frac{\gamma_{tot}}{2}[1\pm \sin2\varphi_i]$ \cite{ramos2014quantum} where $\gamma_{tot} = \gamma_L + \gamma_R + \gamma_{loss}$, the last term of which refers to leakage into nonguided modes and is taken to be zero until Sec. \ref{sec: robustness}.

The
coherent waveguide or spin-chain mediated interaction of the two emitters
with each other is described by the chiral Hamiltonian $H_C$ which takes the
form
\begin{equation}
   H_{C} = \frac{i}{2}(\gamma_{R}e^{-i\phi}-\gamma_{L}e^{i\phi})\sigma_{1}^{\dag}\sigma_{2}+\text{H.c}
  \end{equation}
where $\phi$ here has the same meaning as in does in $c_L$ and $c_R$. We also
drive both emitters with a driving Hamiltonian $H_{\text{drive}}$, which
takes the form
\begin{equation}
  H_\text{drive} = \frac{1}{2}\sum_{\alpha=1}^2 \Omega \sigma_\alpha +
  \Omega^* \sigma_\alpha^\dagger
\end{equation}

Finally, $H_S = \tilde{\Delta}\sum_{\alpha=1}^2 \sigma_{\alpha}^\dagger \sigma_\alpha$, where $\tilde{\Delta}$ is the Lamb-shift-renormalized detuning between the emitters and the channel modes. Its precise form depends on whether the channel is a waveguide or a spin chain.

In what follows, we set $\tilde{\Delta}=0$. Physically, this corresponds to working near the center of the band, where the relevant portion of the spectrum can be treated as effectively flat over the emitter linewidth. Under this assumption, the waveguide and spin-chain realizations can be discussed within the same effective emitter-only framework.

\section{Driven transient entanglement in the Born-Markov model}
\label{section: othermethods}

In this section, we study the scenario where the chiral channel is realised by a photonic waveguide and proceed in three steps: we first review the undriven benchmark, then analyze the driven emitter-only model, and finally examine which features survive in a more microscopic spin-chain treatment.

\subsection{Analytical results for the concurrence in the undriven and weakly-driven Born-Markov regimes}
In the undriven and weakly driven regimes, we can neglect jump terms in the
master equation in \eqref{MasEq} and hence derive some analytical results by
working in the single-excitation subspace using a non-Hermitian Hamiltonian
approach.  More details, including derivations, can be found in
Appendix~\ref{appendix: detailed calculations for undriven and weakly
driven}.

For the case of two undriven emitters, we obtain the concurrence $C(t)$
\begin{equation}
\label{concND}
C(t)=e^{-2\bar\gamma t}\sqrt{\frac{\gamma_{R}}{\gamma_{L}}}\sqrt{\left [\sinh^{2}{\alpha_{1}t}+ \sin^{2}{\alpha_{2}t}          \right ]},
\end{equation}
where $\bar\gamma = (\gamma_L + \gamma_R)/2$, $\alpha_{1}=2\sqrt{\gamma_{L}\gamma_{r}}\cos{kd}$ and
$\alpha_{2}=2\sqrt{\gamma_{L}\gamma_{r}}\sin{kd}$, the result of which was first obtained in \cite{gonzalez2015chiral}. In Appendix \ref{appendix: detailed calculations for undriven and weakly driven}, we present an alternative derivation. We also note that the limit
$\lim_{\gamma_{L} \to 0}C(t)$ exists and is given by the following
expression:
\begin{equation}
C(t)=2\gamma_{R}t e^{-2\bar\gamma t}
\end{equation}
This expression shows that at the time $t_{max}=1/\gamma_{R}$, the concurrence reaches its maximum:
\begin{equation}
    C_{max}=\frac{2}{e}
\end{equation}

\begin{figure}
    \centering
    \includegraphics[width=0.95\linewidth]{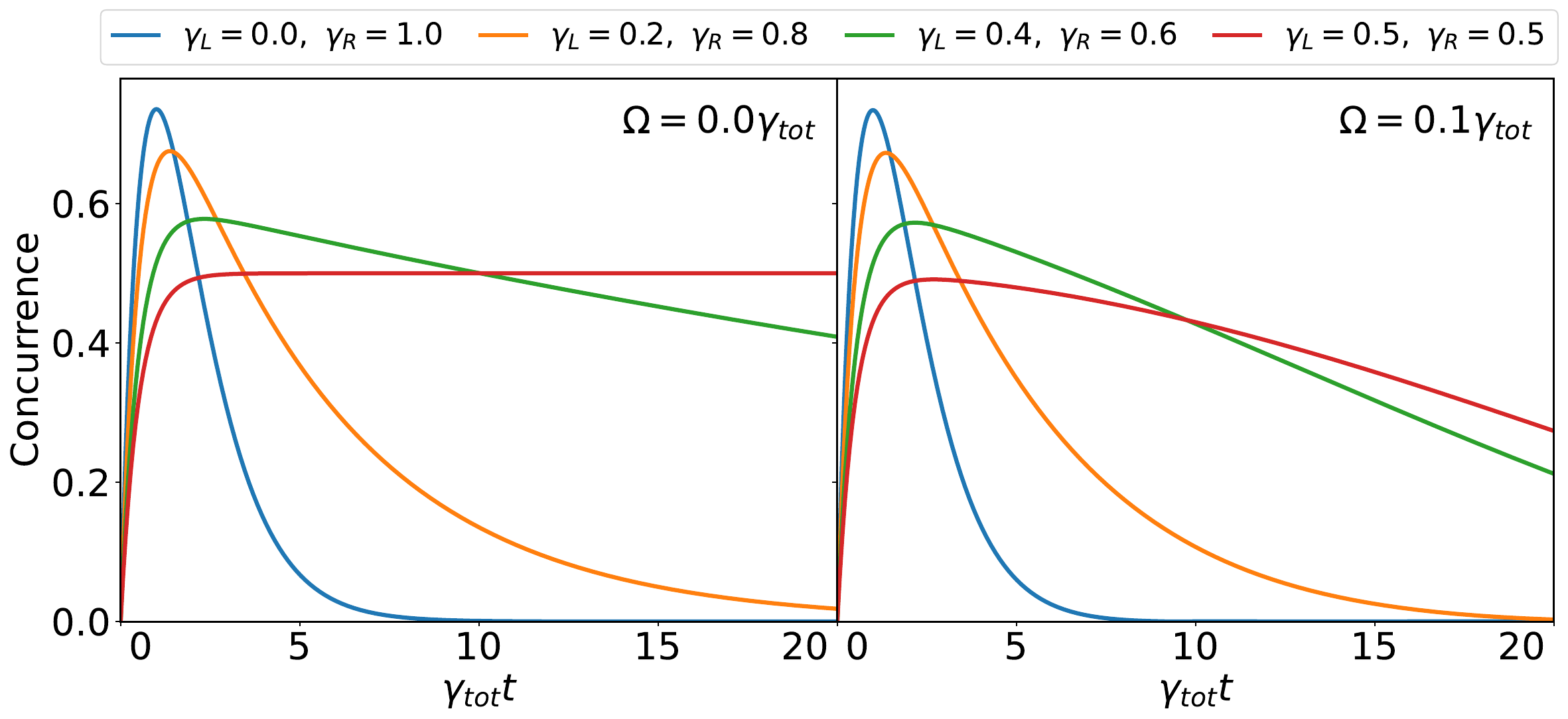}
    \caption{Concurrence $C(t)$ as a function of time obtained from the exact-root solution for four guided-dissipation configurations, $(\gamma_L,\gamma_R)=(0,1)$, $(0.2,0.8)$, $(0.4,0.6)$, and $(0.5,0.5)$, with $\gamma_{tot} \coloneq \gamma_L+\gamma_R=1$. The left panel shows the undriven case, $\Omega=0$, and the right panel the weakly driven case, $\Omega=0.1$. The effects of weak driving are only appreciable for the near-bidirectional regimes beyond the first entanglement bump. Note that the results in (left) coincide with \eqref{concND}.}
    \label{fig:Torr}
\end{figure}

For the case of weakly driven emitters, we have the following analytic results from which \figref{fig:Torr} is derived.
Firstly, for the case of a single weakly driven atom in the perfectly chiral
regime, we have the following expression for the concurrence $C(t)$,
\begin{align}
C(t)=&2|c_{eg}(t)c^{*}_{ge}(t)| \nonumber \\
=&\bigg| 2\gamma_{R}e^{-\bar\gamma t} \left[1-\frac{2\bar\gamma^{2}}{\Omega^2} \right] \left [\frac{1}{2b}\sinh{2bt}-\frac{a}{b^2}\sinh^{2}{bt} \right ] \nonumber \\
&-\frac{8\gamma_{R}\bar\gamma}{\Omega^2}e^{-\frac{3\bar\gamma}{2}t}\left[\cosh{bt}-\frac{a}{b}\sinh{bt}\right ]  \nonumber \\
&+\frac{8\gamma_{R}\bar\gamma}{\Omega^2}e^{-\bar\gamma t}\left [\cosh^{2}{bt}-\frac{a}{2b}\sinh{2bt}\right ] \bigg|,
\end{align}
where $a=\bar\gamma/2$ and $b=\sqrt{(\bar\gamma/2)^{2}-(\Omega/2)^2}$.
At $t=0$, the concurrence vanishes, as expected for an initially separable two-qubit state. In the zero-driving limit, the expression reduces to the familiar value $2/e$.

Second, for a weakly driven atom in the imperfectly chiral regime, the concurrence $C(t)$ takes the form 
\begin{equation}
    C(t)=2|c_{eg}(t)c^{*}_{ge}(t)|
\end{equation}
where 
\begin{align}
c_{eg}(t)=&e^{\tilde{s}_2 t}
+\left[\frac{\tilde{s}_{1}+\bar\gamma}{\tilde{s}_{1}-\tilde{s}_{2}}\right]
\left[e^{\tilde{s}_{1} t}-e^{\tilde{s}_{2} t}\right] \nonumber\\
&+\left[\frac{\tilde{s}_{0}(\tilde{s}_{1}+\bar\gamma)}
{(\tilde{s}_{1}-\tilde{s}_{2})(\tilde{s}_{0}-\tilde{s}_{1})}\right]
\left[e^{\tilde{s}_{0} t}-e^{\tilde{s}_{1} t}\right] \nonumber\\
&-\left[\frac{\tilde{s}_{0}(\tilde{s}_{2}+\bar\gamma)}
{(\tilde{s}_{1}-\tilde{s}_{2})(\tilde{s}_{0}-\tilde{s}_{2})}\right]
\left[e^{\tilde{s}_{0} t}-e^{\tilde{s}_{2} t}\right]
\end{align}

\begin{align}
c_{ge}(t)=&\gamma_{R}e^{ikd}
\left[\frac{1}{\tilde{s}_{1}-\tilde{s}_{2}}\right]
\left[e^{\tilde{s}_{1} t}-e^{\tilde{s}_{2} t}\right] \nonumber\\
&+\left[\frac{\gamma_{R}e^{ikd}\tilde{s}_{0}}
{(\tilde{s}_{1}-\tilde{s}_{2})(\tilde{s}_{0}-\tilde{s}_{1})}\right]
\left[e^{\tilde{s}_{0} t}-e^{\tilde{s}_{1} t}\right] \nonumber\\
&-\left[\frac{\gamma_{R}e^{ikd}\tilde{s}_{0}}
{(\tilde{s}_{1}-\tilde{s}_{2})(\tilde{s}_{0}-\tilde{s}_{2})}\right]
\left[e^{\tilde{s}_{0} t}-e^{\tilde{s}_{2} t}\right]
\end{align}
and $\tilde{s}_0, \tilde{s}_1, \tilde{s}_2$ are the $\mathcal{O}(\Omega^4)$ roots of the cubic expression, 
\begin{equation}
\label{det1}
\det{[M_{T}]}=s(s+\bar\gamma)^2-\gamma_{L}\gamma_{R}s+\frac{\Omega^{2}}{4}(s+\bar\gamma).
\end{equation}

It is instructive to observe from this expressions, that initial conditions
$c_{eg}(0)=1$, ${c_{ge}(0)=0}$ are satisfied which consequently implies that
$C(t=0)=0$, which agrees well with our expectations on having separable state
initially.

\subsection{Numerical results in the Born-Markovian regime beyond the single-excitation ansatz}

\begin{figure}
  \centering
  \includegraphics[width=0.45\textwidth, trim={0cm 0cm 0cm 0cm},clip]{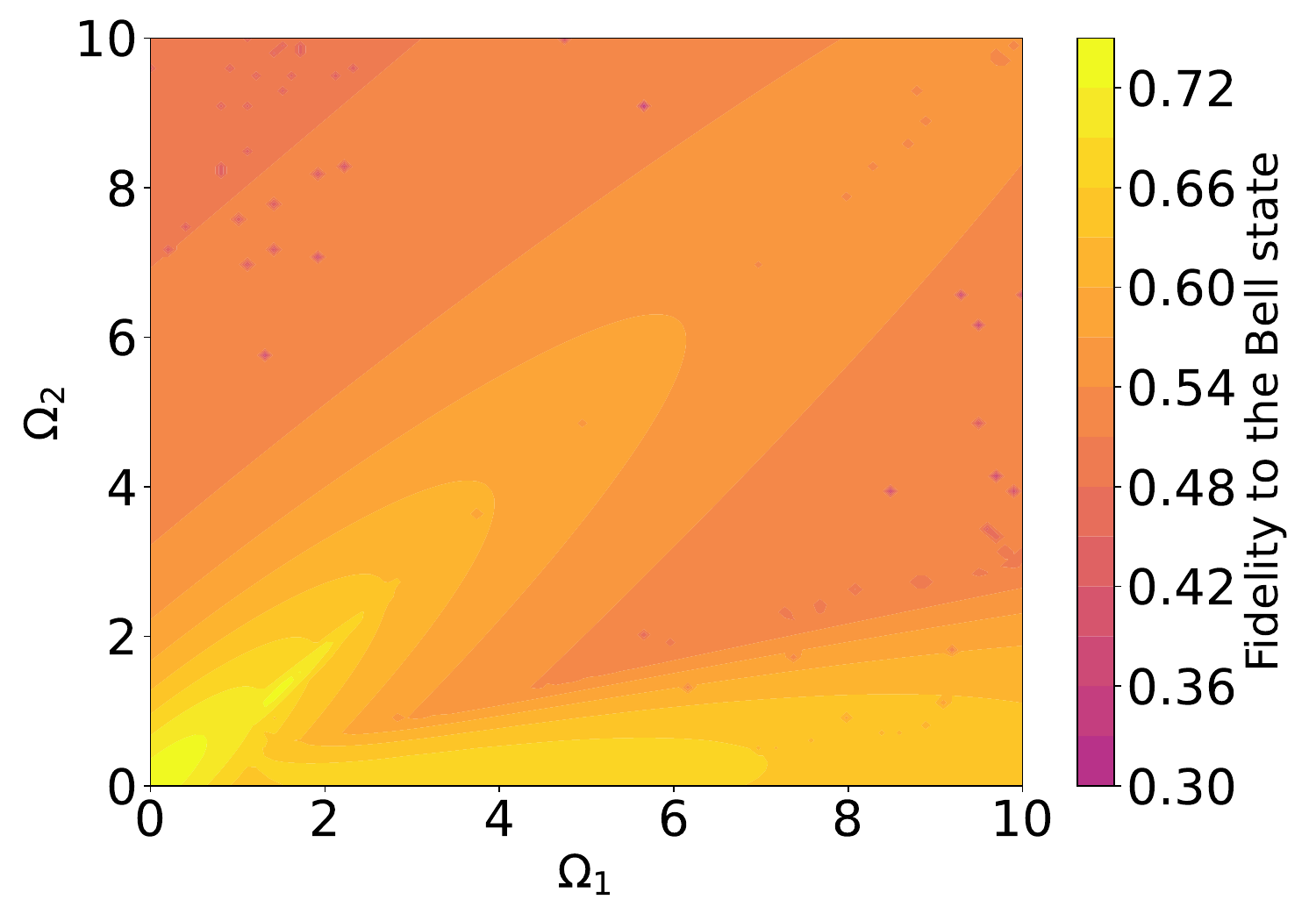}
  \caption[]{Maximum fidelity to the Bell state $(\ket{eg}-\ket{ge})/\sqrt{2}$ as a function of $\Omega_1$ and $\Omega_2$ in the transient regime, starting from $\ket{eg}$. Here we take $d\,\mathrm{mod}\,4 = 0$ and $\gamma_L = 0$. Within the parameter range shown, the largest Bell-state fidelity is $2/e$, attained at $\Omega_1 = \Omega_2 = 0$; in this case, it coincides with the maximal concurrence of the undriven chiral protocol.}
  \label{fig: transient surface plot starting with eg}
\end{figure}

In this section, we relax the weak-driving restriction while staying within the emitter-only Born-Markov description in order to test whether the $2/e$ benchmark survives continuous driving. Since solving \eqref{MasEq} analytically quickly becomes intractable when driving is finite, we allow for arbitrary $0 \leq
\Omega_1 \leq 10\gamma_{tot}$ and $0 \leq \Omega_2 \leq 10\gamma_{tot}$ and we perform numerical
experiments to find the optimal value of concurrence that can be achieved in
time $\tau \ll \gamma_R^{-1}$. Here, we consider the case of perfect
chirality where $\gamma_L = 0$ and we restrict ourselves to the phase-matching condition $\phi \text{ mod
}2\pi=0$. From~\figref{fig: transient surface plot starting with eg}, we see
that if we start with the two quantum nodes in the state $\ket{eg}$, then
similar to the analytical results in the weak driving limit given above, any
driving would cause the concurrence obtained to be lower than the theoretical
$2/e$ limit in the undriven case as previously in \cite{gonzalez2015chiral}.
Though we have only plotted the $\phi \text{ mod } 2\pi = 0$ case for the case
where we have emitters coupled to a 1D chiral spin chain, we also find that
the above is true for all values of $d$ for the values of $\Omega_1$ and
$\Omega_2$ considered for the 1D chiral spin chain.

In contrast, when the emitters are initialized in $\ket{gg}$ and $d\,\mathrm{mod}\,4 = 0$, the optimized concurrence reaches $0.77$, exceeding the undriven benchmark $2/e$. The numerical results are shown in \figref{fig: transient surface plot starting with gg}. Intuitively, three concurrent processes compete and affect Bell-state generation: jump-induced decay from $\ket{eg}$ to $\ket{gg}$ which is negligible in the Markovian regime, coherent driving populating the qubits from $\ket{gg}$ to $\ket{eg}$ and $\ket{ge}$, and a further population of $\ket{ee}$ from the singly-excited states. An initial state $\ket{gg}$ avoids a fixed occupation of the superradiant triplet sector carried by $\ket{eg} = (\ket{T} -\ket{S})/\sqrt{2}$ and hence starts free from bright-state loss. This detail confers slight advantage to the initial $\ket{gg}$ driven scenario over the undriven initial $\ket{eg}$ case. 

\begin{figure}
  \centering
  \includegraphics[width=0.45\textwidth, trim={0cm 0cm 0cm 0cm},clip]{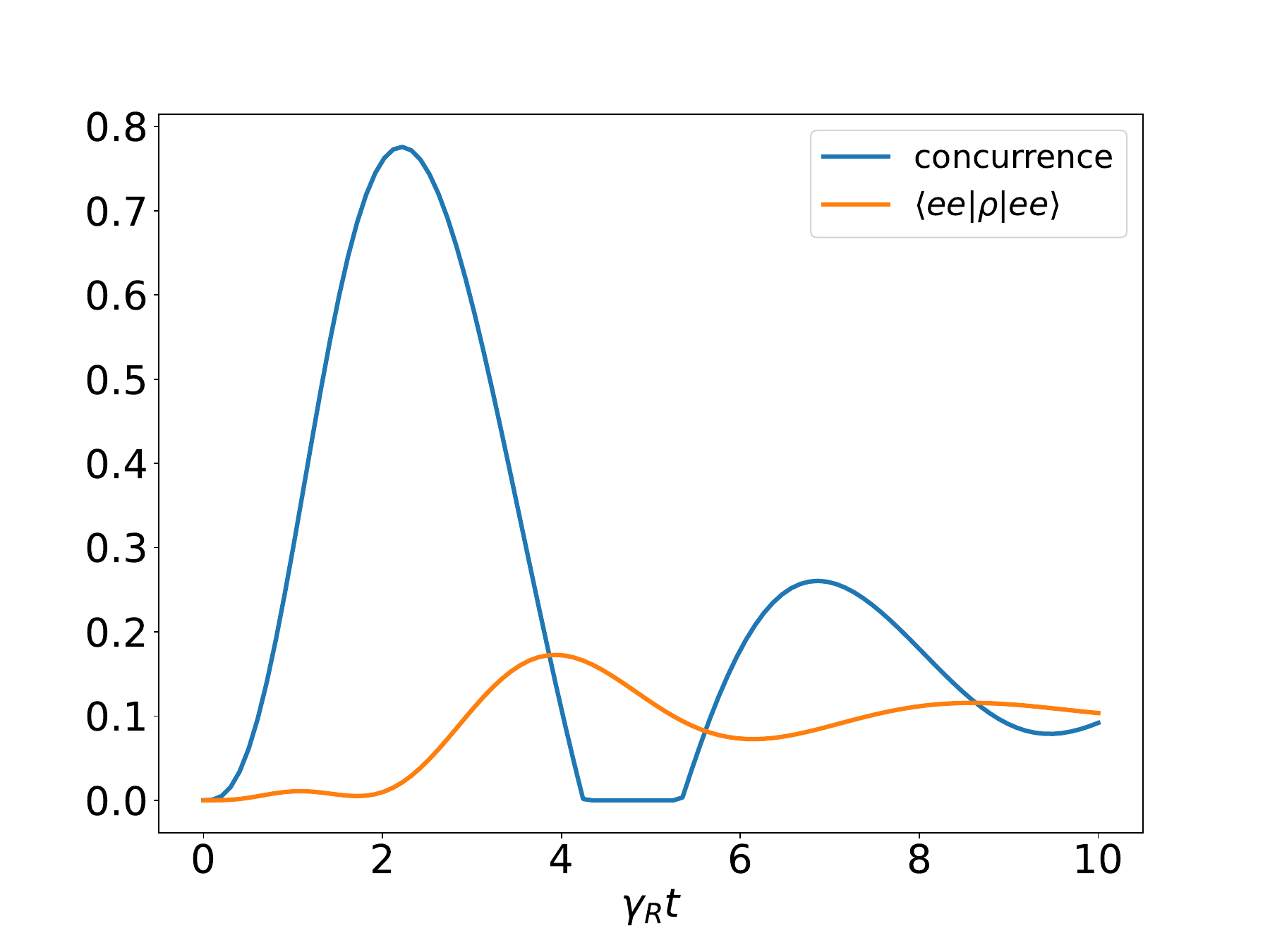}
  \caption[]{Concurrence as a function of $\gamma_R t$ for the values of $\Omega_1$ and $\Omega_2$ that maximize the concurrence in \figref{fig: transient surface plot starting with gg}. Starting from $\ket{gg}$, the concurrence reaches a maximum of $0.77$ when emitter $1$ is driven with $\Omega_1 = 2.05\gamma_R$ and emitter $2$ with $\Omega_2 = 0.74\gamma_R$. In this transient protocol, the concurrence remains below unity because strong driving populates the doubly excited state, leaving a nonzero $\ketbra{ee}{ee}$ component in the two-emitter density matrix.}
  \label{fig: transient plot starting with gg}
\end{figure}

\begin{figure}
  \centering
  \includegraphics[width=0.45\textwidth, trim={0cm 0cm 0cm 0cm},clip]{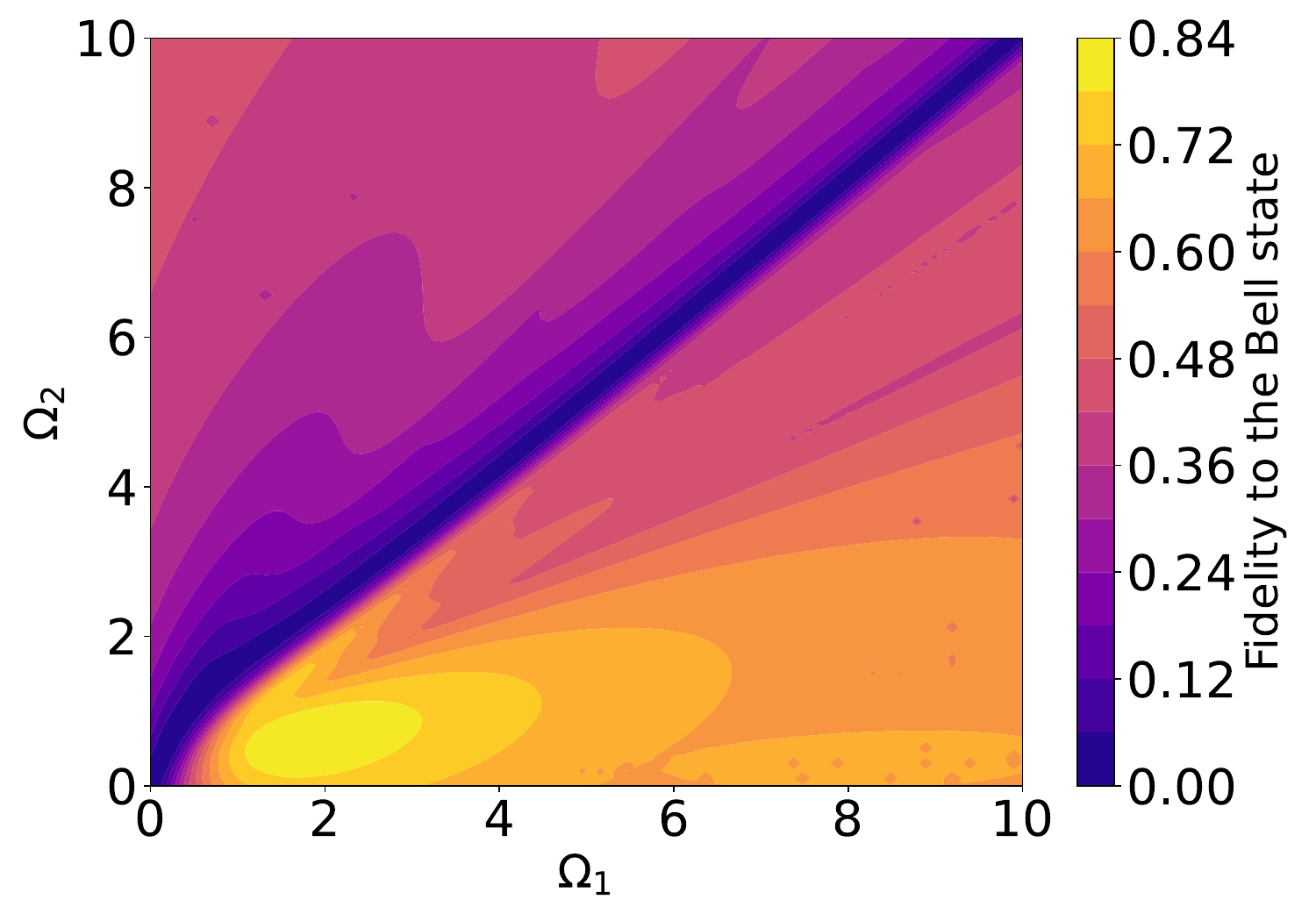}
  \caption[]{Maximal fidelity of the obtained state to the Bell state
  $(\ket{eg} - \ket{ge})/\sqrt{2}$ obtained as a function of
  $\Omega_1$ and $\Omega_2$ in the transient case, starting with the state
  $\ket{gg}$.  Here, we have $\gamma_L = 0$ and $d \text{ mod } 4 = 0$.  We
  see that for these values of $\Omega_1$ and $\Omega_2$ considered, we only
  obtained a maximal fidelity of $0.84$ to the Bell state (corresponding to a
  maximal concurrence of $0.77$), which means that we are unable to perfectly
  obtain the Bell state in the transient case.  The corresponding plot for
  the $\Omega_1, \Omega_2$ values which give maximal concurrence are found in
  \figref{fig: transient plot starting with gg}.}
  \label{fig: transient surface plot starting with gg}
\end{figure}

Despite surpassing the previously derived $2/e$ limit, we note that in the transient case with time-independent driving strengths, we are unable to obtain perfect fidelity to the Bell state within a reasonable range of driving strengths. As shown in \figref{fig: transient plot starting with gg}, a non-zero population of $\ket{ee}$ emerges as the singly-excited manifold gets populated. Moreover, for values of $\phi \text{ mod } 2\pi \neq 0$, the maximal concurrence remains below the previously derived $2/e$ limit, reflecting the sensitivity of the continuously driven protocol to drive-induced phase interference. These limitations motivate us to test the same entanglement-generation mechanism in a more microscopic spin-chain model and to assess, via comparison with MPS dynamics, which features of the transient enhancement remain robust beyond the Born-Markov waveguide treatment.

\section{Microscopic spin-chain model: nonsecular enhancement beyond Born–Markov}
\label{sec: spin-chain}

To test whether the transient enhancement found above survives beyond the emitter-only Born--Markov reduction, we now consider a more microscopic model in which the channel is represented explicitly by an XX spin chain and each emitter couples to two adjacent bath spins, forming a triangular plaquette as shown in \figref{fig: microscopic description of model}. By threading a synthetic flux through each plaquette, one can realize chiral emission into the spin chain through directional interference ~\cite{pichler2015quantum,ramos2016non}.

It is important to note that this section is not a strict one-to-one microscopic continuation of the model used in the previous subsection. In addition to lifting the Born-Markov reduction, we also replace a point-like white-noise waveguide by a structured finite-bandwidth reservoir and a single-point system-bath coupling by a two-point giant-atom-like geometry. The comparison with the earlier $2/e$ benchmark should therefore be read as qualitative rather than literal: our aim is to determine which entanglement-generation mechanisms persist in the microscopic model and which new coherent channels arise only because of the structured bath and plaquette coupling.

\subsection{Analytic approximations with TCL-2}
Given how the previous analysis assumed the Born-Markov approximation from the outset, it is natural to wonder if a similar entanglement generation scheme holds beyond the Born-Markov regime. For clarity, we explicitly define the Born approximation to mean stationarity of the bath state (relative to emitter-relevant timeframes):
\begin{equation}
    \tilde{\rho}_{SB} (t) = \tilde{\rho} (t) \otimes \rho_B (0) + \chi(t) \approx \tilde{\rho}(t) \otimes \rho_B (0),
\end{equation}
such that all system-bath correlations $\chi(t)$ can be neglected. Note that we have assumed the bath and system to be separable at initial time $t=0$ above. Next, a definition for Markovianity can be given in terms of CP-divisibility of evolution \cite{PhysRevLett.105.050403}. Let $\{ \Lambda_{t,0} \}_{t\geq 0}$ be the family of CPTP dynamical maps taking the system state from initial time to time $t$: $\rho(t) = \Lambda_{t,0}[\rho(0)]$. The evolution is CP-divisible if all pairs of times $t \geq s \geq 0$, there exists some intermediate map $\Lambda_{t,s}$ that is CPTP for all $t \geq s$ while adhering to the composition property: $\Lambda_{t,0} = \Lambda_{t,s} \circ \Lambda_{s,0}$. The implication of this is that for all $\Lambda_{t,s}$, the evolution never needs ``extra information" from before time $s$, and so the violation of CP-divisibility arrives as memory effects. In the context of waveguide QED, the Markovian assumption often manifests as the bath having a very short correlation time $\tau_B$ such that its correlation function decays exponentially with respect to it: $G_{\alpha, \beta} (\tau) \sim e^{-\tau/\tau_B}$. By assuming that the evolution time $t >> \tau_B$, this then reflects how the ``short memory" of the bath becomes insignificant to system evolution.

The TCL-ME approach used here can be viewed as the time-local counterpart of projection-operator methods for open-system dynamics. In particular, the Nakajima-Zwanzig construction yields an exact but generally time-nonlocal memory-kernel equation for the reduced-state~\cite{nakajima1958quantum}, whereas the time-convolutionless formulation developed by Breuer \textit{et al.} rewrites the reduced dynamics in terms of a time-dependent local generator~\cite{breuer1999stochastic,breuer2001time,breuer2002theory}. In the present work we truncate this expansion at second order. This TCL-2 equation reduces to the Bloch--Redfield description when truncated to the Markovian limit and provides a practical compromise between analytic tractability and the retention of structured-bath effects.

To explore entanglement dynamics beyond the Born-Markov regime, we make two changes: (a) we replace our Markovian master-equation description for an explicit representation of a hard-core spin chain; (b) in doing so, we replace our emitters with giant-atoms that couple to two adjacent spins in the spin-chain, forming a triangular plaquette, as depicted in \figref{fig: microscopic description of model}. By applying a magnetic flux $\varphi$ through the plaquette, we can invoke a non-zero phase-component in the emitter-spin coupling and realize a notion of chiral coupling to the spin-chain by means of destructive interference for magnons moving in disallowed directions. The introduction of a spin-chain gives us the potential to easily exploit beyond-Born and beyond-Markovian effects respectively; the former via its hard-core two-level structure or the fermionic character of its magnons/spin-waves, and the latter via its capacity as a structured reservoir (i.e. Bessel-valued/algebraically long decay tails of correlation function). It should be noted that in the limit of low spin-excitation number within the chain ($\langle S_+ S_- \rangle \ll 0$), the spin-chain is mappable to a bosonic waveguide using the Holstein-Primakoff transform; in this manuscript, to obtain analytic approximations, we apply such a single-excitation restriction to reduce the spin-chain Hamiltonian to a one-particle hopping problem so we can perform a plane-wave transform and diagonalise the spin-chain into its spin-wave/magnonic $k$-space basis. Subsequently, we employ a variational MPS algorithm with absorbing boundary conditions to probe the hard-core structure of the spins and exactly account for time evolution of the system; the implementation is given in detail in Appendix \ref{appendix: mpsdetails}. 

We begin by employing the time-convolutionless master equation (TCL-ME) approach and obtain an analytic perturbative approximation to study emitters bound to an infinite spin-chain. The TCL-ME in brief is a projective method that yields perturbative time-local equations for the state within the interaction picture. The full system-bath density matrix at time $t$ is $\tilde{\rho}(t)$ and is expressible as:
\begin{equation}
    \frac{d}{dt} \mathcal{P}\tilde{\rho}(t)=\mathcal{K}(t) \mathcal{P}\tilde{\rho}(t),
\end{equation}
where $\mathcal{P}\bullet = \mathrm{Tr}_{B}\{\bullet\} \otimes \rho_B$ projects onto the relevant system degrees of freedom, and $\mathcal{K}(t) = \sum_{n\geq1} \lambda^n \mathcal{K}^{(n)}(t)$ presents a time-dependent generator with a systematic expansion in $\lambda$ (system-bath coupling strength) via ordered cumulants. 

\begin{figure}
    \centering
    \includegraphics[width=0.8\linewidth]{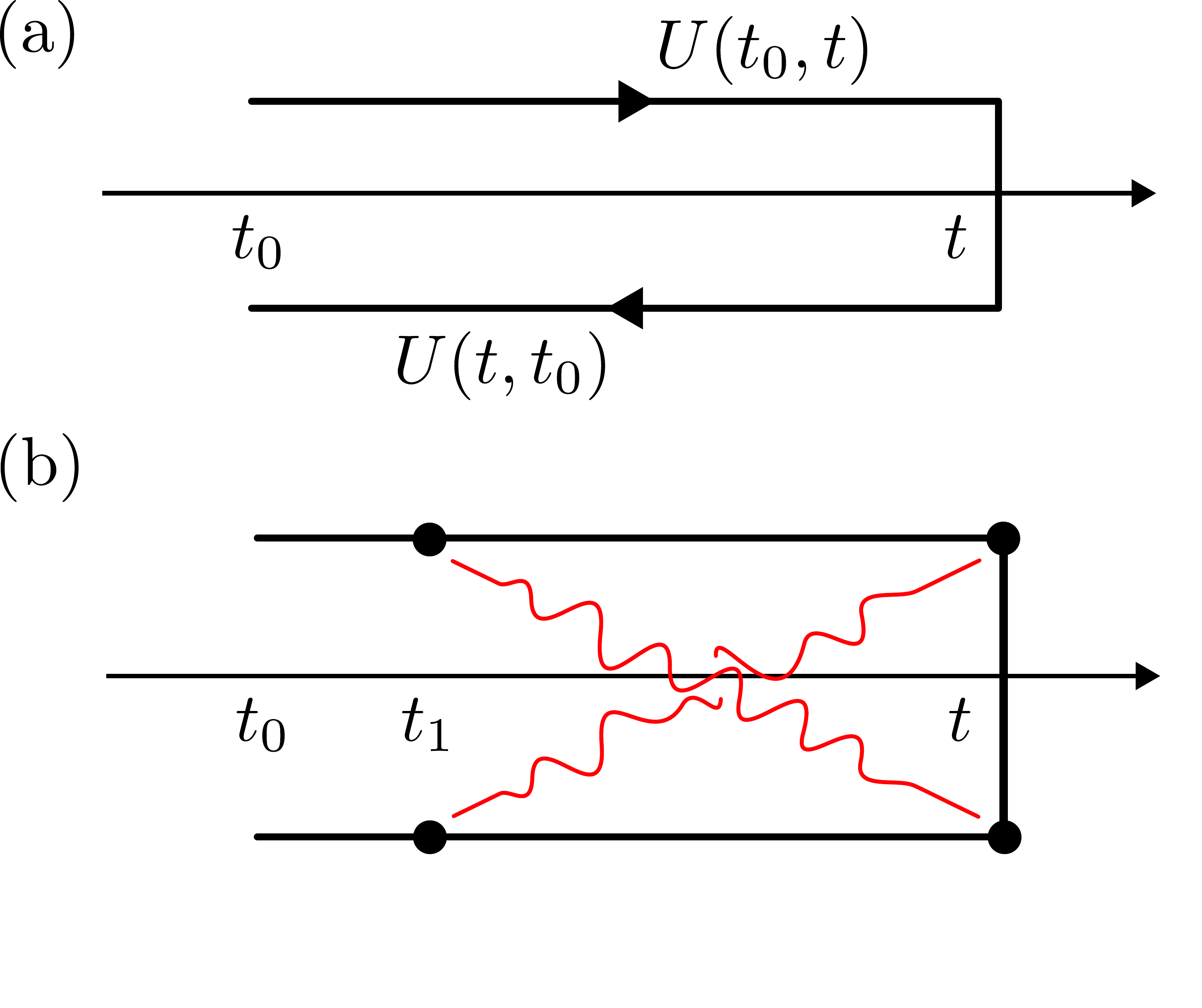}
    \caption{A digrammatic representation of the TCL-ME. (a) The full system-environment density matrix $\rho (t_0)$ undergoes unitary evolution forwards-in-time to $\rho(t)$ as depicted in the upper branch; the converse backwards-time evolution from $\rho(t)$ to $\rho(t_0)$ is given by the lower branch. Together they respectively express the right-action $U(t_0, t)$ and left-action $U(t,t_0) = [U(t_0,t)]^\dagger$ of time-evolution on the density matrix $\rho$. (b) Exact computation of the prior contour under a reduced system description requires time-nonlocal computation. By perturbative expansion, we can express TCL-2 as the sum of all non-zero system-bath interactions occurring between placeholder time $t_1$ and final time $t$, where the vertices are determined by the dressed system operators while the red curved line denotes the associated two-time bosonic correlator through the bath. $t_1$ is then integrated over the interval $[t_0 , t]$.}
    \label{fig:fig5}
\end{figure}

Furthermore, we impose the following assumptions for convenience: system-bath correlations are taken to be negligible at initial time ($\mathcal{P}\tilde{\rho}(t=0) = \tilde{\rho}_S(t=0) \otimes \rho_B$); and a no mean-field condition is also imposed on the bath ($\mathrm{Tr}_B\{\tilde{H}_I(t)\rho_B\}=0 \Rightarrow \mathcal{K}^{(1)}(t) = 0$). The latter is either naturally justified by symmetry of $\rho_B$ or can be enforced by a redefinition of the system Hamiltonian such that the first non-zero dynamical contributions appear at second-order of the expansion, i.e. TCL-2. A full derivation of the cumulant-ordered generator under these conditions can be found in Ref.~\cite{breuer2002theory}; for brevity, we note that the expansion can be neatly illustrated as a diagrammatic expansion over a time-contour as depicted in \figref{fig:fig5}.

The vertices of each diagram are determined by the system operators in the interaction picture. In the case of a driven, on-resonance emitter, it is convenient to work in the dressed-picture basis:
\begin{equation}
    \sigma_- \rightarrow e^{+iH_St}\sigma_- e^{-iH_St} = \frac{1}{2} (A + e^{i\Omega t} B - e^{-i\Omega t} B^{\dagger})
\end{equation}
where $A = \sigma_x ,\; B = \frac{1}{2}(\sigma_z - i\sigma_y)$.

To compute the propagator for an infinite homogeneous chain, the diagonalising k-space transformation can be taken by $c_x=\frac{1}{\sqrt{2\pi}}\int_{-\pi}^{+\pi}dk\,e^{ikx}c_k$, such that the single-magnon propagator is: 
\begin{equation}
    G_d(t)=\langle c_x (t) c_{x'}^\dagger (0) \rangle=e^{-i\Delta t} (-i)^d \mathcal{J}_d(2Jt)
\end{equation}
where $d = x-x'$ and $\mathcal{J}_d$ denotes the Bessel function of the first kind at order $d \in \mathbb{R}$. 

Since the initial bath state is set to the vacuum and assuming the Born approximation, the only nonzero two-time bath correlator is given by $C_{ij} (t) = \langle Q_i(t) Q_j (0) \rangle_B = \sum_{\alpha,\beta} \lambda_{i\alpha} \lambda_{j\beta} G_{i-j}(t)$. Thus we can define the decay rates as a time-integral: 
\begin{equation}
    \begin{split}
        \Gamma_{ij}(t;\omega') &= \int_0^t d\tau \; C_{ij}(\tau) e^{-i\omega'\tau} \\
        &= g_i g_j \sum_{\alpha,\beta} e^{i(s_\alpha \varphi_i - s_\beta \varphi_j)} K_{\alpha,\beta}(t;\omega'),
    \end{split}
\end{equation}
where $\alpha \in \{L[i], R[i]\}$, $\beta \in \{L[j], R[j]\}$ such that $s_{L[\bullet]} = -1$ and $s_{R[\bullet]} = +1$ by convention, while the second term in the product demonstrates how the spin-chain behaves like a time-broadened spectral filter: 
\begin{equation}
    K_{\alpha,\beta}(t;\omega') = \frac{1}{2\pi} \int_{-\pi}^{\pi} dk \; e^{ik(x_{i\alpha} - x_{j\beta})} \frac{1-e^{-i(\omega'+2J\cos k)t}}{i(\omega'+2J\cos k)}
\end{equation}
Summing over all diagrams leads to the following form of the master-equation after suppression of time-indices: $\mathcal{K}^{(2)}\tilde{\rho}_S = \sum_{\chi,\xi,\alpha,\beta} e^{i\chi t} \xi[\alpha\tilde{\rho}_S, \beta]+\mathrm{h.c.}$, where the variable identities are presented in Table \ref{tab:tcl2coefficients}. Additional details can be found in Appendix \ref{appendix: tcl2details}.

\begin{table}
    \centering
    \renewcommand{\arraystretch}{1.2}
    \begin{tabular}{|c||c|}
        \hline
        $\xi$ & $(\alpha, \beta, \chi)$ \\
        \hline
        $\Gamma_{11}(t; 0)$ & $(A, A^\dagger, 0)$, $(-A, B, +2\Omega_1)$, $(A,B^\dagger, -2\Omega_1)$ \\
        $\Gamma_{11}(t; +2\Omega_1)$ & $(B, B^\dagger, 0)$, $(B, A, +2\Omega_1)$, $(-B,B, +4\Omega_1)$ \\
        $\Gamma_{11}(t; -2\Omega_1)$ & $(B^\dagger, B, 0)$, $(-B^\dagger, A, -2\Omega_1)$, $(-B^\dagger, B^\dagger, -4\Omega_1)$ \\
        $\Gamma_{12}(t; 0)$ & $(\sigma_2^-, A, 0)$, $(-\sigma_2^-, B, +2\Omega_1)$, $(\sigma_2^-, B^\dagger, -2\Omega_1)$ \\
        $\Gamma_{21}(t; 0)$ & $(A, \sigma_2^+, 0)$ \\
        $\Gamma_{21}(t; +2\Omega_1)$ & $(B, \sigma_2^+, +2\Omega_1)$ \\
        $\Gamma_{21}(t; -2\Omega_1)$ & $(-B^\dagger, \sigma_2^+, -2\Omega_1)$ \\
        $\Gamma_{22}(t; 0)$ & $(\sigma_2^- ,\sigma_2^+, 0)$ \\
        \hline
    \end{tabular}
    \caption{Nonzero contributions to the TCL-2 generator, $\mathcal{K}^{(2)}\tilde{\rho}_S=\sum_{\chi,\xi,\alpha,\beta} e^{i\chi t}\,\xi[\alpha \tilde{\rho}_S,\beta]+\mathrm{h.c}$. For each kernel $\xi=\Gamma_{ij}(t;\omega')$, the corresponding operator pair $(\alpha,\beta)$ and phase $\chi$ are listed.}
    \label{tab:tcl2coefficients}
\end{table}

As with the Born-Markov waveguide scenario, we performed a numerical sweep over available parameters to determine the conditions for optimal transient entanglement generation and found a maximal concurrence of 0.78 under $(\Omega_1, \Omega_2, g_1, g_2) = (0.063, 0.00, 0.14, 0.30)$ in units of $J$, for initial state $\ket{gg}$, inter-emitter separation of $d = 1$ ($L[2] = R[1]$), and fully-chiral coupling ($\varphi = \pi/4$), i.e. the upstream emitter is subjected to relatively strong driving and is coupled less strongly to the spin-chain than the downstream emitter. The identity of the reduced-state at peak concurrence is close to the phase-shifted singlet $\chi_- = (\ket{eg} -i\ket{ge}) /\sqrt{2}$ with fidelity $0.807$ for $d=1$ and $0.635$ for $d=9$, where the phase-shift arises due to $d \; \textrm{mod} \; 4 = 1$ for both cases. Note that the above quoted concurrence values near $0.78$ and $0.80$ should not be interpreted as direct refinements of the Born-Markov $2/e$ limit because the microscopic model changes both the reservoir structure and the coupling geometry, in conjunction with the optimization or impedance-matching of $g_1$ and $g_2$.

As neither emitter can be considered weakly-coupled to the bath, this motivates cross-comparison with concurrence results from MPS time-evolution. Additionally, we probe the time-local Redfield approximation (extension of the $s$-integral upper limit from $t$ to $+\infty$ within the generator, given $t \gg \tau_B \sim 1/J_B$) and a further GKSL-enforcing secular approximation that forces dissipation to be frequency-diagonal or jump channels to be independent (via removal of all terms cross-frequency terms in the dressed basis, i.e. $e^{i(\omega' - \omega)t} \sim 0$ if $\omega \neq \omega'$) for a fuller view of entanglement dynamics. 

\begin{figure}
    \centering
    \includegraphics[width=0.9\linewidth]{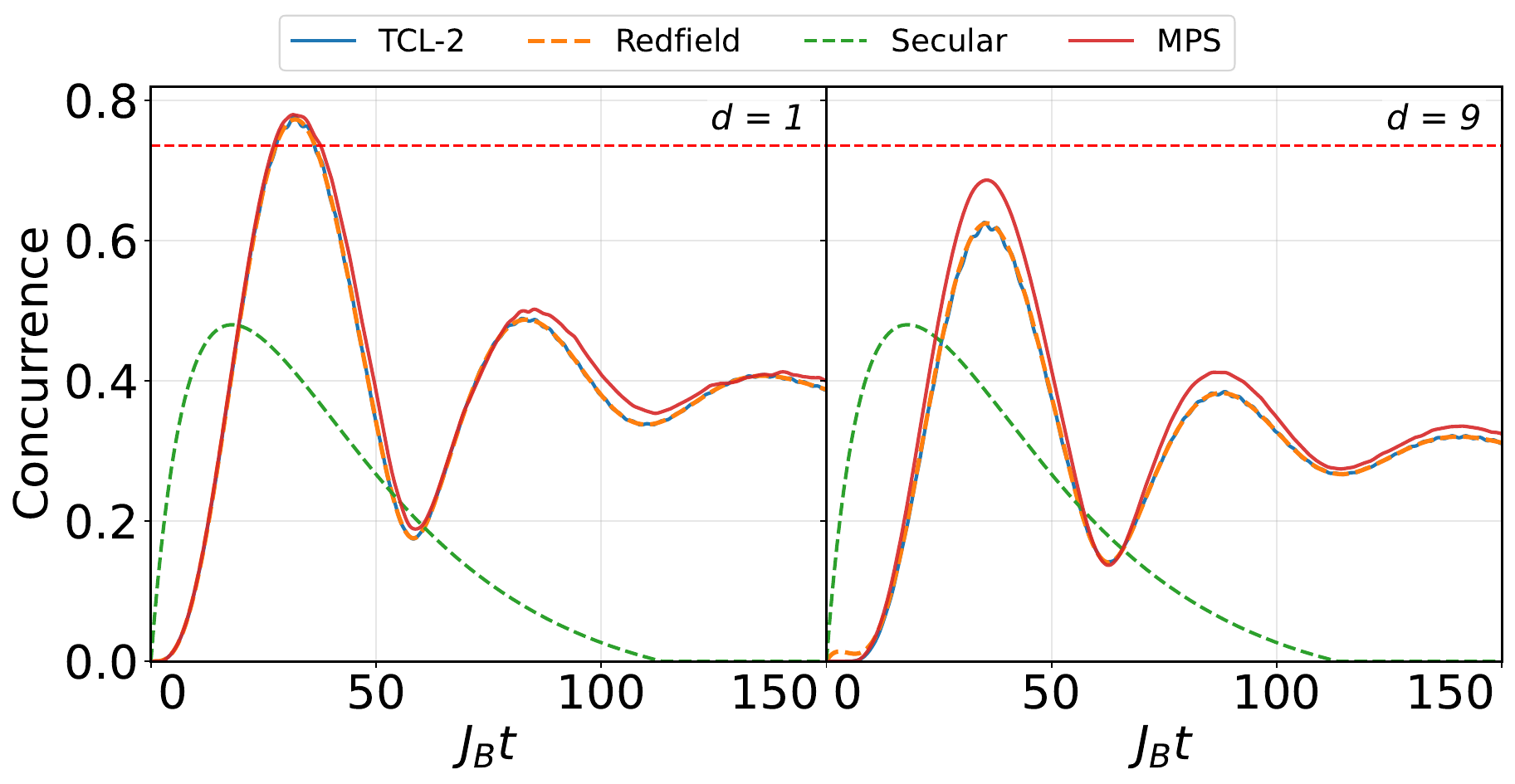}
    \caption{Concurrence $C(t)$ versus $J_Bt$ for inter-emitter separations $d=1$ (left) and $d=9$ (right). The horizontal red dashed line marks $2/e$, used as a benchmark. The non-secular TCL-2 and Redfield (beyond early-time) treatment show qualitative agreement with MPS results for both separation values, while the secular approximation gives distance-invariant results with no oscillatory structure and a broader, weaker concurrence transient.}
    \label{fig:fig6}
\end{figure}

\figref{fig:fig6} shows that non-secular TCL-2 reproduces the overall concurrence envelope and first transient peak seen in the MPS dynamics, while the time-local Redfield limit differs mainly at early times when the inter-plaquette delay is appreciable. By contrast, the secular approximation misses both the oscillatory structure and the peak height. Taken together, this already suggests that the dominant reduced-system correction in the optimal regime is the retention of non-secular coherence-mixing terms rather than strong reduced-state memory effects. For intuition, while a standard bath correlation time, $\tau_B = \int_0^\infty dt \; {|C(t)|}/{|C(0)|}$, diverges since the correlation function associated with the spin-chain contains a power-law tail, an operational bath timescale can nevertheless be identified from the radius of convergence of the small-$s$ expansion used in the Laplace-domain derivation of the Markovian limit ~\cite{ramos2016non}. Since the band edges are at $\omega = \pm2J_B$, the corresponding timescale is $\tau_B^{(op)} \sim \frac{1}{2J_B}$ if we disregard propagation delay. Accordingly given our optimal parameter set, the fastest system timescale is given by $\tau_S \sim 1/\gamma_2 \approx 5.56/J_B$ so $\tau_B^{(op)}/\tau_S \sim 0.09 \ll 1$ indicating a large separation between system and bath interaction timescales, thus accounting for the close-to-Markovian behaviour.

Crucially, however, the near-Markovian character of the reduced dynamics does not rescue the secular approximation. The failure of secularisation indicates only that the bath cannot cleanly resolve the system's distinct dressed transition frequencies on the dissipative timescale, so that oscillatory cross-frequency terms do not average away, as corroborated in similar past work \cite{haikka2010non,haikka2010comparing,farina2019open}. In the present model, this occurs when $|\omega-\omega'|$ is not parametrically larger than the relevant decay scale $\gamma\sim 2g^2/J_B$, where the numerical factor reflects the two-point plaquette coupling. To illustrate, recall that for a homogeneous XX spin-chain of dispersion $\omega(k) = -2J_B \cos k$, its relevant mid-band values are $\omega_{0} = 0$, $k_{0} = \pi/2$, and $v_g(\omega_{0}) = |\frac{d\omega}{dk}|_{k_{0}} = 2J_B$. A simple mid-band estimate for a single emitter with one coupling point to the chain at strength $g$ then gives the decay rate $\gamma^{(1pt)}(\omega_{0}) = 2\pi g^2 D(\omega_{0}) = \frac{2g^2}{v_g(\omega_{0})} = \frac{g^2}{J_B}$, and likewise for two coupling points as with the triangular plaquettes, this becomes $\gamma(\omega_{0}) = \frac{|g(1+e^{ik_{0}})|^2}{J_B} = \frac{2g^2}{J_B}$ assuming negligible delay. From the optimal parameters, we have $\gamma_{R,1} \sim 0.039J_B$ and $\gamma_{R,2} \sim 0.18J_B$ which places the upstream emitter driving strength at order unity to decay rates: $\Omega_1 \sim \gamma_{1,2}$. In this regime, different decay pathways into the spin chain interfere strongly, and the non-secular terms mediate the coherence mixing responsible for the concurrence enhancement. 

Note that this is distinct from the rotating-wave approximation in the interaction picture, where counter-rotating terms in the Hamiltonian are discarded because the bare transition frequency $\omega_b$ is much greater than the other relevant dynamical scales. This hierarchy is well satisfied in optical implementations, where $\omega_b$ lies at optical frequencies while $\{J_B,g,\Omega\}$ govern much slower envelope dynamics \cite{petersen2014chiral, sollner2015deterministic}.

\begin{figure}
    \centering
    \includegraphics[width=0.9\linewidth]{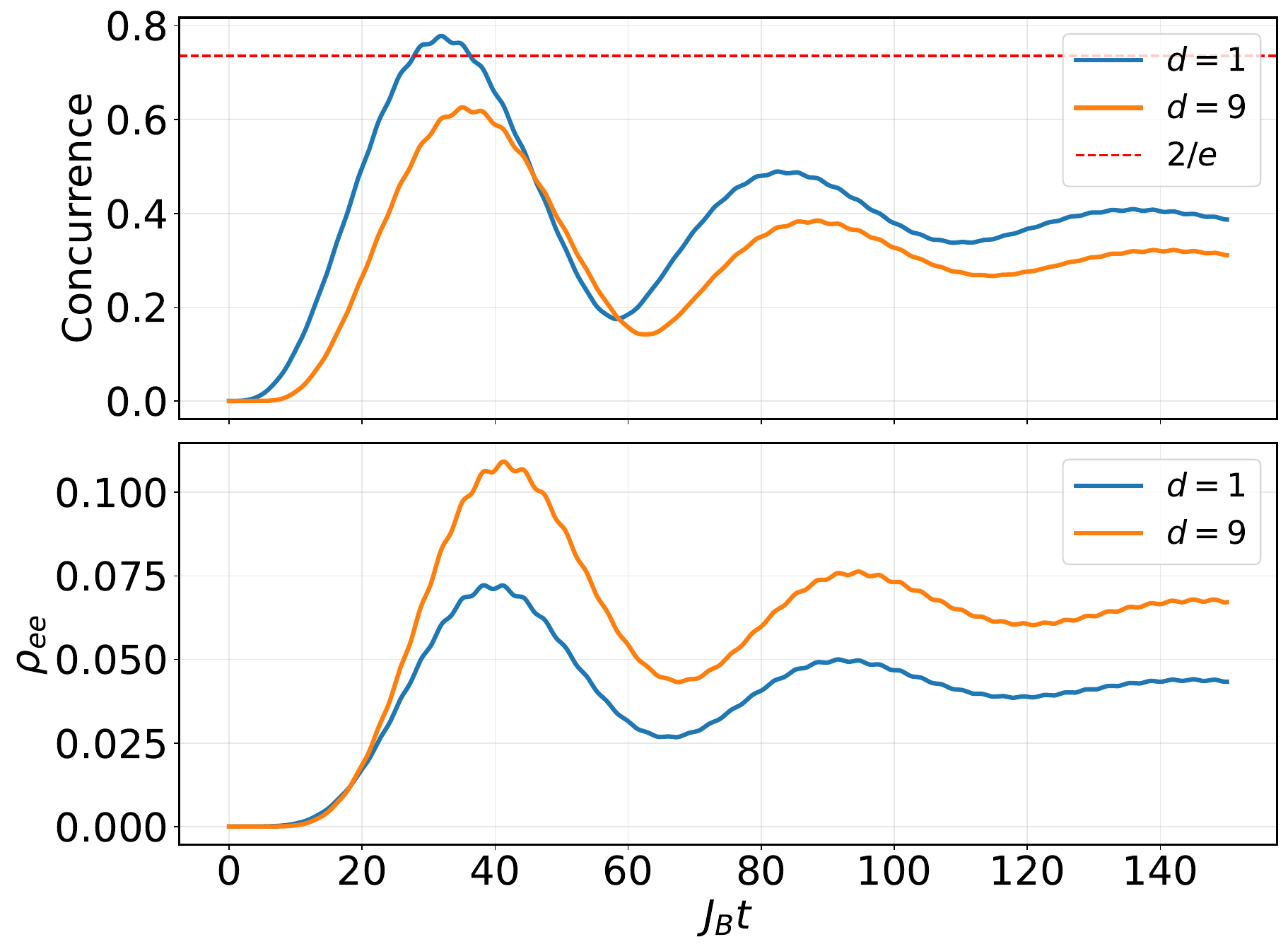}
    \caption{Upper panel: concurrence $C(t)$ for $d=1$ (blue) and $d=9$ (orange), with the horizontal red dashed line indicating $2/e$. Lower panel: population $\rho_{ee}(t)$ of the doubly-excited state for the same two separations. The shorter separation reaches the larger first concurrence maximum, while the larger separation shows systematically larger occupation of $\ket{ee}$ throughout the transient and at late times. All results above were obtained from TCL-2 simulations.}
    \label{fig:fig7}
\end{figure}

Probing the system populations over time in \figref{fig:fig7}, continuous driving of the upstream emitter can be observed to leak population into $\ket{ee}$ which prevents the generation of a clean Bell state as with the Born-Markov waveguide scenario. Despite the similarities in maximum obtainable concurrence across both the ideal Born-Markov waveguide and homogeneous spin-chain scenarios, the optimal parameter regimes for entanglement generation are distinct. In a perfectly chiral Born-Markov waveguide, the waveguide field carries strong which-emitter information and driving only the upstream generates a sequential cascade of $\ket{gg} \rightarrow \ket{eg} \rightarrow \ket{ge}$ with continuous leakage, resulting in a mostly classical mixture upon tracing out the waveguide. To generate coherence from initial state $\ketbra{gg}{gg}$, the which-emitter information must be erased by driving the downstream emitter as well to engineer destructive interference in the bright $\ketbra{ee}{ee}$ channel. For the homogeneous spin-chain, the emitters instead acquire a coherent channel that produces unitary hybridisation of $\ket{eg}$ and $\ket{ge}$. This can be seen when taking the principal-value integral of dissipative spectral function $J_{ij}(\omega) = \mathrm{Re}[\Sigma_{ij}(\omega)]$ is given by,
\begin{equation}
    J_{ij}(t; \omega) = \mathcal{P} \int d\omega' \; \frac{\Gamma_{ij}(t; \omega')}{\omega - \omega'}.
\end{equation}
For the homogeneous spin-chain, the explicitly nonlinear dispersion gives rise to a frequency-dependent dispersive/real part of the propagator even at the band-centre, giving rise to an inter-emitter coherent scale of $|J_{12}| \sim 2g_1g_2/J_B = 0.084J_B$ comparable to the emitter decay rates; in the ideal Born-Markov waveguide or quantum white-noise limit, the flat spectral density instead returns a principal value of zero or a constant renormalizable Lamb shift. Driving only the upstream emitter alone thus prepares a state that develops quickly into a coherent superposition across both emitters, and non-uniform optimisation over emitter-bath coupling strengths boosts concurrence by reducing mismatch-induced reflection and leakage.

\begin{figure}
    \centering
    \includegraphics[width=0.9\linewidth]{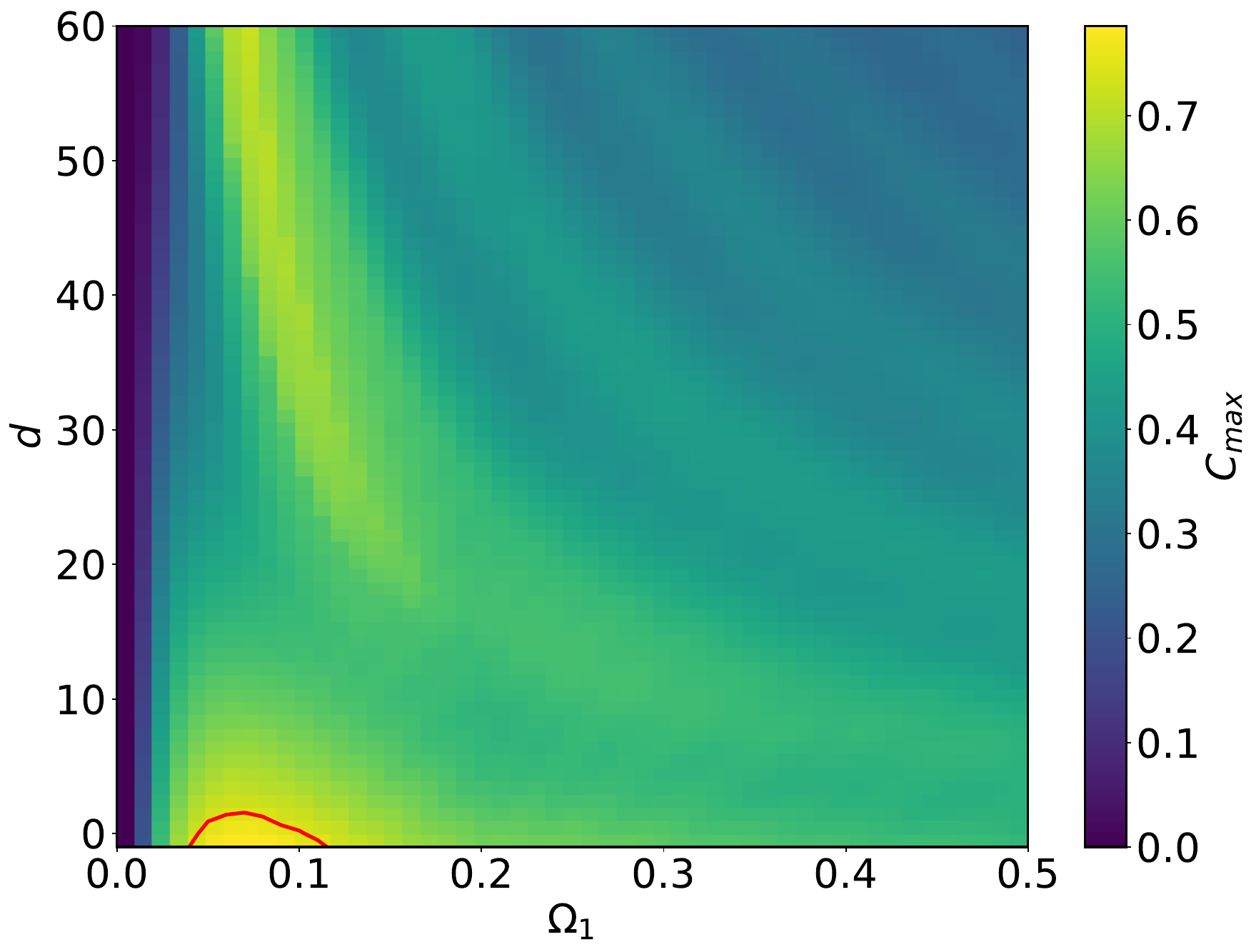}
    \caption{Maximum concurrence $C_{\max}$ obtained from TCL-2 as a function of upstream drive amplitude $\Omega_1$ and inter-emitter separation $d$, for fixed couplings $(g_1,g_2)=(0.14,0.30)$ in units of $J_B$. The color scale gives $C_{\max}$; the red contour denotes $C_{\max}=2/e$. A broad high-concurrence region appears at small $d$ around $\Omega_1 \approx 0.06$--$0.08J_B$, while further ridges extend to larger separations at stronger driving according to the system's interference structure.}
    \label{fig:fig8}
\end{figure}

Additionally, we probe the distance-dependence between the emitters (while maintaining the triangular plaquette geometry they establish with the spin-chain) within the TCL-2 approximation regime to study delay effects on entanglement generation. In \figref{fig:fig8}, we observe both a gradual decrease in maximum concurrence for the optimal driving $\Omega_1 = 0.06J$ over longer inter-emitter delays, and for $\Omega_1 \sim 0.1J$ a resurgence of maximum concurrence at very long emitter distances of $d \sim 40$. Here, the delay $d$ acting as a knob for the delay phase, $\Phi_d \sim \Omega_1d/J$, and destructive cancellation of which-emitter information occurs at $\Phi_d \approx \pi$ (i.e. the retarded drive arrives out-of-phase from the dominant dipole correlations at the dressed splitting and stabilises an entangled superposition across both emitters). For $\Omega_1 \sim 0.1J$, this arises at $d\sim 31$. As observable at higher driving strengths, such resurgence regions arise cyclically at multiples of $d$ but maximum concurrence decreases for more distant regions due to dispersion effects; more details are discussed in Appendix \ref{appendix: tcl2details}.

\subsection{Beyond-Born-Markov effects from MPS numerical evolution}

\begin{figure}
    \includegraphics[width=0.95\linewidth]{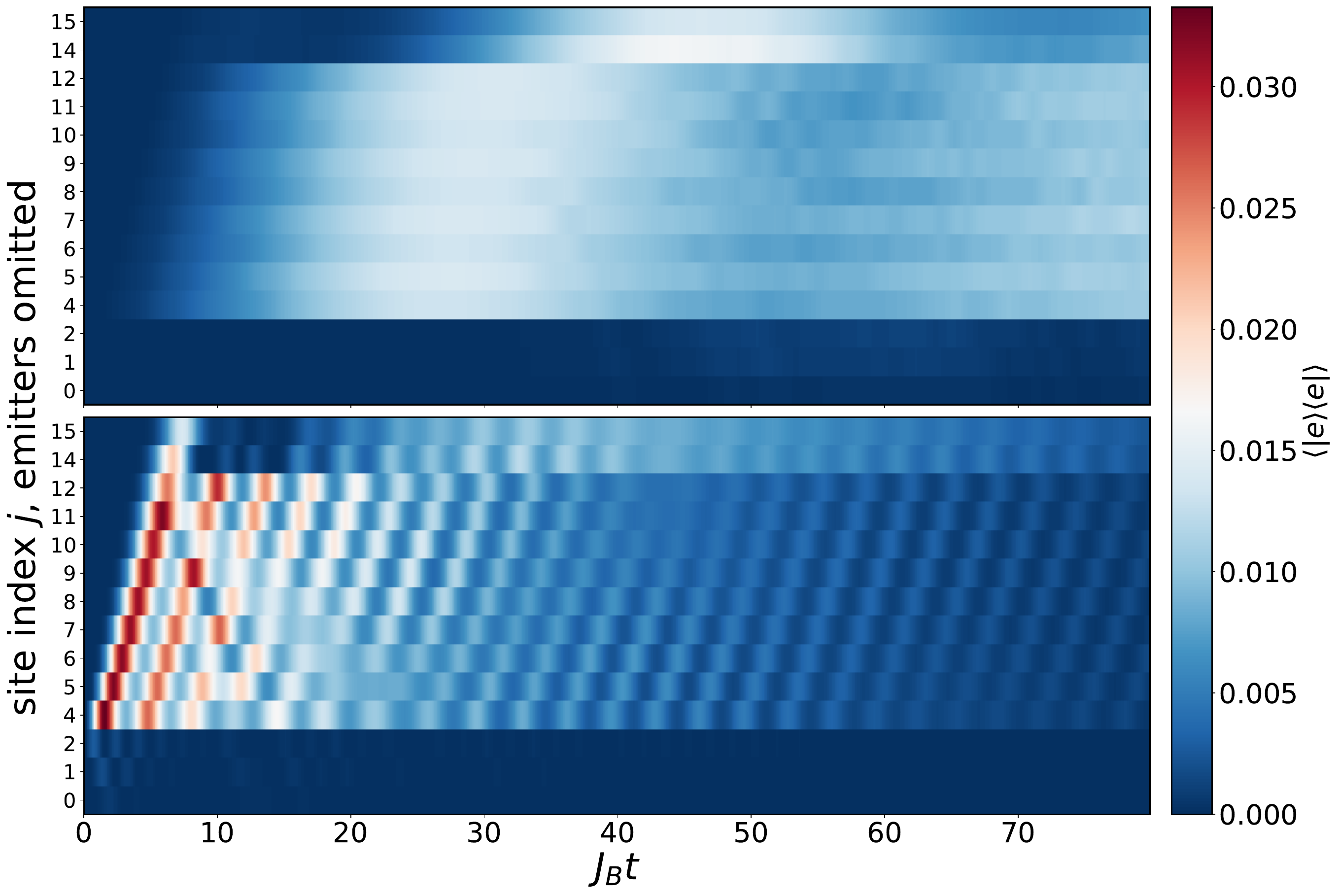}
    \caption{Bath-site excitation density, with emitter sites omitted from the vertical axis, plotted versus site index $j$ and scaled time $J_Bt$. The color scale gives the local bath population. Upper panel: driven evolution from $\ket{gg}$ at the optimal parameter set. Lower panel: undriven evolution from $\ket{eg}$. In the driven case from $\ket{gg}$, the emitted population builds up gradually and forms a comparatively smooth downstream profile, with an appreciable time-delay $\sim20J_B$ at the downstream emitter.}
    \label{fig:fig9}
\end{figure}

\begin{figure}
    \centering
    \includegraphics[width=0.95\linewidth]{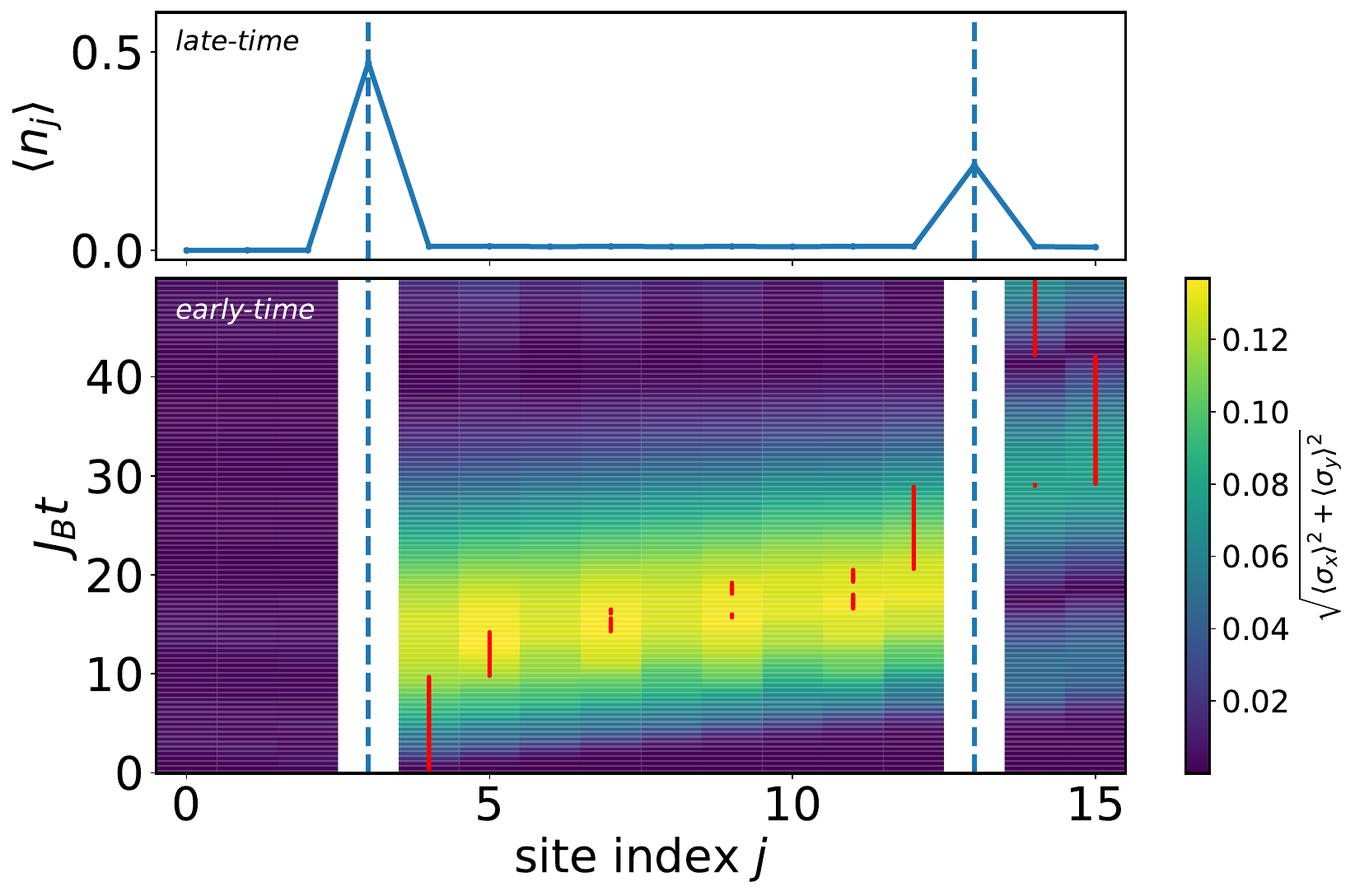}
    \caption{Spatially-resolved bath observables. Upper panel: late-time bath-site occupation $\langle n_j\rangle$ versus site index $j$; the blue dashed vertical lines mark the omitted emitter positions. Lower panel: early-time transverse-coherence single-site magnitude $\sqrt{\langle \sigma_x\rangle^2+\langle \sigma_y\rangle^2}$ shown as a space-time density plot, with the white bands again indicating the omitted emitter sites. The short red vertical markers chart the evolution of the coherence front by indicating the site with the maximum transverse-coherence along time.}
    \label{fig:fig10}
\end{figure}

\begin{figure}
    \centering
    \includegraphics[width=0.95\linewidth]{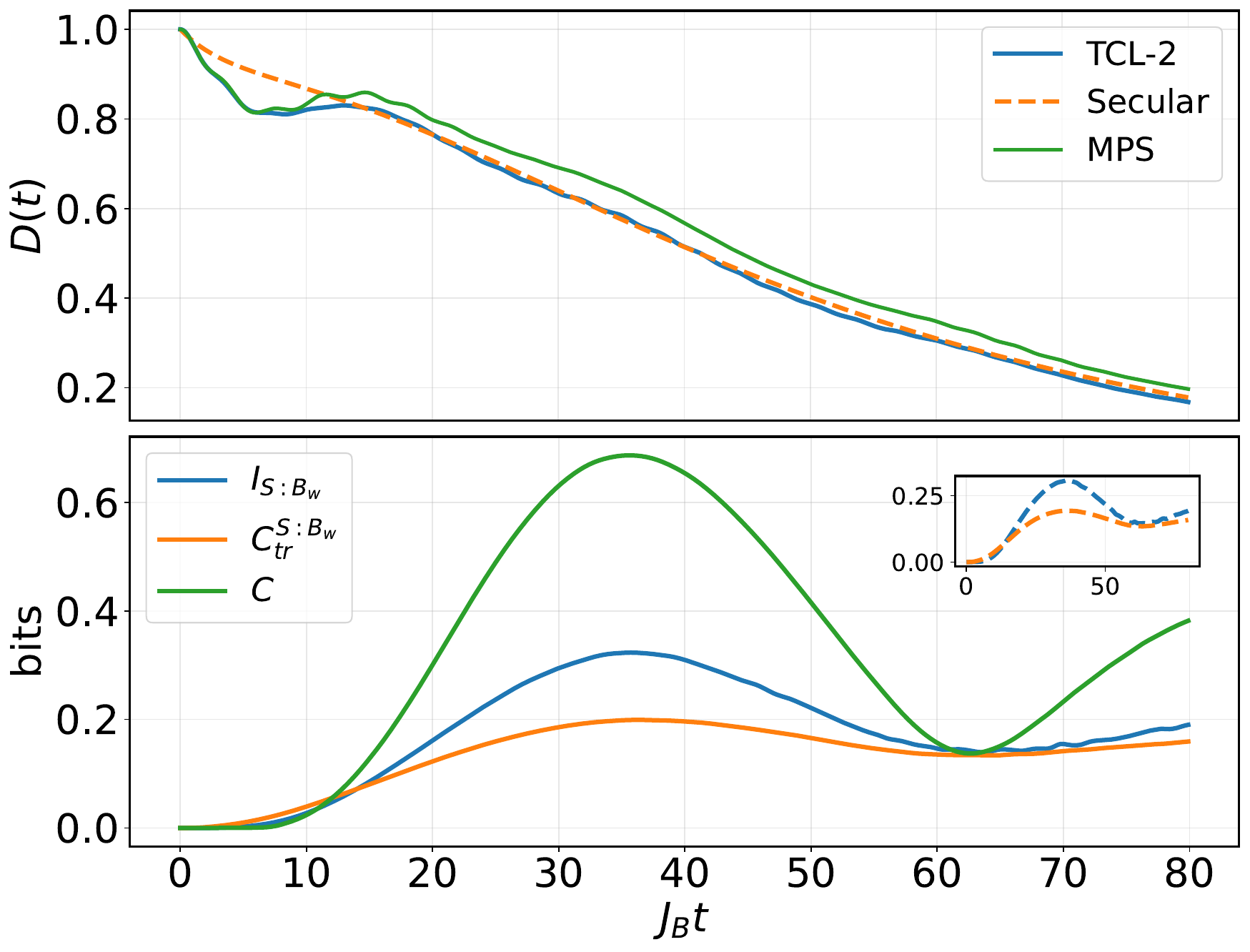}
    \caption{Upper panel: trace distance $D(t)$ between evolutions from initial states $\ket{gg}$ and $\ket{eg}$ (both with driving), comparing TCL-2, secular, and MPS results where $n_1=3,n_2=13, N=16$. Only the MPS results show a distinct increment in the transient regime, alongside TCL-2 results to a much more diminished extent. Lower panel: concurrence $C$ between the emitters, mutual information $I_{S:B_w}$ between the emitters and a finite bath window $B_w$, and the corresponding trace-distance correlation measure $C_{tr}^{S:B_w}$ are compared. The rise of $C$ is accompanied by a simultaneous buildup of both system-bath correlation measures. The main body designates $S=\{n_1,n_2\}$ and $B_w = \{n_i-1, n_i + 1\}, \forall i\in\{1,2\}$; it charts correlations between the emitters and the spins forming the triangular plaquettes. The inset (dashed lines) instead designates the triangular plaquettes as system $S=\{n_i-1, n_i, n_i + 1\}, \forall i\in\{1,2\}$, and a middle section of the spin-chain as the bath window $B_w = \{\left\lfloor\frac{N}{2}\right\rfloor -1, \left\lfloor\frac{N}{2}\right\rfloor, \left\lfloor\frac{N}{2}\right\rfloor + 1\}$; it charts correlations between the plaquettes and the spin-chain bulk.}
    \label{fig:fig11}
\end{figure}

Given that nonsecular TCL-2 gives qualitative rather than quantitative agreement with MPS concurrence dynamics, we next seek to clarify their discrepancy and underlying mechanistic differences. Unless otherwise specified, we performed MPS simulations on a $N=16$ site chain inclusive of emitters at positions $n_1 = 3$ and $n_2 = 13$. Absorbing boundary conditions set at $\zeta = 2J_B$ were employed to suppress reflections and mimic an infinite spin chain. It is worth noting that the difference between does not stem from imperfect approximation of the infinite chiral spin-chain onto an MPS structure. Additionally, the MPS simulations are most reliable in the transient regime relevant to the first concurrence peak: in the driven case, the fixed-bond-dimension one-site TDVP evolution exhibits trace drift that remains controlled up to approximately $t \sim 40J_B$ for the bond dimensions used, with convergence established for concurrence at the first transient peak, as elaborated on in Appendix \ref{appendix: mpsdetails}. In what follows, we thus use MPS in two complementary ways: first, as a microscopic benchmark for the transient concurrence dynamics, especially the first peak in Fig.~\ref{fig:fig6}; and second, as a diagnostic tool for bath buildup and system-bath correlations in Figs.~\ref{fig:fig9} to ~\ref{fig:fig11}.

At a kinematic level, we track the evolution of bath-state populations for initial states $\ket{eg}$ with undriven system and $\ket{gg}$ under optimal driven parameters in \figref{fig:fig9}. For the former undriven case, we see that excitations are transferred into the spin-chain not as a single smooth wavepacket but as a periodic train of packets instead, a feature of non-Markovian interference arising due to the two-point giant-atom-like coupling: a leaky dressed mode with approximate timescale $\gamma^{-1} \sim J_B/2g^2 \sim 25.5 J_B^{-1}$ is formed within the triangular plaquette, and coherent beating between the local mode and propagating chiral chain modes then results in a burst chain with period $T_{plq} \sim 2\pi/\Omega_{plq} = 2\pi/\sqrt{J_B^2 + g_1^2} \sim 3.1 J_B^{-1}$. For the latter driven case, the dynamics is dominated by the upstream build-up time of $\Omega^{-1}$ before it emits strongly into the spin-chain. After the upstream emitter passes its first transient overshoot and becomes phase-locked with the downstream emitter, the downstream emitter follows the incoming narrowband signal adiabatically (in contrast to the $\ket{eg}$ case). The incoming magnon can load the local plaquette mode transiently before amplitude is re-emitted into the chiral chain, and the output-region population thus exhibits a delayed buildup with slight overshoot before the system reaches the driven-dissipative fixed point in which continuous pumping and chiral leakage are balanced. Notably, although the spin-chain bath remains within the low-excitation limit even at late-time as shown in \figref{fig:fig10} (upper) and is hence largely free from hardcore effects or bath nonlinearities, emission into the bath still results in build-up over coherent channels that serve to propagate magnonic excitations downstream as in \figref{fig:fig10} (lower), thus violating the Born approximation. 

Given the disturbance of the bath state and formation of dressed states within the plaquette, it is natural to probe for the growth of system-bath correlations and their role in entanglement generation. To do so, we extract the reduced density operator of the emitters together with a finite bath-window $B_w \in \mathbb{Z}^+$ and evaluate two quantities: mutual information $I_{S:B_w}(t)$ to measure total correlation, and trace-distance correlation $C_{tr}^{S:B_w}(t) \coloneq \frac{1}{2} \left|| \rho_{S\oplus B_w}(t)-\rho_S(t)\otimes\rho_{B_w}(t)\right||_1$ as a direct measure of how far the joint state is from the product of its marginals. The resulting behaviour is shown in \figref{fig:fig11}: factorisation of the joint system-bath state fails at almost all times, and the onset of emitter-emitter entanglement is accompanied by a simultaneous buildup of emitter-bath correlations. The inset in \figref{fig:fig11} (lower) also show that the system-bath correlations are not just due to the formation of dressed states within the plaquettes, but also arise with the inner spin-sites to form an extended transport-induced hybridised state.

It is also useful to separate this point from the question of reduced-state non-Markovianity, and of whether memory effects actually contribute to entanglement generation under our optimal regime. We first consider the trace-distance measure of non-Markovianity, $\mathcal{N}(\Phi) = \max_{\rho_1(0),\rho_2(0)} \int_{\sigma(t)>0} \sigma(t)\, dt$, defined through the trace distance
$\sigma(t) = \frac{d}{dt}D(\rho_1,\rho_2) = \frac{d}{dt}\big[\frac{1}{2}\|\rho_S^{(1)}(t)-\rho_S^{(2)}(t)\|_1\big]$
between two reduced emitter states obtained from different initial conditions $\rho_1(0),\rho_2(0)$ \cite{PhysRevLett.103.210401, PhysRevA.81.062115}. A sustained increase of $D(t)$ over some time interval indicates a temporary recovery of distinguishability and hence information backflow at the level of the reduced emitter dynamics. Strictly speaking, the reduced map $\Phi_{t,0}$ itself remains CPTP; what $\sigma(t)>0$ signals is instead the breakdown of CP divisibility of the intermediate map $\Phi_{t,s}$ in the BLP sense. 

Since here we evaluate $D(t)$ only for the physically motivated initial pair $\ket{eg}$ and $\ket{gg}$, the quantity plotted in \figref{fig:fig11} (upper) should be understood as a witness of BLP non-Markovianity rather than the fully maximized measure $\mathcal{N}(\Phi)$. The numerical results for both the MPS simulations and the TCL-2 treatment then admit a rather natural interpretation: in the MPS data, $D(t)$ decreases monotonically almost everywhere, except for a weak revival in the window $J_Bt \sim[5,20]$, precisely where the downstream emitter in the $\ket{eg}$ evolution has not yet settled into adiabatic following; this is corroborated by the absence of any such increase in the limit $g_2 \rightarrow 0$, where the downstream node is effectively removed. The same feature can be understood equivalently from the TCL-2 side as arising from the non-secular terms shown in \figref{fig:fig11}. In the secular limit, the interference structure of the system is discarded and the resulting reduced dynamics is contractive, so the trace-distance revival disappears and for the initial-state pair considered here, the BLP witness vanishes within numerical precision. Retaining the non-secular terms allows nearby dressed transitions to remain coupled, so that phase information can be stored transiently in the downstream plaquette and returned to the emitter sector at later times. Microscopically, these non-secular terms hence represent the same finite-time storage-and-return process seen as the nonadiabatic lag of the downstream emitter. 

This also clarifies the role of the non-secular terms in enhancing entanglement generation: they open a coherence-mixing channel during the pre-locking transient, allowing the delayed downstream response to feed back coherently into the emitter subspace before the system settles into the more adiabatic, effectively secular CP-divisible regime. Taken together, the main discrepancy between MPS and TCL-2 in the optimal regime is associated with the buildup of system-bath correlations that are discarded in the Born factorization. At the same time, for the initial-state pairs used in our distinguishability diagnostic, the reduced emitter dynamics shows only weak revivals of trace distance, so the corresponding operational signatures of information backflow remain small. In that sense, the dominant correction captured by MPS is beyond-Born rather than strongly beyond-Markovian. Nonsecular terms nevertheless remain essential: their role is to preserve coherence mixing between nearby dressed transitions, and this reduced-system correction is precisely what allows TCL-2 to reproduce the concurrence enhancement at the qualitative level.

\section{Robustness to dissipation and disorder}
\label{sec: robustness}

Thus far we have assumed complete uniformity of parameters, but common sources of experimental noise can affect atomic displacement, detuning patterns, and coupling with nonguided modes. In this section, we assess the sensitivity of the optimal driven protocol to several experimentally relevant imperfections within the nonsecular TCL-2 description. We keep the nominal optimal parameter set fixed and vary one perturbation class at a time. These scans are intended to identify comparative sensitivity trends on the timescale of the first concurrence peak and should not be taken as a full beyond-Born robustness validation of the microscopic dynamics.

\begin{figure}
    \centering
    \includegraphics[width=0.9\linewidth]{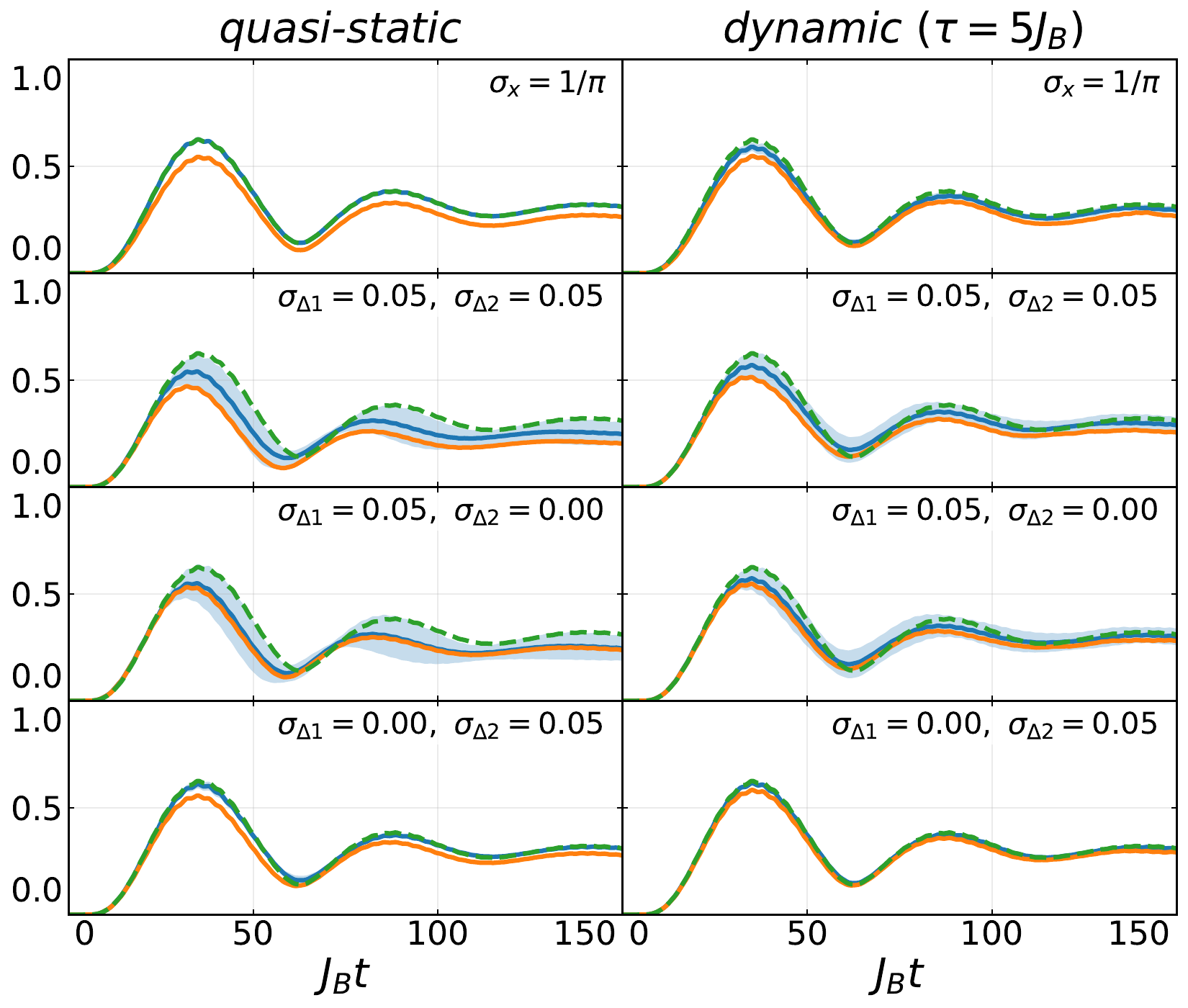}
    \caption{Concurrence dynamics in the presence of positional and detuning disorder over $N=100$ realizations. The left column corresponds to quasi-static disorder and the right column to dynamic disorder with $\tau=5/J_B$, representing moderately slow coloured noise. The top row shows displacement disorder with $\sigma_x=1/\pi$, while the remaining rows show detuning disorder for $(\sigma_{\Delta_1},\sigma_{\Delta_2})=(0.05,0.05)$, $(0.05,0.00)$, and $(0,0.05)$. Non-disordered evolution is represented by green dashed lines, while disordered evolution is given by the single-realization mean in the solid green line, single-realization standard deviation in the blue filled-area, and by the ensemble-averaged quantity in the orange solid line.}
    \label{fig:fig12}
\end{figure}

\figref{fig:fig12} shows the effects of both quasi-static and dynamic inter-nodal displacement on the concurrence. We model the former by sampling from $\mathcal{N}(\mu=0,\sigma)$, which may be interpreted as an error in initializing the trap location, and the latter by an Ornstein-Uhlenbeck process, $dq(t) = -\tau^{-1} q(t) dt + \sqrt{2\sigma_q^2\tau^{-1}} \, dW_t$, which captures thermally induced fluctuations in the trap positions. As expected for a cascaded setup, quasi-static displacement disorder produces essentially no change in the entanglement dynamics, even at the level of individual realizations. Dynamical fluctuations, by contrast, appreciably suppress the concurrence because time-dependent phase scrambling disrupts the nonsecular bath propagation. 

The effects of detuning error are by comparison less pronounced between quasi-static versus dynamic fluctuation than between application on the upstream versus downstream emitter. This is to be expected since the upstream emitter is strongly driven and detuning perturbations directly affect the width of the Mollow triplets, while the downstream emitter only receives excitations that still approximately remain within band-centre. Within the disorder strengths explored in \figref{fig:fig12}, detuning noise is therefore less detrimental than dynamical positional noise, especially when applied only to the downstream emitter.

\begin{figure}
    \centering
    \includegraphics[width=0.95\linewidth]{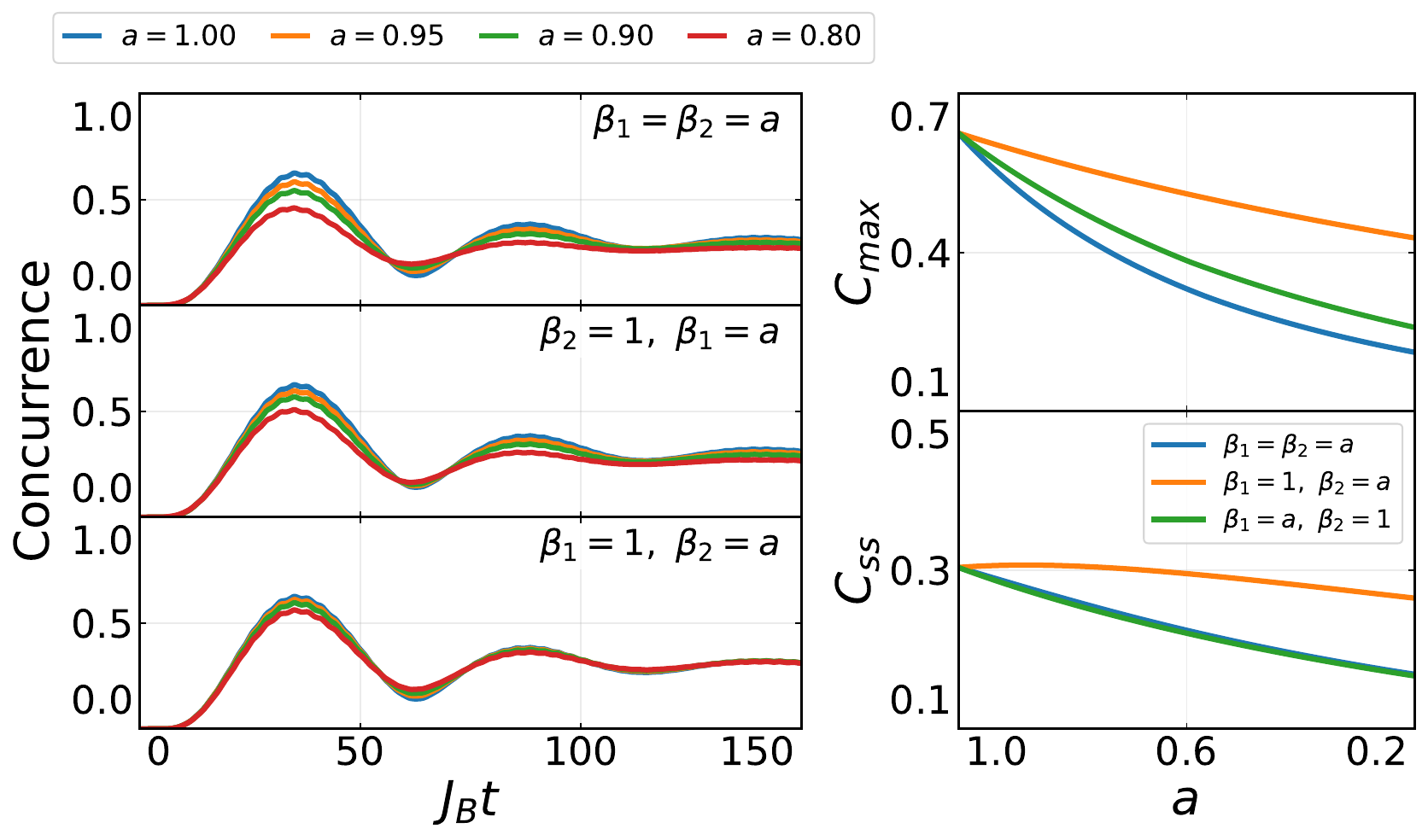}
    \caption{Left: concurrence versus time $J_Bt$ for imperfect coupling to the guided channel, parameterized by $a$ with legend values $a=1.00$, $0.95$, $0.90$, and $0.80$. The three rows correspond to $\beta_1=\beta_2=a$ (upper), $\beta_2=1$, $\beta_1=a$ (middle), and $\beta_1=1$, $\beta_2=a$ (lower). Right: corresponding maximum concurrence $C_{\max}$ (upper) and steady-state concurrence $C_{ss}$ (lower) as functions of $a$ for the same three dissipation configurations. For all cases, $\gamma_{tot}=1$.}
    \label{fig:fig13}
\end{figure}

We also investigated dissipation in the form of emission into nonguided modes (characterised by the $\beta$-factor: $\beta_i \coloneq (\gamma_{tot} - \gamma_{loss})/\gamma_{tot}$), as shown in \figref{fig:fig13}. Loss at the downstream emitter produces the weakest suppression of both the transient concurrence peak and the late-time concurrence-related observables. This is consistent with the physical picture in which the upstream emitter prepares and injects the narrowband excitation that subsequently hybridizes with the downstream plaquette. When only the downstream emitter has imperfect guided-mode coupling, much of that transport channel remains intact provided $\beta_2$ is not too small, consistent with local excitation retention and transient bound-state capture.

\section{Conclusions and Outlook}
\label{section: conclusions}

In summary, we studied transient entanglement generation in driven chiral networks across a hierarchy of models, ranging from an emitter-only Born--Markov description to a microscopic spin-chain treatment benchmarked by MPS simulations. By combining analytical TCL-ME methods with numerical MPS calculations, we characterized the strong-driving regime beyond the usual emitter-only secular picture and beyond the familiar single-excitation setting~\cite{ramos2016non}. Our results demonstrate that within the driven emitter-only model, continuous driving can increase the transient concurrence beyond the undriven $2/e$ benchmark~\cite{gonzalez2015chiral}. It is worth emphasising that this benchmark is tied to an undriven, effectively single-excitation, emitter-only setting; the microscopic spin-chain model should therefore be viewed not as a strict like-for-like continuation of that problem, but as a qualitative test of which enhancement mechanisms survive once the bath is structured and explicit system-bath correlations are allowed.

A central conclusion of this work is that in the strong-driving regime, the key mechanism behind this enhancement is the breakdown of the secular approximation~\cite{haikka2010non,haikka2010comparing,farina2019open}. When nearby dressed transitions are not well separated on the dissipative timescale, the non-secular coherence-population couplings discarded in the secularized treatment become quantitatively important. Rather than degrading performance, these terms constructively mix dressed-state coherences and produce a pronounced increase in the transient concurrence. Comparison between TCL-ME treatments~\cite{breuer1999stochastic,breuer2001time,breuer2002theory} and MPS simulations further clarifies the distinct roles of the Born--Markov and secular approximations: while standard indicators suggest that the reduced dynamics remains close to Markovian, beyond-Born corrections in the microscopic model and non-secular corrections in the emitter description are both essential for accurately capturing the strong-driving regime.

We also performed a systematic optimization over key system parameters, including driving strength, hopping amplitude, and system–bath coupling, and analyzed the robustness of the protocol against experimentally relevant imperfections such as imperfect chirality, fluctuations in inter-node distance, and spontaneous emission. These findings indicate that the proposed scheme is not only theoretically insightful but also suggests experimental viability within current quantum optical platforms. Overall, our work demonstrates that the breakdown of commonly used approximations, such as the secular approximation, can be harnessed as a resource rather than a limitation. This opens new avenues for engineering entanglement in driven-dissipative quantum systems and suggests that carefully controlled departures from standard approximations may play an important role in the design of near-term quantum technologies. 

A natural direction for future work is to generalize the present protocol to the preparation of multipartite entangled states, including GHZ, N00N, and Dicke states. In more realistic implementations of chirality, where the medium responds differently to left- and right-circularly polarized light, an idealized two-level description is no longer sufficient. One should instead consider at least a three-level model, with the relevant transition strengths set by the corresponding Clebsch--Gordan coefficients, as in alkali-atom platforms such as Rb or Cs. It would also be of interest to study the same system in a periodic geometry, where directional propagation around a ring could further reshape the transient interference pattern and possibly increase the achievable entanglement. Related systems, such as spin--cavity configurations, have already shown that larger concurrence values may be attainable in such settings~\cite{you2025generation}.

\section{Acknowledgements}

D.A. would like to acknowledge Dariel Mok, Jian Feng Kong, and Alexia Auffeves for many stimulating discussions about waveguide QED. D.A. also acknowledges the ``fluffy" support he has been receiving from his utterly cute cat, Paco. L.C.K. acknowledges support from the National Research Foundation and the Ministry of Education, Singapore.

\bibliography{biblio}

\newpage

\appendix
\counterwithin{figure}{section}
\counterwithin{table}{section}
\onecolumngrid

\section{Detailed calculations for the undriven emitters and the weakly driven emitters}
\label{appendix: detailed calculations for undriven and weakly driven}
Here we present the derivations for the results obtained in Sec. ~\ref{section: othermethods}.

\subsection{Analytical results for the concurrence in the undriven and weakly driven regimes}
In the undriven and weakly driven regimes, we can neglect jump terms in the
master equation in \eqref{MasEq} and hence derive some analytical results by
working in the single-excitation subspace using a non-Hermitian Hamiltonian
approach.  Here, we follow the method outlined in great detail
in~\cite{mok2020microresonators} where an exact expression for concurrence
for the chirally coupled ring cavities has been obtained by applying the
aforementioned non-Hermitian Hamiltonian approach. Here, we start with the quantum nodes in the state $\ket{eg}$.

First we consider two undriven emitters and obtain analytic results for
the concurrence by writing the effective Hamiltonian and applying it on a
single-excitation ansatz wave function and obtain the Schr\"odinger equation for
the time-dependent amplitudes. We begin by writing the wavefunction in the
single-excitation subspace:
\begin{equation}
|\psi(t)\rangle=c_{gg}|gg \rangle+c_{eg}|eg \rangle+c_{ge}|ge \rangle
\end{equation}
After neglecting jump terms the non-Hermitian Hamiltonian takes the following form:
\begin{align}
    H_{non}=&-i\gamma_{R}e^{ikd}\sigma_{2}^{\dag}\sigma_{1}-i\gamma_{L}e^{ikd}\sigma_{1}^{\dag}\sigma_{2} \nonumber \\
    &-i\frac{(\gamma_{L}+\gamma_{R})}{2}(\sigma_{1}^{\dag}\sigma_{1}+\sigma_{2}^{\dag}\sigma_{2})
\end{align}
After writing the Schr\"odinger equation with the non-Hermitian Hamiltonian, we
obtain the following expressions for the amplitudes:
\begin{align}
\dot{c_{eg}}&=-[\gamma_{L}e^{ikd}c_{ge}+\bar\gamma c_{eg}]  \\
\dot{c_{ge}}&= -[\gamma_{R}e^{ikd}c_{eg}+\bar\gamma c_{ge}]
\end{align}
where $\bar\gamma=(\gamma_{L}+\gamma_{R})/2$.  Solving these equations with the initial conditions $c_{eg}(0)=1$ and $c_{ge}(0)=0$ gives:
\begin{align}
c_{eg}(t)=&e^{-\bar\gamma t}[\cosh{[\alpha t \cos{kd}]}\cos{[\alpha t \sin{k d}]}  \nonumber \\ 
&+ i \sinh{[\alpha t \cos{kd}]}\sin{[\alpha t \sin{kd}}]] \\
c_{ge}(t)=& -\sqrt{\frac{\gamma_{R}}{\gamma_{L}}}e^{-\bar\gamma t}[\sinh{[\alpha t \cos{kd}]}\cos{[\alpha t \sin{k d}]}  \nonumber \\  
&+ i \cosh{[\alpha t \cos{kd}]}\sin{[\alpha t \sin{kd}}]],
\end{align}
where $\alpha=\sqrt{\gamma_{R}\gamma_{L}}$.  Having these expressions it is
easy to obtain the concurrence $C=2|c_{eg}(t)c^{*}_{ge}(t)|$ for the two
two-level emitters:

\begin{equation}
C(t)=2e^{-2\bar\gamma t}\sqrt{\frac{\gamma_{R}}{\gamma_{L}}}\sqrt{\left [\sinh^{2}{\alpha_{1}t}+ \sin^{2}{\alpha_{2}t}          \right ]},
\end{equation}
where $\alpha_{1}=2\sqrt{\gamma_{L}\gamma_{r}}\cos{kd}$ and
$\alpha_{2}=2\sqrt{\gamma_{L}\gamma_{r}}\sin{kd}$.  We comment that limit
$\lim_{\gamma_{L} \to 0}C(t)$ exists and is given by the following
expression:
\begin{equation}
C(t)=2\gamma_{R}t e^{-2\bar\gamma t}
\end{equation}
This expression shows that at the time $t_{max}=1/\gamma_{R}$ Concurrence
obtains its maximal value:
\begin{equation}
    C_{max}=\frac{2}{e}
\end{equation}
We comment, that this result perfectly agrees with the findings in the
Ref.\cite{gonzalez2015chiral}.

\textbf{Non-Hermitian Hamiltonian approach for the weakly driven emitters.}
Next, we obtain some analytical and perturbative results in the limit of weak
driving $\Omega \ll \gamma_{R},\gamma_{L}$.  In this weak driving regime, to
a good approximation, the system stays in the single-excitation subspace
which allows us to use the single-excitation subspace ansatz
\begin{equation}
|\psi(t)\rangle=c_{gg}(t)|gg \rangle+c_{eg}(t)|eg \rangle+c_{ge}(t)|ge \rangle.
\end{equation}
After acting on this state with non-Hermitian Hamiltonian:
\begin{equation}
H=H_{non}+ \frac{1}{2}\sum_{\alpha=1}^{2}{\Omega_{\alpha}\sigma_{\alpha}^{(x)}}
\end{equation}
We obtain the following set of coupled linear differential equations for the probability amplitudes:
\begin{subequations}
\begin{align}
\dot{c_{gg}}&= ic_{eg}\frac{\Omega_{1}}{2}+ic_{ge}\frac{\Omega_{2}}{2}            \\
\dot{c_{eg}}&=  -c_{ge}\gamma_{L}e^{ikd}-\bar\gamma c_{eg}+i\frac{\Omega_{1}}{2}c_{gg}         \\
\dot{c_{ge}}&= -\gamma_{R}e^{ikd}c_{eg}-\bar\gamma c_{ge}+i\frac{\Omega_{2}}{2}c_{gg}
\end{align}
\end{subequations}
For this set of equations Torrey's matrix~\cite{torrey1949transient} is given
by:
\begin{equation}
M_{T}=
\begin{pmatrix}
s & i\frac{\Omega_{1}}{2} & i\frac{\Omega_{2}}{2}  \\
i\frac{\Omega_{1}}{2} & s+\bar\gamma & \gamma_{R}e^{ikd} \\
i\frac{\Omega_{2}}{2} & \gamma_{L}e^{ikd} & s+\bar\gamma \\
\end{pmatrix}
 \end{equation}
As always, it is rather instructive to look at the determinant of this matrix as it sheds the light in which parameter regimes exact analytics or perturbative approach can be implemented.
\begin{align}
\label{det}
\det{[M_{T}]}=&s((s+\bar\gamma)^2-e^{2i k d }\gamma_{L}\gamma_{R})+\frac{1}{4}(s+\bar\gamma)(\Omega_{1}^2+\Omega_{2}^2)  \nonumber \\
&-\frac{1}{4}\Omega_{1}\Omega_{2}e^{ikd}[\gamma_{L}+\gamma_{R}]
\end{align}
In the next section we consider some special cases when the determinant takes easily solvable forms as in general it is given by cubic equation and solutions are not going to look very pleasant.

\subsubsection{Single driven atom in perfectly chiral regime.$$\Omega_{1}=\Omega \neq 0, \Omega_{2}=0,\gamma_{R}\neq 0, \gamma_{L}=0$$}
In this parameter regime it is easy to see that expression for the determinant simplifies significantly, taking the following form:
\begin{equation}
   \det{[M_{T}]}=(s+\bar\gamma)((s+a)^{2}-b^2), 
\end{equation}
where $a=\bar\gamma/2$ and $b=\sqrt{(\bar\gamma/2)^{2}-(\Omega/2)^2}$.
Next for the initial conditions $c_{eg}(0)=1$,$c_{ge(0)}=c_{gg}(0)=0$, the off-diagonal coherences are obtained by calculating the following matrix elements of adjugate matrix:
\begin{align}
c_{eg}(s)&=A_{22}=\frac{s}{((s+a)^{2}-b^2)} \\
c_{ge}(s)&=A_{23}=-\frac{\gamma_{R}se^{ikd}}{(s+\bar\gamma)((s+a)^{2}-b^2)}
\end{align}

After performing the inverse Laplace transformations we obtain the following expressions for the amplitudes:
\begin{align}
c_{eg}(t)&=e^{-at}\left [\cosh{bt}-\frac{a}{b}\sinh{bt}  \right ]   \\
c_{ge}(t)&=-\gamma_{R}e^{ikd-at} \left [ \frac{1}{b}[1-\frac{2\bar\gamma^{2}}{\Omega^2}]\sinh{bt}-\frac{4\bar\gamma}{\Omega^2}[e^{-at}-\cosh{bt}]         \right ]
\end{align}
To obtain the expression for the concurrence all we have to do is to multiply those coherences:
\begin{align}
C(t)=&2|c_{eg}(t)c^{*}_{ge}(t)| \nonumber \\
=&\bigg| 2\gamma_{R}e^{-\bar\gamma t} \left[1-\frac{2\bar\gamma^{2}}{\Omega^2} \right] \left [\frac{1}{2b}\sinh{2bt}-\frac{a}{b^2}\sinh^{2}{bt} \right ] \nonumber \\
&-\frac{8\gamma_{R}\bar\gamma}{\Omega^2}e^{-\frac{3\bar\gamma}{2}t}\left[\cosh{bt}-\frac{a}{b}\sinh{bt}\right ]  \nonumber \\
&+\frac{8\gamma_{R}\bar\gamma}{\Omega^2}e^{-\bar\gamma t}\left [\cosh^{2}{bt}-\frac{a}{2b}\sinh{2bt}\right ] \bigg|
\end{align}
Again we notice that at time $t=0$, concurrence is zero, which agrees with
the fact that we start the system evolution from the separable state of two
qubits.  We also see that the concurrence of $2/e$ is obtained in the
limit of zero driving.

\subsubsection{Single weakly driven atom in non-perfect chiral regime. $$\Omega_{1}=\Omega \neq 0, \Omega_{2}=0,\gamma_{L} \neq 0, \gamma_{R}\neq 0, \gamma_L \neq \gamma_R$$}
In this case the determinant takes a form of cubic equation, and to obtain more insightful expression we take advantage of the fact that we are in the regime of weak driving, which allows us in finding the roots of qubic equation perturbatively. For this case, the determinant takes the form:
\begin{equation}
\label{detApp}
\det{[M_{T}]}=s(s+\bar\gamma)^2-\gamma_{L}\gamma_{R}s+\frac{\Omega^{2}}{4}(s+\bar\gamma)
\end{equation}
Now to find the roots of this equation perturbatively, let us assume the following ansatz:
\begin{equation}
\label{ansatz}
\tilde{s}=s_{O}+A \Omega^2+B \Omega^4
\end{equation}
where $s_{O}$ is a solution of Eq.(\ref{detApp}), when $\Omega=0$, and is given by three different roots as follows:
$s_{0}=0,s_{1,2}=-\bar\gamma \pm \sqrt{\gamma_{L}\gamma_{R}}$. In building this ansatz we were guided by the intuition that physical quantities in quantum optics normally depend on the intensities of external driving fields. Next we substitute this ansatz into the Eq.(\ref{detApp}) and impose that coefficients in front of $\Omega^{2}$ and $\Omega^4$ should be equal to zero. After implementing that we obtain for A and B:

\begin{align}
A&=-\frac{s_{O}+\bar\gamma}{4\left[2s_{O}(s_{O}+\bar\gamma)-\gamma_{L}\gamma_{R}+(s_{O}+\bar\gamma)^2\right]}, \\
B&=-\frac{A^2(3s_{O}+2\bar\gamma)+A/4}{2s_{O}(s_{O}+\bar\gamma)-\gamma_{L}\gamma_{R}+(s_{O}+\bar\gamma)^2}.
\end{align}

In what follows, we keep to the weak driving limit such that $\Omega^4$ terms are discarded (i.e. we assume $B=0$) and we provide the concise expressions for these terms. After some basic algebra we get:

\begin{align}
A(s_{O}=s_{0})&=-\frac{\bar\gamma}{(\gamma_{L}-\gamma_{R})^2}, \\
A(s_{O}=s_{1})&=\frac{1}{4(\sqrt{\gamma_{L}}-\sqrt{\gamma_{R}})^2}, \\
A(s_{O}=s_{2})&=\frac{1}{4(\sqrt{\gamma_{L}}+\sqrt{\gamma_{R}})^2}.
\end{align}

Once that we know the form of $A$ for each root determinant takes very simple form:

\begin{equation}
    \det{[M_{T}]}=(s-\tilde{s_{0}})(s-\tilde{s_{1}})(s-\tilde{s_{2}})+\mathcal{O}(\Omega^4)
\end{equation}

Again using the relevant elements of the adjugate matrix, we obtain the off-diagonal coherences:
\begin{align}
c_{eg}(s)&=A_{22}=\frac{s(s+\bar\gamma)}{(s-\tilde{s_{0}})(s-\tilde{s_{1}})(s-\tilde{s_{2}})} \\
c_{ge}(s)&=A_{23}=\frac{\gamma_{R}se^{ikd}}{(s-\tilde{s_{0}})(s-\tilde{s_{1}})(s-\tilde{s_{2}})}
\end{align}
After doing inverse Laplace transform we obtain:

\begin{align}
c_{eg}(t)=&e^{\tilde{s}_2 t}
+\left[\frac{\tilde{s}_{1}+\bar\gamma}{\tilde{s}_{1}-\tilde{s}_{2}}\right]
\left[e^{\tilde{s}_{1} t}-e^{\tilde{s}_{2} t}\right] \nonumber\\
&+\left[\frac{\tilde{s}_{0}(\tilde{s}_{1}+\bar\gamma)}
{(\tilde{s}_{1}-\tilde{s}_{2})(\tilde{s}_{0}-\tilde{s}_{1})}\right]
\left[e^{\tilde{s}_{0} t}-e^{\tilde{s}_{1} t}\right] \nonumber\\
&-\left[\frac{\tilde{s}_{0}(\tilde{s}_{2}+\bar\gamma)}
{(\tilde{s}_{1}-\tilde{s}_{2})(\tilde{s}_{0}-\tilde{s}_{2})}\right]
\left[e^{\tilde{s}_{0} t}-e^{\tilde{s}_{2} t}\right]
\end{align}

\begin{align}
c_{ge}(t)=&\gamma_{R}e^{ikd}
\left[\frac{1}{\tilde{s}_{1}-\tilde{s}_{2}}\right]
\left[e^{\tilde{s}_{1} t}-e^{\tilde{s}_{2} t}\right] \nonumber\\
&+\left[\frac{\gamma_{R}e^{ikd}\tilde{s}_{0}}
{(\tilde{s}_{1}-\tilde{s}_{2})(\tilde{s}_{0}-\tilde{s}_{1})}\right]
\left[e^{\tilde{s}_{0} t}-e^{\tilde{s}_{1} t}\right] \nonumber\\
&-\left[\frac{\gamma_{R}e^{ikd}\tilde{s}_{0}}
{(\tilde{s}_{1}-\tilde{s}_{2})(\tilde{s}_{0}-\tilde{s}_{2})}\right]
\left[e^{\tilde{s}_{0} t}-e^{\tilde{s}_{2} t}\right]
\end{align}

To obtain the expression for the concurrence all we have to do is to multiply those coherences:
\begin{equation}
    C(t)=2|c_{eg}(t)c^{*}_{ge}(t)|.
\end{equation}
It is also useful to note that the initial conditions $c_{eg}(0)=1$ and $c_{ge}(0)=0$ are satisfied, which implies $C(0)=0$, as expected for an initially separable state.  We also see that the maximal concurrence of $2/e$ is obtained in the limit of zero driving.  Thus, we conclude that starting from the state
$\ket{eg}$, it is better to leave the two emitters undriven than to weakly
drive them.

\section{Master equation approach for the undriven case}
\textbf{Torrey's solution.} To solve this equations we follow Torrey's
solution~\cite{torrey1949transient} outlined in great detail in the textbook
by DA.  Steck \cite{steck2007quantum}(For more details See Section 5.5.2.2).
The main idea here is to use Laplace transform on the Liovillian $L$(which is a matrix $n\times n$) of the system and after obtaining the inverse of so called Torrey's matrix to implement the inverse Laplace transform to obtain time dependent elements of the density matrix. In short this can be captured  through the following formula:
\begin{equation}
\rho(t)= \mathcal{L}^{-1} \left [ \frac{1}{s\delta_{ij}-L}\right ]\rho(0)
\end{equation}
where $\mathcal{L}^{-1}$ means inverse Laplace transformation. This expression has few advantages. Firstly it is very easy to change the initial conditions(as final solution is merely matrix multiplication between the inverse of the Torrey's matrix and the column of initial conditions) and secondly calculations can be performed by making us of matrix algebra. In Particular, one only needs to just by find the inverse of the Torrey's matrix $M_{T}=s\delta_{ij}-L$. Here, for calculating the inverse of Torrey's matrix we use the adjugate matrix approach, with which the matrix elements of the inverse matrix are given by:
\begin{equation}
A_{ij}=(-1)^{(i+j)}\frac{M_{ji}}{\det{[M_{T}]}}
\end{equation}
The adjugate matrix $M_{ji}$ is given by the determinant of the $(n-1)\times(n-1)$matrix which one obtains after crossing out from Torrey's matrix $j$-th row and $i$-th column. Same approach can be used for obtaining solutions for time dependent amplitudes only in that case we solve Schr\"odinger's equations for non-Hermitian Hamiltonian. In that case instead of Torrey's matrix one gets $M_{T}=s\delta_{ij}-H$. Later this Hamiltonian generated matrix we will also be calling as Torrey's matrix, since all the methodology developed for Torrey's matrix holds for it. \\
\textbf{Chiral master equations}. It is instructive to also derive concurrence using full master equation approach where we do not neglect jump terms and obtain differential equation for the density matrix elements. Since we are still working in the single-excitation subspace our basis is given by the following vectors: $|0 \rangle|=|gg \rangle;1 \rangle=|eg \rangle;|2 \rangle=|ge \rangle $. Master equation (\ref{MasEq}) gives the following set of coupled differential equations:
\begin{subequations}
\begin{align}
    \dot{\rho_{11}}&=-2\bar\gamma\rho_{11}-\gamma_{L}e^{-ikd}\rho_{12}-\gamma_{L}e^{ikd}\rho_{21} \\
    \dot{\rho_{22}}&= -2\bar\gamma\rho_{22}-\gamma_{R}e^{ikd}\rho_{12}-\gamma_{R}e^{-ikd}\rho_{21}\\
    \dot{\rho_{12}}&=-2\bar\gamma\rho_{12}-\gamma_{R}e^{-ikd}\rho_{11}-\gamma_{L}e^{ikd}\rho_{22}
\end{align}
\end{subequations}
In order to solve this system of equations let us introduce some new variables which will result into 4 coupled linear differential equations due to the fact that off diagonal terms of density matrix are complex numbers(In general,for $N$ dimensional density matrix number of density matrix elements is given by $N-1$ real diagonal elements and by $N(N-1)/2$ complex numbered coherences).

At first we introduce $Z=e^{-ikd}\rho_{12}$,complex number with  real and imaginary parts $Z_{1}$ and $Z_{2}$. Then we introduce the following variables:
\begin{subequations}
\begin{align}
\rho_{11}&=e^{-2\bar\gamma t}\tilde{\rho_{11}} \\
\rho_{22}&= e^{-2\bar\gamma t}\tilde{\rho_{22}}\\
Z_{1}&= e^{-2\bar\gamma t}\tilde{Z_{1}} \\
Z_{2}&= e^{-2\bar\gamma t}\tilde{Z_{2}} 
\end{align}
\end{subequations}
For new variables the master equations can be written as:
\begin{subequations}
\begin{align}
\dot{\tilde{\rho_{11}}}&=-2\gamma_{L}\tilde{Z_{1}} \\
\dot{\tilde{\rho_{22}}}&= -2\gamma_{R}\cos{2kd}\tilde{Z_{1}}+2\gamma_{R}\sin{2kd}\tilde{Z_{2}}\\
\dot{\tilde{Z_{1}}}&= -\gamma_{R}\cos{2kd}\tilde{\rho_{11}}-\gamma_{L}\tilde{\rho_{22}} \\
\dot{\tilde{Z_{2}}}&= \gamma_{R}\sin{2kd}\tilde{\rho_{11}}
\end{align}
\end{subequations}
 For this set of master equations Torrey's matrix takes the following form:
 \begin{equation}
M_{T}=
\begin{pmatrix}
s & 0 & 2\gamma_{L} & 0 \\
0 & s & 2\gamma_{R}\cos{2kd} & -2\gamma_{R}\sin{2kd}\\
\gamma_{R}\cos{2kd} & \gamma_{L} & s & 0 \\
-\gamma_{R}\sin{2kd} & 0 & 0 & s
\end{pmatrix}
 \end{equation}
 Determinant of Torrey's matrix is given by:
 \begin{equation}
 \det{M_{T}}=(s^2-\alpha_{1}^2)(s^2+\alpha_{2}^2),
 \end{equation}
where $\alpha_{1}=2\sqrt{\gamma_{L}\gamma_{r}}|\cos{kd}|$ and $\alpha_{2}=2\sqrt{\gamma_{L}\gamma_{r}}|\sin{kd}|$. For finding the Concurrence we are going to need the following matrix elements of the inverse matrix:
\begin{align}
    \tilde{Z_{1}}(s)&=A_{31}= \frac{-\gamma_{R}[s^2 \cos{2kd}+2\gamma_{R}\gamma_{L}\sin^{2}{2kd}]} {(s^2-\alpha_{1}^2)(s^2+\alpha_{2}^2)}       \\
    \tilde{Z_{2}}(s)&=A_{41}= \frac{\gamma_{R}\sin{2kd}[s^2-2\gamma_{R}\gamma_{L}\cos{2kd})}{(s^2-\alpha_{1}^2)(s^2+\alpha_{2}^2]}
\end{align}
After implementing the inverse Laplace transformation with initial conditions $\rho_{11}=1,\rho_{22}=0,Z_{1}=0,Z_{2}=0$ we obtain:
\begin{align}
Z_{1}(t)=&\frac{-e^{-2\bar\gamma t}}{4\gamma_{L}}[\frac{(2\gamma_{R}\gamma_{L}\sin^2{2kd}+\alpha_{1}^{2}\cos{2kd})}{\alpha_{1}}\sinh{\alpha_{1} t} \nonumber \\
&+\frac{(-2\gamma_{R}\gamma_{L}\sin^{2}{2kd}+\alpha_{2}^{2}\cos{2kd})}{\alpha_{2}}\sin{\alpha_{2} t}] \\
Z_{2}(t)=&-\frac{\sin{2kd}}{4\gamma_{L}}e^{-2\bar\gamma t}[\frac{(-2\gamma_{R}\gamma_{L}\cos{2kd}+\alpha_{1}^{2})}{\alpha_{1}}\sinh{\alpha_{1} t} \nonumber \\
&+\frac{(2\gamma_{R}\gamma_{L}\cos{2kd}+\alpha_{2}^{2})}{\alpha_{2}}\sin{\alpha_{2} t}]
\end{align}
After substituting the expression for $\alpha_{1}$ and $\alpha_{2}$ and using basic trigonometry more elegant expressions for coherences can be obtained:
\begin{align}
Z_{1}(t)&=\sqrt{\frac{\gamma_{R}}{4\gamma_{L}}}e^{-2\bar\gamma t}[\cos{kd}\sinh{\alpha_{1} t} 
-\sin{kd}\sin{\alpha_{2} t}] \\
Z_{2}(t)&=\sqrt{\frac{\gamma_{R}}{4\gamma_{L}}}e^{-2\bar\gamma t}[\sin{kd}\sinh{\alpha_{1} t} 
+\cos{kd}\sin{\alpha_{2} t}]
\end{align}

Concurrence can be obtained using expressions for $Z_{1}(t)$ and $Z_{2}(t)$:
\begin{align}
C^{2}(t)&=4|\rho_{12}|^2=|Z_{1}(t)|^2+|Z_{2}t)|^2 \\
    &=\frac{\gamma_{R}}{\gamma_{L}}e^{-4\bar\gamma t}  \left [\sinh^2{\alpha_{1}t}+\sin^2{\alpha_{2}t}\right ]
\end{align}
Notice that at $t=0$, concurrence is zero, as expected of an initially separated state. As shown in the previous section, the limit of perfectly chiral case ($\gamma_{L}=0$) exists and it can be shown again that maximum value of concurrence is given by $C_{max}=2/e$, which agrees with the non-Hermitian Hamiltonian approach studied in previous section.

\section{Details on TCL-2 model}
\label{appendix: tcl2details}
Here, we aim to derive the second-order time-convolutionless master equation and obtain the form of several related objects: the bath correlation function $C_{ij}(\tau)$ and the time-dependent kernels $\Gamma_{ij}(t;\omega)$ that contributes to the explicit operator content summarized in Table~\ref{tab:tcl2coefficients}. Throughout, we take $\hbar=1$.

\subsection{Introduction to the TCL protocol}

To briefly describe, the TCL protocol is a perturbative projection-operator technique that can be used to derive analytic equations-of-motion within the transient regime. Taking $\mathcal{P}$ and $\mathcal{Q}$ to be the projection superoperators of the relevant and irrelevant parts of the system respectively, we begin with the exact time-local equation:

\begin{equation}
    \frac{\partial}{\partial t} \mathcal{P} \rho(t) = \mathcal{K}(t)\mathcal{P}\rho(t) + \mathcal{I}(t)\mathcal{Q}\rho(t_0)
\end{equation}

where $\mathcal{K}$ and $\mathcal{I}$ are superoperators that respectively generate dynamics in the relevant part of the system and account for inhomogeneities, the former of which is defined as:

\begin{equation}
    \begin{split}
        \mathcal{K}(t) & = \alpha \mathcal{P} \mathcal{L}(t) [1-\Sigma(t)]^{-1} \mathcal{P} \\
        & = \alpha \mathcal{P} \mathcal{L}(t) \left(\sum_{n=0}^{\infty} [\Sigma(t)]^{n} \right) \mathcal{P} \\
        & = \sum_{n=1}^{\infty} \alpha^{n} \mathcal{K}_{n}(t),
    \end{split}
\end{equation}
where $\Sigma(t)$ is an auxiliary superoperator defined as:
\begin{equation}
    \Sigma(t) = \alpha \int_{t_0}^{t} ds \; \mathcal{G}(t,s) \mathcal{Q} \mathcal{L}(s) \mathcal{P} \mathcal{U}^{-1}(t,s).
\end{equation}
where $\mathcal{G}(t,s) = T_+ \exp[\alpha \int_s^t d\tau \mathcal{Q} \mathcal{L}(\tau)]$ is the propagator restricted to the $\mathcal{Q}$-subspace, and $\mathcal{U}^{-1}(t,s) = T_- \exp[\alpha \int_s^t d\tau \mathcal{L}(\tau)]$ is the full backwards-in-time propagator, while $T_\pm$ denotes forwards/backwards Dyson time-ordering and $\alpha$ is the dimensionless system-bath coupling parameter defined by: $H = H_0 + \alpha H_{SB}$.

Physically, the $\Sigma(t)$ superoperator thus represents the return influence of $\mathcal{Q}$ on the effective $\mathcal{P}$-space dynamics. Its resolvent factor $[1-\Sigma(t)]^{-1}$ arises within the TCL derivation upon solving for the $\mathcal{Q}$-part in terms of the $\mathcal{P}$-part; more details can be found in \cite{lidar2019lecture, breuer2002theory}. For the inverse to exist, we thus require $\norm{\Sigma(t)}<1$. Implicitly, we see that $\alpha$ must be small (i.e. weak system-bath coupling should hold).

To determine contributions to $n$-order, we can then expand it as a Neumann series: $\Sigma(t)=\sum_{m=1}^{\infty} \alpha^{m}\Sigma_{m}(t)$. This can be neatly interpreted as a resummation of repeated memory-feedback events, where each $n$-th reads as $n$ excursions into the irrelevant sector. Thus, although TCL keeps a time-local form, it retains memory effects within the time-dependent $[1-\Sigma(t)]^{-1}$ factor. 

This also makes it physically clear why the TCL protocol can fail beyond a ``strong-enough" system-bath coupling: when $[1-\Sigma(t)]^{-1}$ is undefined, $\mathcal{Q}\rho(t)$ is no longer uniquely defined by $\mathcal{P}\rho(t)$, and hence the dynamical map on $\mathcal{P}\rho(t)$ is non-invertible in some finite time (i.e. some direction in reduced-state space / some system degree-of-freedom has been removed during bath interaction).

Setting factorised initial conditions such that $\mathcal{Q}\rho(t_{0}) = 0$, we can apply the above to yield the cumulant expansion:

\begin{equation}
    \mathcal{K}(t) = \mathcal{P} \mathcal{L}(t) \left[\alpha I + \sum_{m=1}^{\infty} \alpha^{m+1} \Sigma_{m}(t) + \sum_{m,m'=1}^{\infty} \alpha^{m+m'+1} \Sigma_{m}(t) \Sigma_{m'}(t) + \cdots \right] \mathcal{P}
\end{equation}

Then, collating terms to the fourth order of $\alpha$:
\begin{flalign}
    &\begin{aligned}
        &\qquad\qquad \alpha^{1} : \mathcal{K}_{1}(t) =\alpha \mathcal{P} \mathcal{L} \mathcal{P} = 0 \\
        &\qquad\qquad \alpha^{2} : \mathcal{K}_{2}(t) =\alpha \mathcal{P} \mathcal{L} \Sigma_{1}(t) \mathcal{P} \\
        &\qquad\qquad \alpha^{3} : \mathcal{K}_{3}(t) =\alpha \mathcal{P} \mathcal{L} \left( \Sigma_{1}^{2}(t) + \Sigma_{2}(t) \right) \mathcal{P} = 0 \\
        &\qquad\qquad \alpha^{4} : \mathcal{K}_{4}(t) =\alpha \mathcal{P} \mathcal{L} \left( \Sigma_{1}^{3}(t) + \{\Sigma_{1}(t), \Sigma_{2}(t)\} + \Sigma_{3}(t) \right) \mathcal{P}  \\
    \end{aligned}&&
\end{flalign}
Note that the odd-indexed orders vanish because we additionally assume that the bath state is centered (i.e. zero mean value) and is parity-symmetric (e.g. Gaussian).

This gives the perturbative TCL master equation up to fourth-order:

\begin{equation}
    \begin{split}
        \frac{\partial}{\partial t} \mathcal{P} \rho(t) & = \biggl\{ \alpha^2 \int_{0}^{t_0} \mathrm{d}t_1 \>\> \mathcal{P} \mathcal{L}(t_0) \mathcal{L}(t_1) \mathcal{P} \\
        & \quad + \alpha^4 \int_{0}^{t_0} \mathrm{d}t_1 \int_{0}^{t_1} \mathrm{d}t_2 \int_{0}^{t_2} \mathrm{d}t_3 \>\> \mathcal{P} \mathcal{L}(t_0) \mathcal{L}(t_1) \mathcal{L}(t_2) \mathcal{L}(t_3) \mathcal{P} - \mathcal{P} \mathcal{L}(t_0) \mathcal{L}(t_1) \mathcal{P} \mathcal{L}(t_2) \mathcal{L}(t_3) \mathcal{P} \\
        & \qquad\qquad\qquad\qquad\qquad\qquad - \mathcal{P} \mathcal{L}(t_0) \mathcal{L}(t_2) \mathcal{P} \mathcal{L}(t_1) \mathcal{L}(t_3) \mathcal{P} - \mathcal{P} \mathcal{L}(t_0) \mathcal{L}(t_3) \mathcal{P} \mathcal{L}(t_1) \mathcal{L}(t_2) \mathcal{P} \biggr\} \>\> \mathcal{P}\rho(t)
    \end{split}
\end{equation}

Our choice of projector will determine the accuracy of this perturbative expansion to its exact form. Generally for convenience, the standard projector, $\mathcal{P}\rho = \mathrm{Tr}_{B}(\rho) \otimes \rho_{B}$, is used where the bath is assumed to be time-independent, thus giving:
\begin{equation}
    \frac{\partial}{\partial t} \rho_{S}(t) = -\int_{0}^{t_0} \mathrm{d}t_1 \>\> \mathrm{Tr}_{B} [H_{I}(t_0),[H_{I}(t_1), \rho_{S}(t) \otimes \rho_{B}]] + \cdots
    \label{eq:tcl2firstform}
\end{equation}
for $\alpha = 1$. 

Higher-order TCL truncations are not pursued here. Historically, their practical use has been limited by the rapidly increasing cost of evaluating the higher-order generators, although recent work has renewed interest in systematic benchmarking of TCL-4 and even TCL-6 corrections~\cite{chen2025benchmarking,kumar2025asymptotic,crowder2024invalidation,lampert2025sixth}.

\subsection{Microscopic model and low-excitation reduction}

Having discussed the TCL protocol, we begin calculation proper from an XX spin chain and then specialize to its one-magnon sector. Let $x\in\mathbb Z$ label the bath sites. In the rotating frame relevant to the main matter, the bath Hamiltonian may be written as:
\begin{equation}
    H_B
    =
    \Delta\sum_x \sigma_{c_x}^+\sigma_{c_x}^-
    +
    J\sum_x
    \left(
        \sigma_{c_{x+1}}^+\sigma_{c_x}^-
        +
        \sigma_{c_x}^+\sigma_{c_{x+1}}^-
    \right),
    \label{eq:appD_spinchain}
\end{equation}
where $\Delta$ is the uniform on-site detuning and $J$ is the nearest-neighbour hopping/exchange scale. In the zero- and one-excitation sector of the chain, the hard-core nature of the bath remains insignficant, and one may replace the spin-flip operators by ordinary annihilation and creation operators, $\sigma_{c_x}^-, \sigma_{c_x}^+ \rightarrow c_x, c_x^\dagger$, so that:
\begin{equation}
    H_B = \Delta\sum_x c_x^\dagger c_x + J\sum_x \left(c_{x+1}^\dagger c_x + c_x^\dagger c_{x+1}\right).
    \label{eq:appD_bosonisedHB}
\end{equation}

The two emitters are labelled by $i\in\{1,2\}$, and the system Hamiltonian in the rotating frame is taken to be:
\begin{equation}
    H_S
    =
    \frac{\delta_1}{2}\sigma_1^z
    +
    \frac{\delta_2}{2}\sigma_2^z
    +
    \frac{\Omega_1}{2}\sigma_1^x
    +
    \frac{\Omega_2}{2}\sigma_2^x,
    \label{eq:appD_HS_general}
\end{equation}
and we set the emitters to be on-resonance with the bath and only drive the first emitter: $\delta_1=\delta_2=0, \Omega_2=0$.

Each emitter couples to the chain at two spatially separated points. We denote the two attachment points of emitter $i$ by $L[i]$ and $R[i]$, located at sites $x_{iL}$ and $x_{iR}$. The interaction Hamiltonian is:
\begin{equation}
    H_I
    =
    \sum_{i=1}^2
    \left(
        \sigma_i^+ Q_i + \sigma_i^- Q_i^\dagger
    \right),
    \qquad
    Q_i
    \equiv
    \sum_{\alpha\in\{L[i],R[i]\}}
    \lambda_{i\alpha}\, c_{x_{i\alpha}}.
    \label{eq:appD_HI_realspace}
\end{equation}
The complex amplitudes $\lambda_{i\alpha}$ encode the synthetic gauge phase of the triangular giant-atom plaquette. We parameterize them as $\lambda_{i\alpha}=g_i e^{is_\alpha\varphi_i}$ and $s_{L[i]}=-1, s_{R[i]}=+1$. Thus the left and right coupling points of emitter $i$ carry phases $e^{-i\varphi_i}$ and $e^{+i\varphi_i}$ respectively.

\subsection{Propagator and emitter-resolved kernel of the bath}

The infinite homogeneous chain is diagonalized by the plane-wave transform, $c_x = \frac{1}{\sqrt{2\pi}}\int_{-\pi}^{\pi}dk\, e^{ikx}c_k$, to give:
\begin{equation}
    H_B
    =
    \int_{-\pi}^{\pi}dk\, \varepsilon_k\, c_k^\dagger c_k,
    \qquad
    \varepsilon_k = \Delta + 2J\cos k,
    \qquad
    v_g(k)
    =
    \frac{d\varepsilon_k}{dk}
    =
    -2J\sin k.
    \label{eq:appD_dispersion}
\end{equation}

The interaction-picture evolution of the bath operator is thus:
\begin{equation}
    c_x(t)
    \equiv
    e^{+iH_B t}c_x e^{-iH_B t}
    =
    \frac{1}{\sqrt{2\pi}}\int_{-\pi}^{\pi}dk\,
    e^{ikx}e^{-i\varepsilon_k t}c_k.
    \label{eq:appD_cxt}
\end{equation}

We begin now by taking the bath to start in the (stationary) vacuum, $\rho_B=\ket{\mathrm{vac}}\bra{\mathrm{vac}}.$ Then, the only non-vanishing two-point site correlator is:
\begin{equation}
    \langle c_x(t)c_{x'}^\dagger(0)\rangle_B
    =
    \frac{1}{2\pi}\int_{-\pi}^{\pi}dk\,
    e^{ik(x-x')}e^{-i(\Delta+2J\cos k)t}.
    \label{eq:appD_sitecorrelator1}
\end{equation}
Defining $d=x-x'$, we can thus evaluate the exact single-magnon propagator as:
\begin{equation} \label{eq:appD_sitecorrelator2}
    \begin{split}
    G_d(t) \equiv
    \langle c_x(t)c_{x'}^\dagger(0)\rangle_B & =
    e^{-i\Delta t} \frac{1}{2\pi}\int_{-\pi}^{\pi}dk\, e^{ikd}e^{-i2Jt\cos k} \\
    & = e^{-i\Delta t} \sum_n (-i)^n \mathcal{J}_n(2Jt)
    \frac{1}{2\pi}\int_{-\pi}^{\pi}dk\, e^{ik(d+n)} \\
    & = e^{-i\Delta t} (-i)^d \mathcal{J}_d (2Jt).
    \end{split}
\end{equation}

We now turn to the collective bath operators $Q_i$ appearing in the interaction Hamiltonian. From ~\eqref{eq:appD_HI_realspace},
\begin{equation}
    Q_i(t)
    =
    e^{+iH_B t}Q_i e^{-iH_B t}
    =
    \sum_{\alpha\in\{L[i],R[i]\}}
    \lambda_{i\alpha} c_{x_{i\alpha}}(t).
    \label{eq:appD_Qitreal}
\end{equation}
Substituting in the form of the bath operator in ~\eqref{eq:appD_cxt}, we arrive at:
\begin{equation}
    Q_i(t)
    =
    \frac{1}{\sqrt{2\pi}}\int_{-\pi}^{\pi}dk\,
    F_i(k)\, e^{-i\varepsilon_k t} c_k,
    \label{eq:appD_Qitk}
\end{equation}
where we introduced the giant-atom form factor:
\begin{equation}
    F_i(k)
    \equiv
    \sum_{\alpha\in\{L[i],R[i]\}}
    \lambda_{i\alpha} e^{ikx_{i\alpha}}
    =
    g_i
    \left(
        e^{-i\varphi_i}e^{ikx_{iL}}
        +
        e^{+i\varphi_i}e^{ikx_{iR}}
    \right).
    \label{eq:appD_formfactor}
\end{equation}
This is the direct giant-atom analogue of the momentum-dependent coupling function in the waveguide treatment. The directionality of the triangular plaquette is encoded in the interference structure of $F_i(k)$.

Because the bath is initially set in the vacuum, the only non-vanishing contraction is:
\begin{equation}
    \langle c_k(t)c_{k'}^\dagger(0)\rangle_B
    =
    e^{-i\varepsilon_k t}\delta(k-k').
\end{equation}
Therefore, the only non-vanishing emitter-resolved bath correlator is:
\begin{equation}
    \begin{split}
    C_{ij}(\tau)
    \equiv
    \langle Q_i(\tau)Q_j^\dagger(0)\rangle_B
    & =
    \frac{1}{2\pi}\int_{-\pi}^{\pi}dk\,
    F_i(k)F_j^*(k)e^{-i\varepsilon_k \tau} \\
    & =
    \frac{1}{2\pi}\int_{-\pi}^{\pi}dk\,
    \sum_{\alpha,\beta}
    \lambda_{i\alpha}\lambda_{j\beta}^*
    e^{ik(x_{i\alpha}-x_{j\beta})}
    e^{-i\varepsilon_k\tau} \\
    & =
    \sum_{\alpha\in\{L[i],R[i]\}}
    \sum_{\beta\in\{L[j],R[j]\}}
    \lambda_{i\alpha}\lambda_{j\beta}^*
    G_{x_{i\alpha}-x_{j\beta}}(\tau) \\
    & = 
    g_i g_j
    \sum_{\alpha,\beta}
    e^{i(s_\alpha\varphi_i-s_\beta\varphi_j)}
    G_{x_{i\alpha}-x_{j\beta}}(\tau).
    \end{split}
    \label{eq:appD_Cdef}
\end{equation}

\subsection{Definition and explicit evaluation of the TCL-2 kernel}

With the bath correlator identified, we truncate \eqref{eq:tcl2firstform} to the first term (i.e. TCL-2) and move to the interaction picture for convenience, thus receiving:
\begin{equation}
    \frac{d}{dt}\tilde{\rho}_S(t)
    =
    -\int_0^t d\tau\,
    \mathrm{Tr}_B
    \left[
        \tilde{H}_I(t),
        \left[
            \tilde{H}_I(t-\tau),
            \tilde{\rho}_S(t)\otimes \rho_B
        \right]
    \right],
    \label{eq:appD_TCL2start}
\end{equation}
where all interaction picture objects have been accented with a tilde, and:
\begin{equation}
    \tilde{H}_I(t)
    =
    \sum_{i=1}^2
    \left[
        \tilde{\sigma}_i^+(t)Q_i(t)
        +
        \tilde{\sigma}_i^-(t)Q_i^\dagger(t)
    \right].
    \label{eq:appD_HIint}
\end{equation}

At this point, it is convenient to note that because the reference bath state is in vacuum, the only non-vanishing contraction is: $\mathrm{Tr}_B\!\left[Q_i(t)Q_j^\dagger(t-\tau)\rho_B\right] = \langle Q_i(t)Q_j^\dagger(t-\tau)\rangle_B$.
All terms containing $Q_i^\dagger Q_j$, $Q_iQ_j$, or $Q_i^\dagger Q_j^\dagger$ vanish. Furthermore, by stationarity of the bath correlator, we can then simplify the two-time dependence:
\begin{equation}
    \langle Q_i(t)Q_j^\dagger(t-\tau)\rangle_B
    =
    \langle Q_i(\tau)Q_j^\dagger(0)\rangle_B
    =
    C_{ij}(\tau).
\end{equation}

Expanding the commutators and evaluating the trace over the bath in  ~\eqref{eq:appD_TCL2start} thus gives:
\begin{equation}
\begin{aligned}
    \frac{d}{dt}\tilde{\rho}_S(t)
    =
    \sum_{i,j=1}^2 \int_0^t d\tau
    \Bigl[
    &
    C_{ij}(\tau)
    \Bigl(
        \tilde{\sigma}_j^-(t-\tau)\tilde{\rho}_S(t)\tilde{\sigma}_i^+(t)
        -
        \tilde{\sigma}_i^+(t)\tilde{\sigma}_j^-(t-\tau)\tilde{\rho}_S(t)
    \Bigr)
    \\
    +
    &
    C_{ji}^*(\tau)
    \Bigl(
        \tilde{\sigma}_i^-(t)\tilde{\rho}_S(t)\tilde{\sigma}_j^+(t-\tau)
        -
        \tilde{\rho}_S(t)\tilde{\sigma}_j^+(t-\tau)\tilde{\sigma}_i^-(t)
    \Bigr)
    \Bigr].
\end{aligned}
    \label{eq:appD_TCL2expanded}
\end{equation}

Next, to cast Eq.~\eqref{eq:appD_TCL2expanded} into a more compact and physically tractable form, we decompose each interaction-picture lowering operator into components of definite Bohr frequency with respect to $H_S$ to extract their time-dependency:
\begin{equation}
    \tilde{\sigma}_i^-(t)
    =
    \sum_{\omega\in\mathcal{F}_i}
    e^{-i\omega t}S_i(\omega),
    \qquad
    \tilde{\sigma}_i^+(t)
    =
    \sum_{\omega\in\mathcal{F}_i}
    e^{+i\omega t}S_i^\dagger(\omega).
    \label{eq:appD_freqdecomp}
\end{equation}
Then:
\begin{equation}
    \tilde{\sigma}_j^-(t-\tau)
    =
    \sum_{\omega\in\mathcal{F}_j}
    e^{-i\omega(t-\tau)}S_j(\omega)
    =
    \sum_{\omega\in\mathcal{F}_j}
    e^{-i\omega t}e^{+i\omega\tau}S_j(\omega),
\end{equation}
and:
\begin{equation}
    \tilde{\sigma}_i^+(t)
    =
    \sum_{\omega'\in\mathcal{F}_i}
    e^{+i\omega' t}S_i^\dagger(\omega').
\end{equation}
Substituting into the first line of ~\eqref{eq:appD_TCL2expanded}, we obtain:
\begin{equation}
\begin{aligned}
    \sum_{i,j}\int_0^t d\tau\,
    C_{ij}(\tau)
    \Bigl(
        \tilde{\sigma}_j^-(t-\tau)\tilde{\rho}_S(t)\tilde{\sigma}_i^+(t)
        -
        \tilde{\sigma}_i^+(t)\tilde{\sigma}_j^-(t-\tau)\tilde{\rho}_S(t)
    \Bigr)
    \\
    =
    \sum_{i,j}
    \sum_{\omega\in\mathcal{F}_j}
    \sum_{\omega'\in\mathcal{F}_i}
    e^{i(\omega'-\omega)t}
    \left(
        \int_0^t d\tau\, C_{ij}(\tau)e^{+i\omega\tau}
    \right)
    \left[
        S_j(\omega)\tilde{\rho}_S(t), S_i^\dagger(\omega')
    \right].
\end{aligned}
\end{equation}
To absorb the time-integral within a kernel, we define:
\begin{equation}
    \Gamma_{ij}(t;\omega)
    \equiv
    \int_0^t d\tau\, C_{ij}(\tau)e^{-i\omega\tau},
    \label{eq:appD_Gammadef}
\end{equation}
the same expression is obtained after the conventional replacement $\omega\mapsto -\omega$ in the set of Bohr frequencies. Adopting that convention throughout, one arrives at the compact TCL-2 equation
\begin{equation}
    \frac{d}{dt}\tilde{\rho}_S(t)
    =
    \sum_{i,j=1}^2
    \sum_{\omega\in\mathcal{F}_j}
    \sum_{\omega'\in\mathcal{F}_i}
    e^{i(\omega'-\omega)t}
    \Gamma_{ij}(t;\omega)
    \left[
        S_j(\omega)\tilde{\rho}_S(t),
        S_i^\dagger(\omega')
    \right]
    + \mathrm{H.c.}
    \label{eq:appD_TCL2compact}
\end{equation}
where the main text then compresses all explicit operator content into the schematic notation, $e^{i\chi t}\,\xi\,[\alpha\tilde{\rho}_S,\beta]+\mathrm{H.c.}$, with fixed numerical factors absorbed into the displayed coefficients as displayed in Tab. \ref{tab:tcl2coefficients}.

Lastly, we note that the TCL kernel can be evaluated further for physical understanding. Substituting ~\eqref{eq:appD_Cdef}, it follows that:
\begin{equation}
    \begin{split}
    \Gamma_{ij}(t;\omega)
    & =
    \frac{1}{2\pi}\int_{-\pi}^{\pi}dk\,
    F_i(k)F_j^*(k)
    \int_0^t d\tau\, e^{-i(\omega+\varepsilon_k)\tau} \\
    & =
    \frac{1}{2\pi}\int_{-\pi}^{\pi}dk\,
    F_i(k)F_j^*(k)
    \frac{1-e^{-i(\omega+\varepsilon_k)t}}{i(\omega+\varepsilon_k)}.
    \end{split}
    \label{eq:appD_Gammamomentum}
\end{equation}
Expanding the form factors once more yields
\begin{equation}
    \Gamma_{ij}(t;\omega)
    =
    g_i g_j
    \sum_{\alpha,\beta}
    e^{i(s_\alpha\varphi_i-s_\beta\varphi_j)}
    K_{\alpha\beta}(t;\omega),
    \label{eq:appD_Gammasum}
\end{equation}
with:
\begin{equation}
    K_{\alpha\beta}(t;\omega)
    \equiv
    \frac{1}{2\pi}\int_{-\pi}^{\pi}dk\,
    e^{ik(x_{i\alpha}-x_{j\beta})}
    \frac{1-e^{-i(\omega+\Delta+2J\cos k)t}}{i(\omega+\Delta+2J\cos k)}.
    \label{eq:appD_Kkernel}
\end{equation}
On resonance, $\Delta=0$, this reduces to the form used in Sec.~III.C.1. ~\eqref{eq:appD_Gammadef}--\eqref{eq:appD_Kkernel} show explicitly how the structured spectrum of the chain enters the time-local generator: the chain does not provide a white bath, but a frequency- and time-dependent filter. While this is for all intents and purposes the simplest analytic form needed, Appendix \ref{appendix: analyticSpectralFilter} showcases proper closed-form expressions at the early- and late-time limits.

\subsection{Redfield limit and secular approximation}

The time-local Redfield equation is obtained by taking the long-time limit of the TCL-2 coefficients,
\begin{equation}
    \Gamma_{ij}^{\mathrm{R}}(\omega)
    \equiv
    \lim_{t\rightarrow\infty}\Gamma_{ij}(t;\omega)
    =
    \int_0^\infty d\tau\, C_{ij}(\tau)e^{-i\omega\tau}.
    \label{eq:appD_Redfielddef}
\end{equation}
Using ~\eqref{eq:appD_Kkernel} and the distributional identity:
\begin{equation}
    \lim_{t\rightarrow\infty}
    \frac{1-e^{-ixt}}{ix}
    =
    \pi\delta(x)-i\,\mathcal{P}\frac{1}{x},
\end{equation}
we obtain:
\begin{equation}
\begin{aligned}
    \Gamma_{ij}^{\mathrm{R}}(\omega)
    =
    g_i g_j
    \sum_{\alpha,\beta}
    e^{i(s_\alpha\varphi_i-s_\beta\varphi_j)}
    \frac{1}{2\pi}
    \int_{-\pi}^{\pi}dk\,
    e^{ik(x_{i\alpha}-x_{j\beta})}
    \left[
        \pi\delta(\omega+\Delta+2J\cos k)
        -
        i\,\mathcal{P}\frac{1}{\omega+\Delta+2J\cos k}
    \right].
\end{aligned}
    \label{eq:appD_Redfieldsplit}
\end{equation}
It is therefore natural to split between real and imaginary components: $\Gamma_{ij}^{\mathrm{R}}(\omega) = \frac{\gamma_{ij}(\omega)}{2} + iJ_{ij}(\omega)$, where the principal-value part is given by:
\begin{equation}
    J_{ij}(\omega)
    =
    -
    g_i g_j
    \sum_{\alpha,\beta}
    e^{i(s_\alpha\varphi_i-s_\beta\varphi_j)}
    \frac{1}{2\pi}
    \mathcal{P}\int_{-\pi}^{\pi}dk\,
    \frac{e^{ik(x_{i\alpha}-x_{j\beta})}}{\omega+\Delta+2J\cos k}.
    \label{eq:appD_Jij}
\end{equation}
This approximation thus essentially takes the Markovian limit by eliminating the memory kernel.

The secular approximation (not to be conflated with the rotating-wave approximation) can then be applied to enforce a GKSL form, thereby implicitly returning a physical CPTP map. Starting from ~\eqref{eq:appD_TCL2compact} or from its Redfield version, one discards all terms with $\omega\neq\omega'$ (i.e. all terms multiplying non-unity oscillatory factors $e^{i(\omega'-\omega)t}$). Physically, this can be interpreted as discarding interference terms between different transitions of the system, and it is only physically valid when $    |\omega-\omega'| \gg |\Gamma_{ij}^{\mathrm{R}}(\omega)|$ (i.e. when oscillatory factors average to zero at a much faster timescale than system evolution due to interaction with the bath); secularising in inappropriate regimes is akin to assuming rapid dephasing between decay channels and risks neglecting non-trivial coherence-transfer effects.

\subsection{Mid-band estimate of the delay phase}

The simple estimate for the delay phase quoted in Sec.~III.C.1 follows directly from the spin-chain dispersion. We first note that the resonant momenta satisfy $\omega+\Delta+2J\cos k_\omega=0$. Additionally, at the band centre and on resonance, $\Delta=0$ and $k_0=\pi/2$. Expanding about $k_0$ thus gives $ \cos k \approx \cos\!\left(\frac{\pi}{2}+\delta k\right) \approx -\delta k$, so that:
\begin{equation}
    \omega - 2J\delta k \approx 0
    \qquad\Longrightarrow\qquad
    k_\omega \approx \frac{\pi}{2}+\frac{\omega}{2J}.
    \label{eq:appD_komega}
\end{equation}
The corresponding group velocity at the band centre is $v_g(k_0)=2J$, so propagation over a distance $d$ corresponds to a delay time of $\tau_d \approx {d}/{2J}$. The phase picked up by a component of frequency $\omega$ after propagating over that distance is thus $e^{ik_\omega d} \approx e^{i\pi d/2}e^{i\omega d/(2J)}=e^{i\pi d/2}e^{i\omega\tau_d}$. For the two dressed sidebands $\omega=\pm\Omega_1$, the relative phase is therefore:
\begin{equation}
    \Phi_d
    =
    (k_{+\Omega_1}-k_{-\Omega_1})d
    \approx
    \frac{2\Omega_1 d}{2J}
    =
    2\Omega_1\tau_d
    \approx
    \frac{\Omega_1 d}{J}.
    \label{eq:appD_delayphase}
\end{equation}

\section{Analytic results for the time-broadened spectral filter function}
\label{appendix: analyticSpectralFilter}

We consider the integral:
\begin{equation}
I_n(t)=\frac{1}{2\pi}\int_{-\pi}^{\pi} dk\, e^{ikn}
\frac{1-e^{-i(\omega'+2J\cos k)t}}{i(\omega'+2J\cos k)},
\end{equation}
where
\begin{equation}
n := x_{i\alpha}-x_{j\beta}, \qquad m := |n|, \qquad a := 2J.
\end{equation}

We note that the exact Bessel-function representation form is particularly useful for deriving the long-time and short-time limits:
\begin{equation}
I_n(t)=(-i)^m\int_0^t ds\, e^{-i\omega' s}J_m(as).
\label{eq:bessel-rep}
\end{equation}

\subsection{Band center: $\omega'=0$}

At the band center, ~\eqref{eq:bessel-rep} reduces to:
\begin{equation}
I_n(t)=(-i)^m\int_0^t ds\, J_m(2Js).
\label{eq:band-center}
\end{equation}

\paragraph{Short-time behavior.}
For small argument,
\begin{equation}
J_m(z)\sim \frac{1}{m!}\left(\frac{z}{2}\right)^m,
\qquad z\to 0.
\end{equation}
Substituting $z=2Js$ into ~\eqref{eq:band-center}, we obtain:
\begin{equation}
J_m(2Js)\sim \frac{(Js)^m}{m!},
\end{equation}
and therefore
\begin{align}
I_n(t)
&\sim (-i)^m\int_0^t ds\,\frac{(Js)^m}{m!}
= (-i)^m\frac{J^m}{m!}\frac{t^{m+1}}{m+1}, \quad t\to 0.
\end{align}
Note that this short-time result holds independently of the regime in $\omega'$.\\

\paragraph{Long-time behavior.}
For fixed integer $m\ge 0$, one has the standard integral:
\begin{equation}
\int_0^\infty du\, J_m(u)=1.
\end{equation}
After the change of variables $u=2Js$, this gives:
\begin{equation}
\int_0^\infty ds\, J_m(2Js)=\frac{1}{2J}.
\end{equation}
Thus the long-time limit exists as:
\begin{equation}
I_n(\infty)=\frac{(-i)^m}{2J}.
\end{equation}
Therefore, at the band center, the asymptotic kernel has constant magnitude $\frac{1}{2J}$ and only a distance-dependent phase.

\subsection{Inside the band: $|\omega'|<2J$}

The long-time limit here is most cleanly derived from the Abel-regularized integral:
\begin{equation}
I_n^{(\infty)}(\omega')
:=(-i)^m\lim_{\eta\to 0^+}\int_0^\infty ds\, e^{-(\eta+i\omega')s}J_m(2Js).
\label{eq:abel}
\end{equation}
This allows us to use the Laplace transform identity:
\begin{equation}
\int_0^\infty ds\, e^{-ps}J_m(as)
=
\frac{\big(\sqrt{p^2+a^2}-p\big)^m}{a^m\sqrt{p^2+a^2}},
\qquad \Re p>0.
\end{equation}
Setting $p=\eta+i\omega'$ and $a=2J$, ~\eqref{eq:abel} becomes:
\begin{equation}
I_n^{(\infty)}(\omega')
=
(-i)^m
\lim_{\eta\to0^+}
\frac{\big(\sqrt{(\eta+i\omega')^2+(2J)^2}-(\eta+i\omega')\big)^m}
{(2J)^m\sqrt{(\eta+i\omega')^2+(2J)^2}}.
\label{eq:inside-general}
\end{equation}

Now for $|\omega'|<2J$,
\begin{equation}
\sqrt{(\eta+i\omega')^2+(2J)^2}\xrightarrow[\eta\to0^+]{}\sqrt{4J^2-\omega'^2}.
\end{equation}
Inside the band, it is natural to introduce a real wave number $q\in(0,\pi)$ by $\omega'=-2J\cos q.$ Thus,
\begin{equation}
\sqrt{4J^2-\omega'^2}=2J\sin q,
\qquad
-\omega'=2J\cos q.
\end{equation}
Hence:
\begin{align}
\frac{\sqrt{4J^2-\omega'^2}-i\omega'}{2J}
&=\sin q+i\cos q
=e^{\,i(\frac{\pi}{2}-q)}.
\end{align}
Substituting back into ~\eqref{eq:inside-general} gives:
\begin{align}
I_n(\infty)
&=
(-i)^m
\frac{e^{\,im(\frac{\pi}{2}-q)}}{\sqrt{4J^2-\omega'^2}}
=
\frac{e^{-imq}}{\sqrt{4J^2-\omega'^2}}.
\end{align}
Therefore,
\begin{equation}
I_n(\infty)=\frac{e^{-i|n|q}}{\sqrt{4J^2-\omega'^2}},
\qquad
q=\arccos\!\left(-\frac{\omega'}{2J}\right),
\qquad |\omega'|<2J.
\end{equation}

\paragraph{Approach to the asymptotic limit.}
We write:
\begin{equation}
I_n(t)=I_n(\infty)-R_n(t),
\qquad
R_n(t)=(-i)^m\int_t^\infty ds\, e^{-i\omega' s}J_m(2Js).
\end{equation}
For large $s$,
\begin{equation}
J_m(2Js)\sim \sqrt{\frac{1}{\pi Js}}
\cos\!\left(2Js-\frac{m\pi}{2}-\frac{\pi}{4}\right).
\end{equation}
Away from the band edge, the phase remains oscillatory, so the tail integral is of stationary/non-stationary oscillatory type with amplitude $s^{-1/2}$. Therefore,
\begin{equation}
I_n(t)=I_n(\infty)+O(t^{-1/2}),
\qquad |\omega'|<2J,\ \omega'\neq \pm 2J.
\end{equation}
Thus inside the band, the kernel approaches a finite oscillatory limit algebraically.

\subsection{Band edge: $|\omega'|=2J$}

This is the singular case. Consider first $\omega'=2J$. Then ~\eqref{eq:bessel-rep} gives:
\begin{equation}
I_n(t)=(-i)^m\int_0^t ds\, e^{-i2Js}J_m(2Js).
\label{eq:edge-plus}
\end{equation}
Using the large-$s$ asymptotic:
\begin{equation}
J_m(2Js)\sim \sqrt{\frac{1}{4\pi Js}}
\left[
e^{\,i(2Js-\frac{m\pi}{2}-\frac{\pi}{4})}
+
e^{-i(2Js-\frac{m\pi}{2}-\frac{\pi}{4})}
\right],
\end{equation}
we obtain:
\begin{align}
e^{-i2Js}J_m(2Js)
&\sim
\sqrt{\frac{1}{4\pi Js}}
\left[
e^{-i(\frac{m\pi}{2}+\frac{\pi}{4})}
+
e^{-i(4Js-\frac{m\pi}{2}-\frac{\pi}{4})}
\right].
\end{align}
The second term is oscillatory and contributes only $O(1)$ after integration, while the first term is nonoscillatory and yields:
\begin{equation}
\int_0^t ds\, s^{-1/2}=2\sqrt{t}.
\end{equation}
Hence,
\begin{equation}
I_n(t)\sim
(-i)^m
\frac{e^{-i(\frac{m\pi}{2}+\frac{\pi}{4})}}{\sqrt{\pi J}}\sqrt{t}
+O(1),
\qquad t\to\infty.
\end{equation}
Therefore:
\begin{equation}
I_n(t)\sim C_n\sqrt{t},
\qquad |\omega'|=2J,
\end{equation}
with a phase-dependent coefficient $C_n$.

A completely analogous argument applies for $\omega'=-2J$, with the corresponding phase structure. The important point is that at the band edge the long-time limit no longer exists as a finite constant. Instead, the integral grows like $\sqrt{t}$. This is the time-domain signature of the one-dimensional van Hove singularity.

\subsection{Near the band edge: $\omega'=\pm 2J+\Delta$, $|\Delta|\ll 2J$}

We now consider the crossover regime close to the band edge.\\

\paragraph{Approach from inside the band.}
Suppose $\Delta<0$ and write $\omega'=2J+\Delta$ with $|\Delta|\ll 2J$. Then
\begin{equation}
4J^2-\omega'^2 =-(4J\Delta+\Delta^2) \sim 4J|\Delta|,
\end{equation}
since $\Delta<0$. 
Hence from the inside-band formula,
\begin{equation}
I_n(\infty)=\frac{e^{-imq}}{\sqrt{4J^2-\omega'^2}}
\sim \frac{e^{-imq}}{\sqrt{4J|\Delta|}}.
\end{equation}
Thus the amplitude diverges as:
\begin{equation}
I_n(\infty)\sim \frac{1}{\sqrt{4J|\Delta|}},
\qquad \Delta\to 0^-.
\end{equation}

\paragraph{Approach from outside the band.}
Suppose instead $\Delta>0$, so that $|\omega'|>2J$.  \\

We first define $\kappa=\operatorname{arcosh}\!\left(|\omega'|/(2J)\right).$ Then for small positive detuning, using $\cosh\kappa=1+\kappa^2/2+\mathcal{O}(\kappa^4)$,
\begin{equation}
\frac{|\omega'|}{2J}=1+\frac{\Delta}{2J},
\end{equation}
so
\begin{equation}
\kappa\sim \sqrt{\frac{\Delta}{J}}.
\end{equation}
Since the outside-band solution behaves as $e^{-m\kappa}/\sqrt{\omega'^2-4J^2}$, and  $\omega'^2-4J^2\sim 4J\Delta$, we obtain:
\begin{equation}
I_n(\infty)\sim \frac{e^{-m\kappa}}{\sqrt{4J\Delta}}.
\end{equation}
Therefore:
\begin{equation}
I_n(\infty)\sim \frac{e^{-|n|\kappa}}{\sqrt{4J|\Delta|}},
\qquad
\kappa\sim \sqrt{\frac{|\Delta|}{J}},
\qquad \Delta\to 0^+.
\end{equation}
The associated localization length diverges as $\xi=\kappa^{-1}\sim \sqrt{J/|\Delta|}$. Thus the band edge controls the crossover from oscillatory propagating behavior to evanescent localized behavior.

\subsection{Outside the band: $|\omega'|>2J$}

Outside the band, the square root in ~\eqref{eq:inside-general} becomes imaginary. We define:
\begin{equation}
\Lambda:=\sqrt{\omega'^2-4J^2},
\qquad
\kappa:=\operatorname{arcosh}\!\left(\frac{|\omega'|}{2J}\right).
\end{equation}
Then
\begin{equation}
e^{-\kappa}=\frac{|\omega'|-\Lambda}{2J}.
\end{equation}

\paragraph{Case $\omega'>2J$.}
Using the Abel-regularised prescription in ~\eqref{eq:inside-general}, we find:
\begin{equation}
\sqrt{(\eta+i\omega')^2+(2J)^2}\xrightarrow[\eta\to0^+]{} i\Lambda.
\end{equation}
Hence
\begin{equation}
\frac{i\Lambda-i\omega'}{2J}
=
-i\,\frac{\omega'-\Lambda}{2J}
=
-i\,e^{-\kappa}.
\end{equation}
Raising this to the $m$-th power and multiplying by $(-i)^m$ gives:
\begin{equation}
(-i)^m(-i)^m e^{-m\kappa}=(-1)^m(-i)^{2m}e^{-m\kappa}=(-1)^m e^{-m\kappa}.
\end{equation}
Keeping track of the prefactor $1/(i\Lambda)=-i/\Lambda$, we obtain:
\begin{equation}
I_n(\infty)=
-\,i\,(-1)^m\frac{e^{-m\kappa}}{\sqrt{\omega'^2-4J^2}},
\qquad \omega'>2J.
\end{equation}

\paragraph{Case $\omega'<-2J$.}
A similar calculation gives
\begin{equation}
I_n(\infty)=
+\,i\,\frac{e^{-m\kappa}}{\sqrt{\omega'^2-4J^2}},
\qquad \omega'<-2J.
\end{equation}

Thus outside the band the asymptotic kernel is exponentially localized in distance:
\begin{equation}
I_n(\infty)\propto e^{-|n|\kappa},
\qquad
\kappa=\operatorname{arcosh}\!\left(\frac{|\omega'|}{2J}\right).
\end{equation}

\paragraph{Large but finite time.}
The finite-time remainder is
\begin{equation}
R_n(t)=(-i)^m\int_t^\infty ds\, e^{-i\omega' s}J_m(2Js).
\end{equation}
Using again the large-$s$ asymptotic of $J_m(2Js)$, the amplitude decays like $s^{-1/2}$ and the phase is non-resonant. Therefore
\begin{equation}
R_n(t)=O(t^{-1/2}),
\end{equation}
without the singular enhancement that occurs at the band edge. So the outside-band regime is much better behaved than the edge itself.

\subsection{Summary of regimes}

The asymptotic behavior of $I_n(t)$ is controlled by the position of $\omega'$ relative to the tight-binding band. In particular, the band center is a simple special case of the inside-band regime, while the band edge separates oscillatory propagating behavior from exponentially localized evanescent behavior. The full picture is summarized in Table \ref{tab: app}.

\begin{table*}[t]
\centering
\small
\renewcommand{\arraystretch}{1.18}
\setlength{\tabcolsep}{6pt}
\setlength{\arrayrulewidth}{0.8pt}
\setcellgapes{3pt}
\makegapedcells

\begin{tabular}{|l|c|c|l|}
\hline
\textbf{Regime}
& \textbf{Parameterization}
& \textbf{Large-$t$ behavior}
& \textbf{Spatial behavior} \\
\hline

Band center
& $\omega'=0$
& $I_n(\infty)=(-i)^{|n|}/(2J)$
& \makecell[l]{Pure phase,\\ constant magnitude} \\
\hline

Inside band
& \makecell[c]{$|\omega'|<2J$\\ $\omega'=-2J\cos q$}
& \makecell[c]{$I_n(\infty)=\dfrac{e^{-i|n|q}}{\sqrt{4J^2-\omega'^2}}$\\
$I_n(t)=I_n(\infty)+O(t^{-1/2})$}
& Oscillatory in $|n|$ \\
\hline

Band edge
& $|\omega'|=2J$
& $I_n(t)\sim C_n\sqrt{t}$
& Critical / singular \\
\hline

Outside band
& \makecell[c]{$|\omega'|>2J$\\
$\kappa=\operatorname{arcosh}\!\bigl(|\omega'|/(2J)\bigr)$}
& \makecell[c]{$I_n(\infty)\propto\dfrac{e^{-|n|\kappa}}{\sqrt{\omega'^2-4J^2}}$\\
$I_n(t)=I_n(\infty)+O(t^{-1/2})$}
& Evanescent in $|n|$ \\
\hline
\end{tabular}
\caption{Distinguished analytic regimes of the lattice integral.}
\label{tab: app}
\end{table*}

\section{Details on MPS algorithm and benchmarking diagnostics}
\label{appendix: mpsdetails}

In this supplementary section, we describe the framework used to represent our model in matrix-product-state (MPS) simulations. In brief, we manifest all spins in the model as two-level sites treated directly at the level of the density operator. This is because we need to implement absorbing boundary conditions (i.e. dissipative loss) at the ends of the chain to eliminate boundary reflections and simulate an infinitely long chain. The dynamics is evolved in Liouville space by means of an MPS representation of the density matrix and a matrix product operator (MPO) representation of the Liouvillian. The chain contains $N$ sites labelled by
$i=0,1,\dots,N-1$ and two distinguished sites within this chain, denoted by $n_1, n_2 \in i$, are identified as the emitter sites. Of note, due to the linear topology of MPS, a triangular plaquette formed between sites $\{n_1-1,n_1,n_1+1\}$ for example can no longer be fully cast using nearest-neighbour bonds; instead a next-nearest-neighbour bond connects sites $\{n_1-1,n_1+1\}$.

\subsection{Model on linear topology with absorbing boundary conditions}

Assuming a rotating frame, the coherent part of the dynamics can be described by the Hamiltonian:
\begin{equation}
\begin{split}
H
&=
\sum_{i=0}^{N-1}\Delta_i \sigma_i^+\sigma_i^-
+\sum_{i=0}^{N-1}\left(\Omega_i \sigma_i^-+\Omega_i^\ast \sigma_i^+\right)\\
&\quad
+\sum_{i=0}^{N-2}
\left(
J_{1,i}\sigma_{i+1}^+\sigma_i^- + J_{1,i}^\ast \sigma_i^+\sigma_{i+1}^-
\right)\\
&\quad
+\sum_{i=0}^{N-3}
\left(
J_{2,i}\sigma_{i+2}^+\sigma_i^- + J_{2,i}^\ast \sigma_i^+\sigma_{i+2}^-
\right),
\end{split}
\end{equation}
where $\sigma_i^\pm$ are the raising and lowering operators of site $i$. The coefficients $\Delta_i$, $\Omega_i$, $J_{1,i}$, and $J_{2,i}$ denote respectively onsite detunings, local coherent drives, nearest-neighbour hopping amplitudes, and next-nearest-neighbour hopping amplitudes.

Upon designating the two emitter sites $n_1 , n_2 \in i$ in an otherwise uniform chain, we implement standard settings of no-detuning among all spins $(\Delta_i = 0)$ and coherent driving only on emitter sites $(\Omega_i = 0, i \notin \{n_1, n_2\})$. The nearest-neighbour couplings in the bulk are uniform as a rule $(J_{1,i}=J_B)$; we take this to define the time-scale of the spin-chain.

To implement the triangular plaquette, the bonds adjacent to the emitter are replaced with scaled complex amplitudes $(J_{1,n_1-1}=J_{1,n_1}=g_1 e^{-i\varphi_1}; \; 
J_{1,n_2-1}=J_{1,n_2}=g_2 e^{-i\varphi_2})$, while the next-nearest-neighbour bypass couplings are introduced across the emitter positions $(J_{2,n_1-1}=J_{2,n_2-1}=J_B)$ with all other $J_{2,i}$ taken to vanish. Physically, these terms provide a direct channel across each emitter position while preserving the chain geometry.

Next, dissipation is incorporated through a local-loss Lindblad master equation,
\begin{equation}
\label{eq:lindblad}
\partial_t \rho
=
-i[H,\rho]
+
\sum_{i=0}^{N-1}\zeta_i
\left(
\sigma_i^- \rho \sigma_i^+
-\frac{1}{2}\left\{\sigma_i^+\sigma_i^-,\rho\right\}
\right),
\end{equation}
where $\zeta_i$ is the decay rate at site $i$. It is convenient to introduce the non-Hermitian effective Hamiltonian
\begin{equation}
\label{eq:Heff}
H_{\mathrm{eff}}
=
H-\frac{i}{2}\sum_{i=0}^{N-1}\zeta_i \sigma_i^+\sigma_i^-,
\end{equation}
in terms of which the equation of motion may be rewritten as
\begin{equation}
\label{eq:lindblad_heff}
\partial_t \rho
=
i\left(\rho H_{\mathrm{eff}}^\dagger-H_{\mathrm{eff}}\rho\right)
+
\sum_{i=0}^{N-1}\zeta_i \sigma_i^- \rho \sigma_i^+.
\end{equation}

The dissipation profile is chosen to represent lossy boundaries only at the edge sites rather than a graduated loss profile $(\zeta_0 =\zeta_{N-1}=2J_B)$, with the emitter sites and all other bath sites kept lossless $(\zeta_i=0, i\notin\{0,N-1\})$.
The resulting model is thus an open chain with boundary losses, two locally driven emitter sites, emitter-modified nearest-neighbour couplings, and bypass couplings across the emitter positions.

\subsection{Liouville-space formulation and MPS/MPO representations}

Since the evolution is dissipative, it is natural to represent the density operator directly rather than work with wavefunctions. The density matrix is vectorized according to
\begin{equation}
|\rho\rangle\rangle = \operatorname{vec}(\rho),
\end{equation}
so that the master equation becomes a linear equation in Liouville space,
\begin{equation}
\label{eq:LiouvilleEq}
\partial_t |\rho\rangle\rangle
=
\mathcal L |\rho\rangle\rangle.
\end{equation}
Using the standard identity under the column-stacking convention
\begin{equation}
\operatorname{vec}(A\rho B^\dagger)
=
(B^\ast\otimes A)\operatorname{vec}(\rho),
\end{equation}
the Liouvillian corresponding to Eq.~(\ref{eq:lindblad_heff}) takes the form
\begin{equation}
\label{eq:Liouvillian}
\mathcal L
=
i\left(H_{\mathrm{eff}}^\ast\otimes I - I\otimes H_{\mathrm{eff}}\right)
+
\sum_{i=0}^{N-1}\zeta_i\, \sigma_i^- \otimes \sigma_i^- .
\end{equation}

With this, the vectorized density operator is represented as an open-boundary MPS,
\begin{equation}
\label{eq:MPSrho}
|\rho\rangle\rangle
=
\sum_{\mu_0,\dots,\mu_{N-1}=1}^{4}
\sum_{a_1,\dots,a_{N-1}}
M^{\mu_0}_{1,a_1}[0]
M^{\mu_1}_{a_1,a_2}[1]
\cdots
M^{\mu_{N-1}}_{a_{N-1},1}[N-1]
|\mu_0\mu_1\cdots\mu_{N-1}\rangle\rangle.
\end{equation}
Here $\mu_i$ is the local Liouville index at site $i$, and the virtual indices $a_k$ have dimensions determined adaptively, subject to a prescribed maximum bond dimension $D_{\max}$.

Additionally, the Liouvillian of Eq.~(\ref{eq:Liouvillian}) is represented as an MPO of fixed bond dimension $(\chi_{\mathrm{MPO}}=10)$ accounting for all onsite, nearest-neighbour, and next-nearest-neighbour terms.

To make the structure explicit, it is useful to define the Liouville-space operator channels
\begin{equation}
S_1=\sigma^-\otimes I,
\qquad
S_2=\sigma^+\otimes I,
\qquad
S_3=I\otimes \sigma^-,
\qquad
S_4=I\otimes \sigma^+,
\end{equation}
together with their corresponding closing operators
\begin{equation}
E_1=+i\,\sigma^+\otimes I,
\qquad
E_2=+i\,\sigma^-\otimes I,
\qquad
E_3=-i\,I\otimes \sigma^+,
\qquad
E_4=-i\,I\otimes \sigma^-.
\end{equation}
The onsite Liouville contribution at site $i$ is
\begin{equation}
\label{eq:Lonsite}
\begin{split}
\mathcal L_i^{\mathrm{on}}
&=
i\Bigl[
\bigl(\Delta_i+\tfrac{i}{2}\zeta_i\bigr)\sigma_i^+\sigma_i^- \otimes I
+\Omega_i^\ast \sigma_i^- \otimes I
+\Omega_i \sigma_i^+ \otimes I
\Bigr]\\
&\quad
-i\Bigl[
I\otimes \bigl(\Delta_i-\tfrac{i}{2}\zeta_i\bigr)\sigma_i^+\sigma_i^-
+I\otimes \Omega_i \sigma_i^-
+I\otimes \Omega_i^\ast \sigma_i^+
\Bigr]
+\zeta_i\, \sigma_i^- \otimes \sigma_i^-.
\end{split}
\end{equation}

With these definitions, the left-boundary MPO tensor can be written schematically as:
\begin{equation}
\label{eq:left_mpo_block}
W^{[0]}
=
\begin{bmatrix}
\mathcal L_0^{\mathrm{on}} &
J_{1,0}^\ast S_1 &
J_{1,0} S_2 &
J_{1,0} S_3 &
J_{1,0}^\ast S_4 &
J_{2,0}^\ast S_1 &
J_{2,0} S_2 &
J_{2,0} S_3 &
J_{2,0}^\ast S_4 &
I\otimes I
\end{bmatrix}.
\end{equation}
For an interior site $i=1,\dots,N-2$, the MPO tensor has the block structure
\begin{equation}
\label{eq:bulk_mpo_block}
W^{[i]}
=
\begin{pmatrix}
I\otimes I & 0 & 0 & 0 & 0 & 0 & 0 & 0 & 0 & 0 \\
E_1 & 0 & 0 & 0 & 0 & 0 & 0 & 0 & 0 & 0 \\
E_2 & 0 & 0 & 0 & 0 & 0 & 0 & 0 & 0 & 0 \\
E_3 & 0 & 0 & 0 & 0 & 0 & 0 & 0 & 0 & 0 \\
E_4 & 0 & 0 & 0 & 0 & 0 & 0 & 0 & 0 & 0 \\
0 & I\otimes I & 0 & 0 & 0 & 0 & 0 & 0 & 0 & 0 \\
0 & 0 & I\otimes I & 0 & 0 & 0 & 0 & 0 & 0 & 0 \\
0 & 0 & 0 & I\otimes I & 0 & 0 & 0 & 0 & 0 & 0 \\
0 & 0 & 0 & 0 & I\otimes I & 0 & 0 & 0 & 0 & 0 \\
\mathcal L_i^{\mathrm{on}} &
J_{1,i}^\ast S_1 &
J_{1,i} S_2 &
J_{1,i} S_3 &
J_{1,i}^\ast S_4 &
J_{2,i}^\ast S_1 &
J_{2,i} S_2 &
J_{2,i} S_3 &
J_{2,i}^\ast S_4 &
I\otimes I
\end{pmatrix}.
\end{equation}
The right-boundary MPO tensor thus closes the chain:
\begin{equation}
\label{eq:right_mpo_block}
W^{[N-1]} =
\begin{bmatrix}
I\otimes I & E_1 & E_2 & E_3 & E_4 & 0 & 0 & 0 & 0 & \mathcal L_{N-1}^{\mathrm{on}}
\end{bmatrix}^T.
\end{equation}

The interpretation of these channels is straightforward. One channel propagates the identity. Four channels encode nearest-neighbour hopping strings. Four further channels encode next-nearest-neighbour hopping strings by allowing the interaction string to remain open for one extra site before being closed. The final channel accumulates onsite and dissipative contributions and carries the MPO to completion. 

\subsection{Dynamical evolution via the time-dependent variational principle in Liouville space}
\label{TDVP}

Our time-evolution scheme is the one-site projector-splitting time-dependent variational principle (TDVP) algorithm for finite
MPS, adapted from Hamiltonian real-time evolution to Liouville-space evolution of a
vectorized density operator; see
Refs.~\cite{Haegeman2011TDVP,Haegeman2016Unifying}
and, for the tangent-space formulation, Ref.~\cite{Haegeman2013TangentSpace}.
The only substantive change is that the state being propagated is the vectorized density
matrix $|\rho\rangle\rangle$, and the generator is the Liouvillian $\mathcal L$ rather
than $-iH$.

The exact equation of motion is
\begin{equation}
\partial_t |\rho(t)\rangle\rangle = \mathcal L |\rho(t)\rangle\rangle,
\end{equation}
where $|\rho\rangle\rangle \in \bigotimes_{i=0}^{N-1}\mathbb C^4$ and the inner product
is the Hilbert--Schmidt inner product
\begin{equation}
\langle\langle X|Y\rangle\rangle = \mathrm{Tr}(X^\dagger Y).
\end{equation}
Restricting the dynamics to the manifold $\mathcal M_D$ of MPS with fixed maximum bond
dimension $D$ gives the TDVP flow
\begin{equation}
\partial_t |\rho[A(t)]\rangle\rangle
=
P_{T_{|\rho[A(t)]\rangle\rangle}\mathcal M_D}\,
\mathcal L |\rho[A(t)]\rangle\rangle,
\label{eq:tdvp_liouville_projected}
\end{equation}
that is, the Liouvillian vector field projected onto the tangent space of the MPS
manifold.

As in the standard finite-size one-site TDVP construction, the MPS is brought into mixed
canonical form with a single orthogonality centre. If the centre is placed at site $i$,
the vectorized density matrix is written as
\begin{equation}
|\rho\rangle\rangle
=
\sum_{\alpha_{i-1},\alpha_i,\mu_i}
A^{[i]\mu_i}_{C,\alpha_{i-1}\alpha_i}\,
|\Phi^{[0:i-1]}_{L,\alpha_{i-1}}\rangle\rangle
\otimes
|\mu_i\rangle\rangle
\otimes
|\Phi^{[i+1:N-1]}_{R,\alpha_i}\rangle\rangle ,
\end{equation}
with orthonormal left and right block states. The tangent-space projector then has the
usual one-site form for finite MPS
\begin{equation}
P_T
=
\sum_{i=0}^{N-1}
P_L^{[0:i-1]}\otimes I_i \otimes P_R^{[i+1:N-1]}
-
\sum_{i=0}^{N-2}
P_L^{[0:i]}\otimes P_R^{[i+1:N-1]},
\label{eq:liouville_tangent_projector}
\end{equation}
where the first term generates the forward evolution of the site-centre tensor and the
second term produces the backward evolution of the bond matrix in the
projector-splitting algorithm
\cite{Haegeman2016Unifying}.

Let $W^{[i]}$ denote the local MPO tensor of the Liouvillian. Given the standard left
and right environments built from the canonical MPS and the Liouvillian MPO, one obtains
at each site an effective one-site Liouvillian
$\mathcal L_{\mathrm{eff}}^{[i]}$ acting on the centre tensor $A_C^{[i]}$, and an
effective bond generator $K_{\mathrm{eff}}^{[i]}$ acting on the bond matrix $C^{[i]}$.
The resulting local equations are the direct Liouville-space analogue of the standard
one-site TDVP equations:
\begin{equation}
\partial_t A_C^{[i]} = \mathcal L_{\mathrm{eff}}^{[i]} A_C^{[i]},
\qquad
\partial_t C^{[i]} = -\,K_{\mathrm{eff}}^{[i]} C^{[i]}.
\label{eq:local_tdvp_liouville}
\end{equation}
Compared with Hamiltonian evolution, the factor of $-i$ is simply replaced by the
Liouvillian generator, while the minus sign in the bond evolution is inherited from the
subtraction term in Eq.~(\ref{eq:liouville_tangent_projector}).

We use the standard symmetric one-site projector-splitting sweep
\cite{Haegeman2016Unifying}. Starting with the orthogonality centre at the left
boundary, one performs a left-to-right half sweep in which each site-centre tensor is
evolved forward by $\Delta t/2$, factorized as $A_C^{[i]}=A_L^{[i]}C^{[i]}$, and the
bond matrix is then evolved backward by $\Delta t/2$ before being absorbed into the next
site. At the right boundary, the final centre tensor is evolved for a full time step
$\Delta t$. One then performs the mirror right-to-left half sweep using the right
canonical factorization $A_C^{[i]}=C^{[i-1]}A_R^{[i]}$. This yields a second-order
integrator for the projected Liouvillian dynamics on the fixed-bond-dimension MPS
manifold.

Some remarks are in order to discuss the differences between the Hamiltonian and Liouvillian cases. Firstly, while trace preservation is exact for the full Liouvillian evolution,
\begin{equation}
\langle\langle I|\mathcal L = 0,
\end{equation}
it need not be preserved exactly after tangent-space projection at finite bond
dimension.
Secondly, because $\mathcal{L}$ is non-Hermitian and the evolution is dissipative, the Hamiltonian-language statements usually associated with TDVP (such as symplecticity, energy conservation, or exact conservation of the MPS norm) no
longer apply in their standard form. Nevertheless, the current scheme still gives a controlled projection of the Liouvillian flow onto the fixed-bond-dimension MPS manifold while preserving the mixed canonical structure throughout the sweep.
Finally, in the present implementation the evolution is a one-site TDVP scheme: once the initial MPS has been prepared, the bond dimensions remain fixed during the propagation, and the numerical accuracy is controlled by the chosen maximum bond dimension $D_{max}$ together with the time step $\Delta t$.

\subsection{Effectiveness of chiral implementation and absorbing boundary conditions}

\begin{figure}
    \centering
    \includegraphics[width=0.4\linewidth]{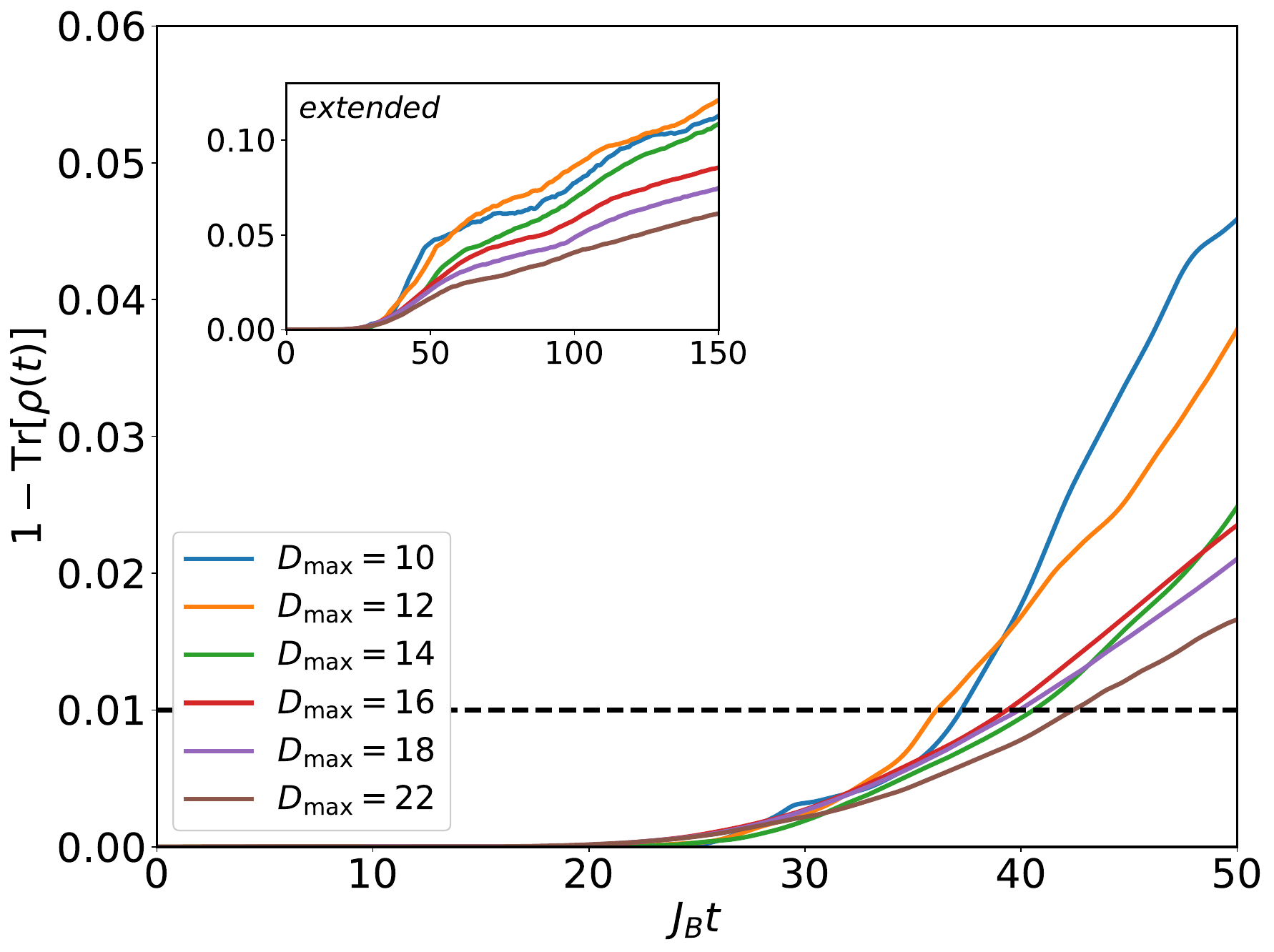}
    \includegraphics[width=0.4\linewidth]{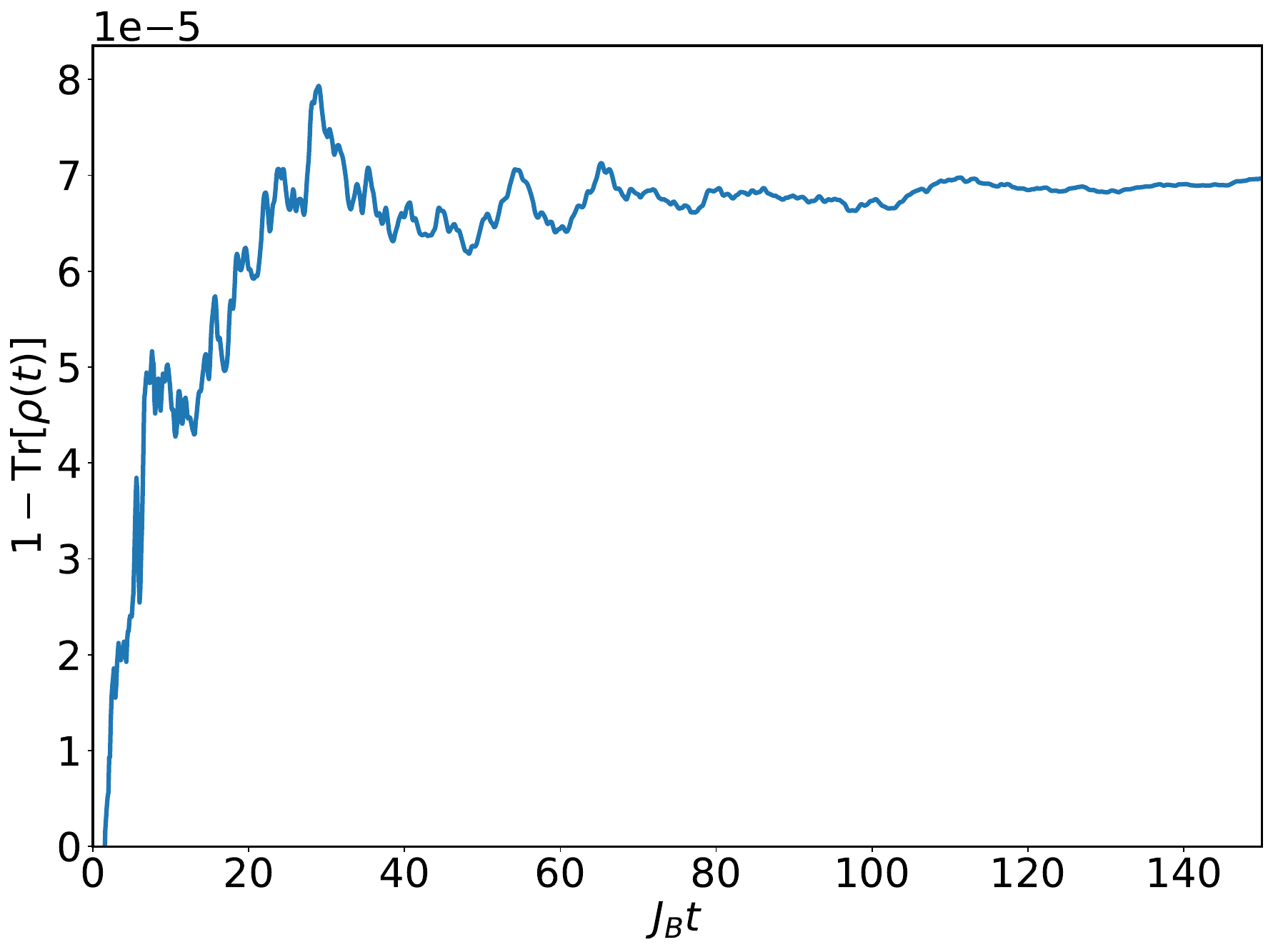}
    \caption{Trace leakage over the course of MPS evolution for the driven case ($\Omega_1 \neq 0$) with initial state $\ket{gg}$ (left) and for the undriven case ($\Omega_1 = 0$) with initial state $\ket{eg}$ at $D_{max}=18$ (right). The inset in (left) follows the same axes as the main plot.}
    \label{fig:figAp1}
\end{figure}

\begin{figure}
    \centering
    \includegraphics[width=0.4\linewidth]{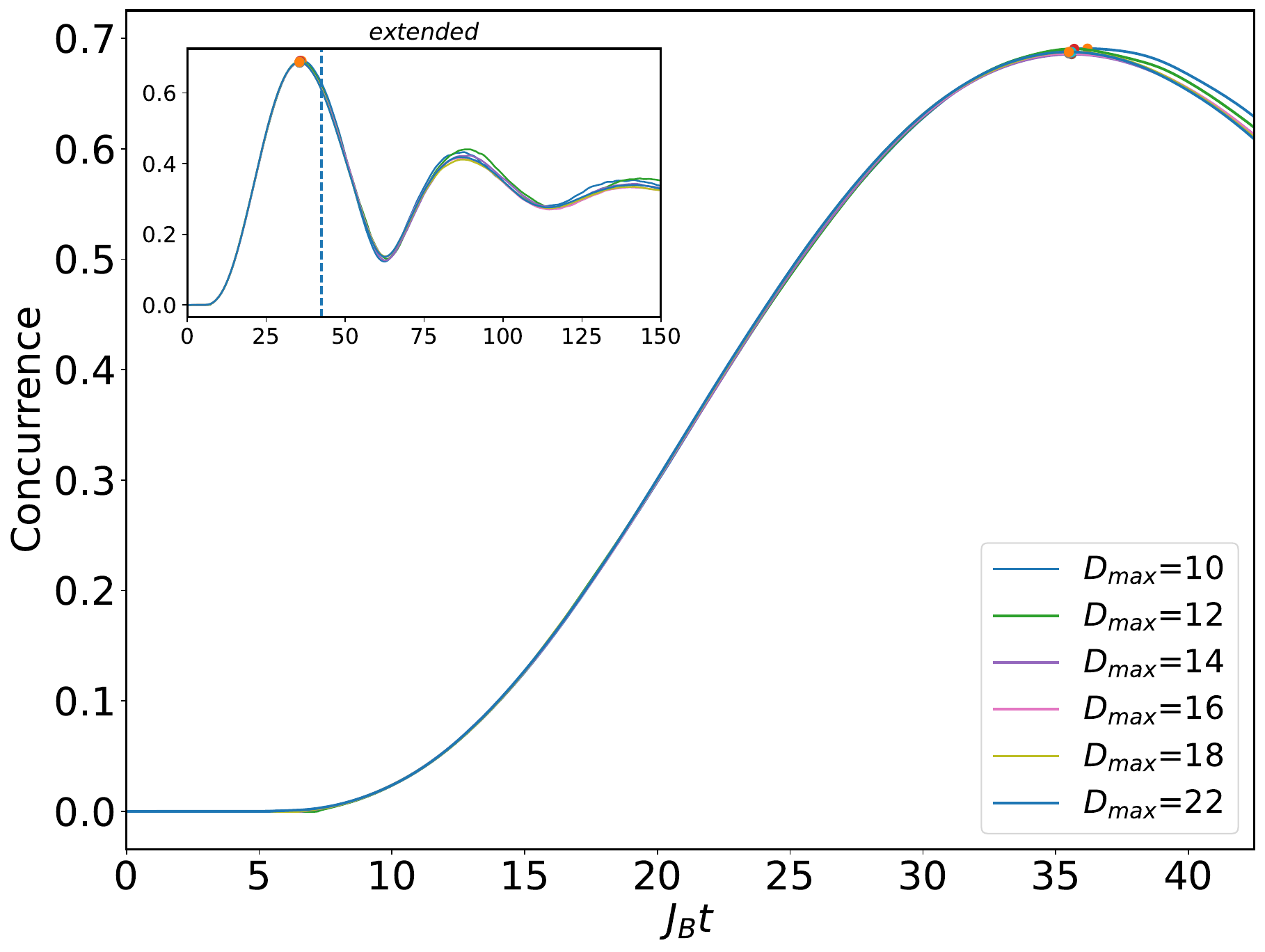}
    \includegraphics[width=0.4\linewidth]{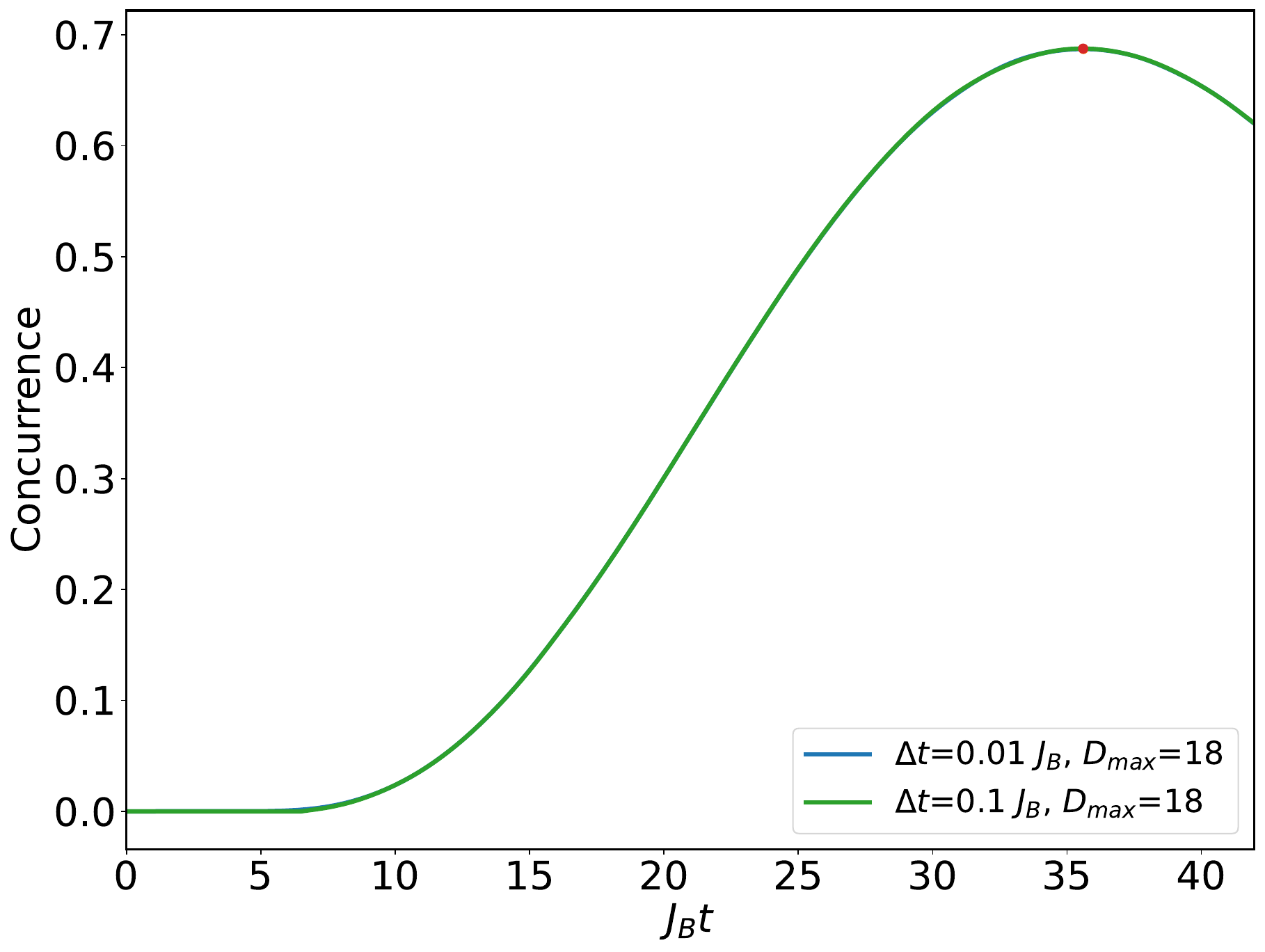}
    \caption{Concurrence over time for different $D_{max}$ values at $\Delta t = 0.1 J_B$ (left) or different $\Delta t$ values at $D_{max} =18$. The inset in (left) follows the same axes as the main plot. The filled points represent the point of maximum concurrence.}
    \label{fig:figApConv}
\end{figure}

For all MPS simulations used in the main matter, we set $D_{max} = 18, \; \Delta t = 0.1 J_B$. To justify the parameter choice, we assign $\{N,n_1,n_2\} =\{16,3,13\}$ and inspect the trace drift in \figref{fig:figAp1} with varying $D_{max}$ and find that trace drift is contained up to $10^{-2}$ within the transient regime up to $t\sim 40 /J_B$ for $D_{max} \gtrsim 14$. However, the trace drift increases linearly with time at steady-state; we find that this is directly due to continuous driving (contrast with negligible trace-drift of the undriven evolution with initial state $\ket{eg}$.) which creates long-lived entanglement within the triangular plaquette manifesting as bound states. Due to computational cost, simulations could not be run with a high enough $D_{max}$ to eliminate trace-drift throughout the entire time-evolution. Nevertheless, we find that concurrence between the emitters converges well with our chosen $D_{max}$ (\figref{fig:figApConv} (left) and Table \ref{tab:tabApConv}) up to the first transient peak, while fine-graining $\Delta t$ has insignificant effect on convergence (\figref{fig:figApConv} (right)). This thus demonstrates the physical reliability of our simulations in the transient regime.

\begin{table}[]
    \centering
    \begin{tabular}{|c|c|c|}
    \hline
         $D_{max}$ & $t_{peak}$ & $C_{peak}$ \\
         \hline
         10 & 36.2 & 0.6907  \\
         12 & 35.7 & 0.6904  \\
         14 & 35.6 & 0.6856  \\
         16 & 35.5 & 0.6863  \\
         18 & 35.6 & 0.6876  \\
         22 & 35.5 & 0.6876  \\
         \hline
    \end{tabular}
    \caption{Details of maximum concurrence points for \figref{fig:figApConv} (left).}
    \label{tab:tabApConv}
\end{table}

\begin{figure}
    \centering
    \includegraphics[width=0.4\linewidth]{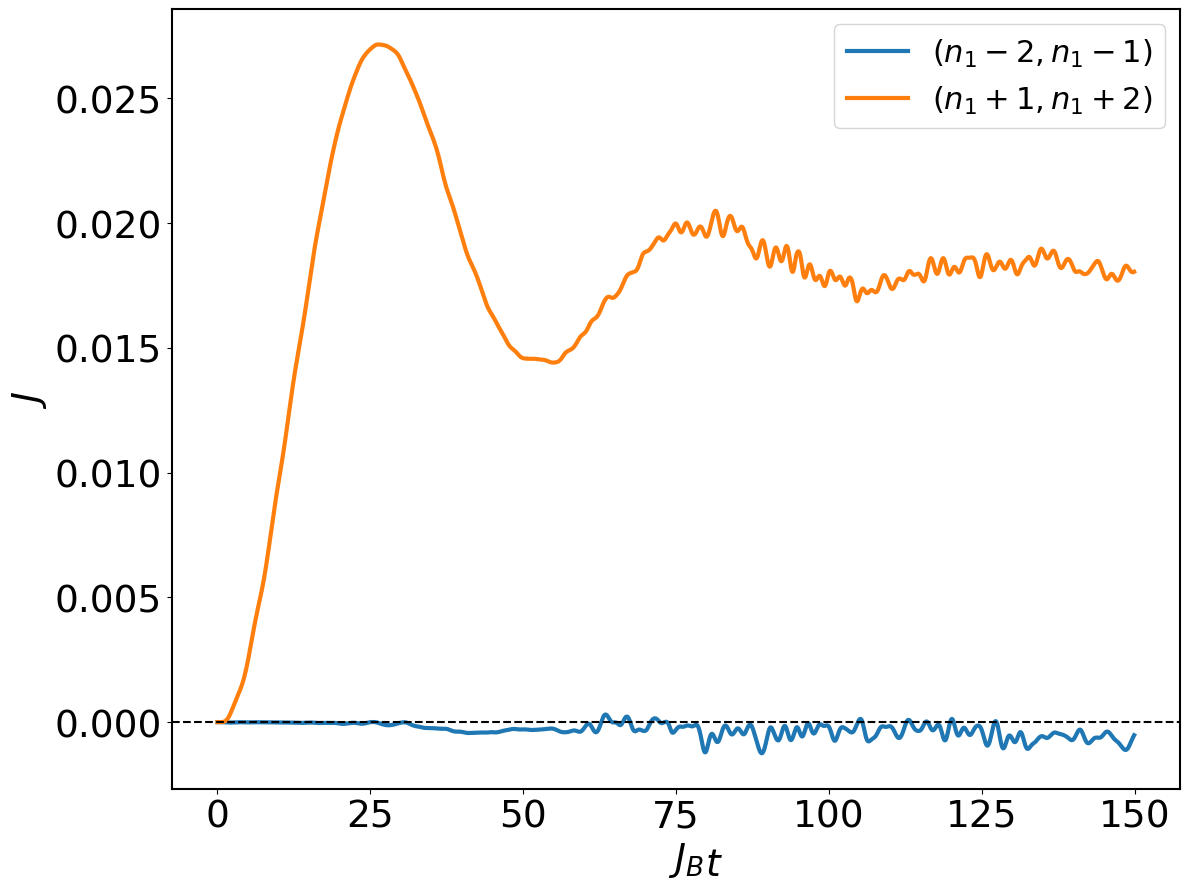}
    \includegraphics[width=0.4\linewidth]{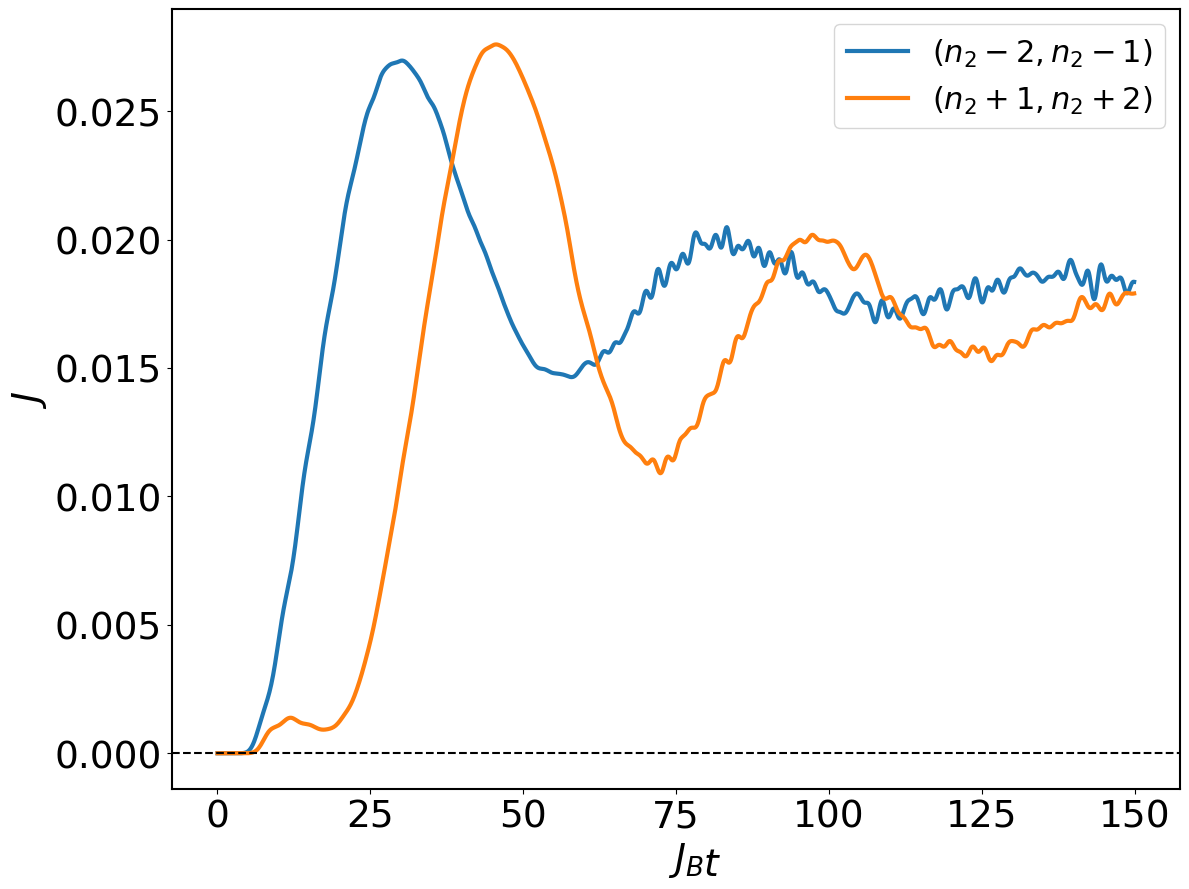}
    \caption{Bond current in the vicinity of $n_1$ (left) and $n_2$ (right).}
    \label{fig:figAp2}
\end{figure}

\begin{figure}
    \centering
    \includegraphics[width=0.6\linewidth]{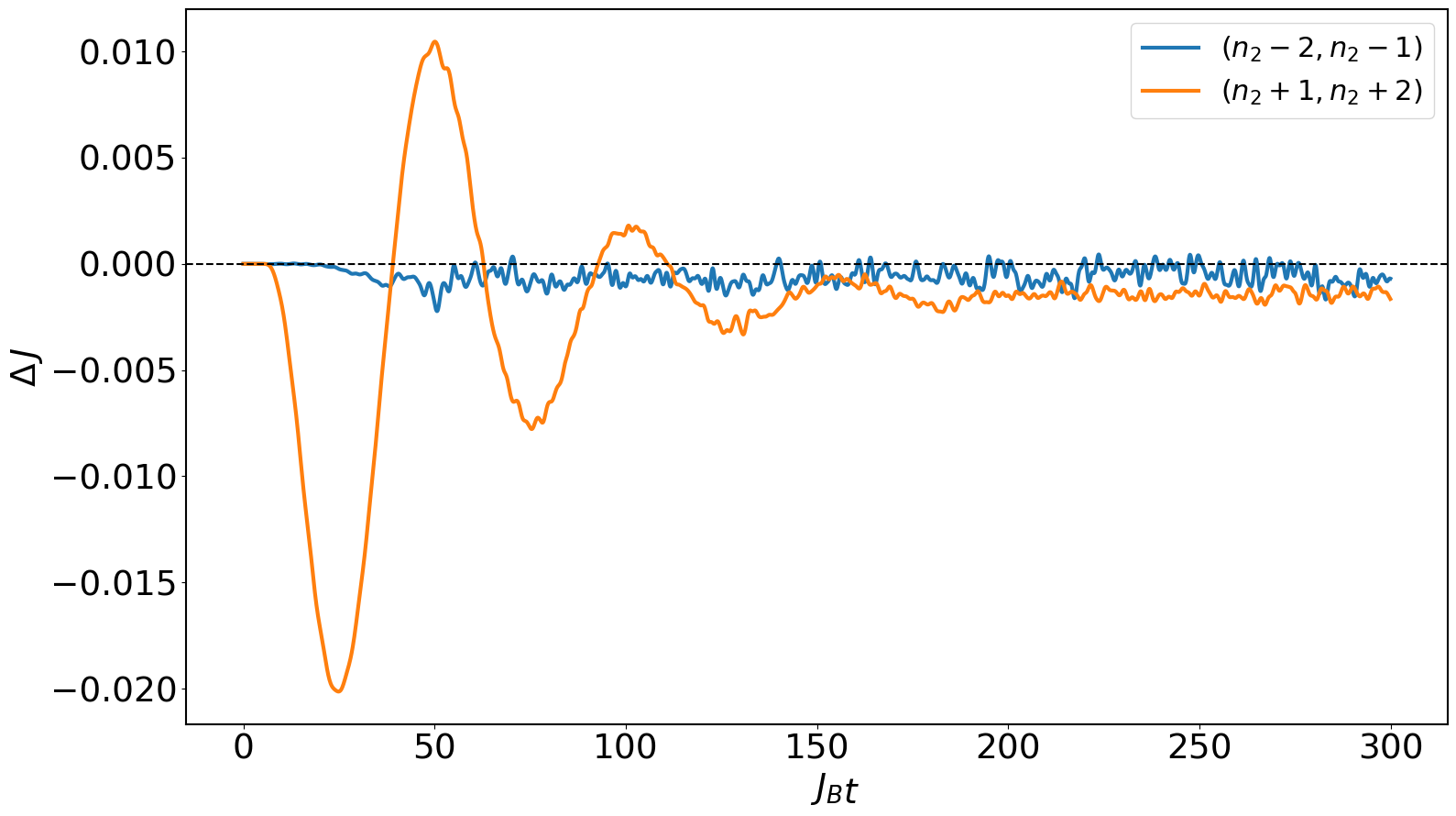}
    \caption{Bond current difference between $g_2 = 0.30$ and $g_2 = 0$ systems in the vicinity of $n_2$.}
    \label{fig:figAp3}
\end{figure}

To determine whether the model remains fully chiral under the MPS approximation regime, we first probe the outer regions of the chain by calculating the outgoing fluxes in the vicinity of the emitters, defined by the bond current on $(i,i+1)$: $J_{i \rightarrow i+1}(t) \coloneq 2\mathrm{Im}[J_B \langle \sigma_i^+ \sigma_{i+1}^- \rangle]$. As shown in \figref{fig:figAp2}, the current around $n_1$ is highly unidirectional with $J_{n_1 -2 \rightarrow n_1 -1}(t) \sim 0$, while the current around $n_2$ is overall right-moving and is still inconclusive. We then probe the emergence of back-reflections from the downstream emitter $n_2$, comparing the flux difference between an evolution with and an evolution without $n_2$ (done by setting $g_2=0$), i.e. $\Delta J_{i \rightarrow i+1}(t) \coloneq J_{i \rightarrow i+1}(t) - J_{i \rightarrow i+1}^{g_2 = 0}(t)$; we find a lack of significant back-reflection beyond numerical noise in \figref{fig:figAp3}. The results thus verify that the system can be taken as fully chiral.

\end{document}